\documentclass[10pt,a4paper,reqno]{amsart}
\usepackage{latexsym}
\usepackage{amssymb,amsmath,amsthm}
\usepackage{a4wide}
\usepackage{verbatim}

\newtheorem{theorem}{Theorem}
\newtheorem{proposition}[theorem]{Proposition}
\newtheorem{lemma}[theorem]{Lemma}
\newtheorem{claim}[theorem]{Claim}

\theoremstyle{definition}\newtheorem{definition}{Definition}[section]
\theoremstyle{remark}\newtheorem{remark}{Remark}[section]

\usepackage{graphicx}
\usepackage{amsmath, amsxtra, amsfonts}
\usepackage{color}
\usepackage{eucal, url}
\usepackage{mdwtab}
\usepackage{algorithmic}
\usepackage{amssymb}

\newcounter{exampleno}
\newenvironment{example}{\begin{small}\refstepcounter{exampleno}%
\smallbreak\noindent%
{\sc Example~\theexampleno.} }%
{\dotfill\hbox{~$\square$}\end{small}\smallbreak}

\renewcommand{\Phi}{G}

\def\cRp{\mathcal{R}^{\circ}}
\newcommand{\cacher}[1]{}

\def\Bern{\mathrm{Bern}}
\newcommand{\cA}{\mathcal{A}}
\newcommand{\cX}{\mathcal{X}}
\newcommand{\cY}{\mathcal{Y}}
\newcommand{\wh}{\widehat}
\newcommand{\cE}{\mathcal{E}}
\newcommand{\cM}{\mathcal{M}}

\newcommand{\cP}{\mathcal{P}}

\newcommand{\cQ}{\mathcal{Q}}
\newcommand{\cT}{\mathcal{T}}
\newcommand{\cF}{\mathcal{F}}
\newcommand{\cFp}{\mathcal{F}^{\circ}}
\newcommand{\cL}{\mathcal{L}}
\newcommand{\cMv}{\mathcal{M}_{\mathrm{v}}}
\newcommand{\cMe}{\mathcal{M}_{\mathrm{e}}}
\newcommand{\cFv}{\mathcal{F}_{\mathrm{v}}}
\newcommand{\cFe}{\mathcal{F}_{\mathrm{e}}}
\newcommand{\cFo}{\mathcal{F}_{\Omega}}
\newcommand{\cFop}{\mathcal{F}_{\Omega}^{\circ}}
\newcommand{\cFos}{\mathcal{F}_{\Omega}^{\cst}}
\newcommand{\cEo}{\mathcal{E}_{\Omega}}
\newcommand{\cEop}{\mathcal{E}_{\Omega}^{\circ}}
\newcommand{\cEos}{\mathcal{E}_{\Omega}^{\cst}}

\newcommand{\vbar}{(s_i,t_i)_{i\geq 1}}

\def\cAp{\mathcal{A}^{\circ}}
\def\cAo{\mathcal{A}_{\Omega}}
\def\cAop{\mathcal{A}_{\Omega}^{\circ}}
\def\cR{\mathcal{R}}
\def\cRo{\mathcal{R}_{\Omega}}
\newcommand{\cRop}{\mathcal{R}_{\Omega}^{\circ}}
\def\wt{\widetilde}
\def\wh{\widehat}
\def\Geom{\mathrm{Geom}}
\def\Pois{\mathrm{Pois}}
\def\Loga{\mathrm{Loga}}
\def\vecc{(s_1,t_1;s_2,t_2;\ldots)}
\newcommand{\cB}{\mathcal{B}}
\newcommand{\cEp}{\mathcal{E}^{\circ}}
\newcommand{\cEs}{\mathcal{E}^{\cst}}
\newcommand{\cBp}{\mathcal{B}^{\circ}}

\def\Set{\textsc{Set}}
\def\Seq{\textsc{Seq}}
\def\Cyc{\textsc{Cyc}}
\def\Setk{\textsc{Set}^{[k]}}
\def\Seqk{\textsc{Seq}^{[k]}}
\def\Cyck{\textsc{Cyc}^{[k]}}

\def\oZ{\overline{Z}}
\def\ds{\displaystyle}
\def\Seto{\Set_{\Omega}}
\def\Cyco{\Cyc_{\Omega}}
\def\Setop{\Set_{\Omega}^{\circ}}
\def\Cycos{\Cyc_{\Omega}^{\cst}}

\def\Setos{\Set_{\Omega}^{\cst}}
\def\fomeg{f_{\Omega}}
\def\romeg{r_{\Omega}}
\def\eomeg{e_{\Omega}}

\def\cGs{\cG^{\cst}}
\def\cBs{\cB^{\cst}}
\def\Mell{M_{(\ell)}}
\def\Bell{B_{(\ell)}}
\def\cMs{\cM^{\cst}}
\def\cH{\mathcal{H}}
\def\cHp{\mathcal{H}^{\circ}}
\def\cK{\mathcal{K}}
\def\cKp{\mathcal{K}^{\circ}}
\def\cGsv{(\mathcal{G}^{\cst})_{\mathrm{v}}}
\def\cGsB{(\mathcal{G}^{\cst})_{\mathrm{B}}}
\def\cGv{\mathcal{G}_{\mathrm{v}}}
\def\cGB{\mathcal{G}_{\mathrm{B}}}
\def\oS{\overline{S}}
\def\oeta{\overline{\eta}}

\def\opi{\overline{\pi}}

\def\cFs{\cF^{\cst}}

\def\cFvs{\cFv^{\cst}}
\def\cFes{\cFe^{\cst}}

\def\cMvs{\cMv^{\cst}}

\def\cMes{\cMe^{\cst}}
\def\cMep{\cMe^{\circ}}

\def\Seqominusone{\Seq_{\Omega-1}^{\circ}}

\newcommand\cst{{\circledast}}

\def\Sym{\mathrm{Sym}}
\def\RSym{\mathrm{RSym}}

\newcommand{\bcdot}{\boldsymbol{\cdot}}

\def\cF{\mathcal{F}}

\newcommand{\cC}{\mathcal{C}}

\def\cS{\mathcal{S}}

\newcommand{\cG}{\mathcal{G}}
\newcommand{\cGp}{\mathcal{G}^{\circ}}

\def\cZ{\mathcal{Z}}
\def\cX{\mathcal{X}}

\newcommand{\margin}[1]{\marginpar{\footnotesize #1}}

\newcommand{\ignore}[1]{}

\begin{document}
\title{Boltzmann Samplers, P\'olya Theory, and Cycle Pointing}

\author{Manuel Bodirsky}
\address{Manuel Bodirsky:  LIX (CNRS UMR 7161), \'Ecole Polytechnique, 92128 Palaiseau, 
France.}

\author{\'Eric Fusy}
\address{\'Eric Fusy: LIX (CNRS UMR 7161), \'Ecole Polytechnique, 92128 Palaiseau, 
France.}

\author{Mihyun Kang}
\address{Mihyun Kang: Institut f\"ur Mathematik, Technische Universit\"at Berlin, Stra{\ss}e des 17. Juni 136,
D-10623 Berlin, Germany.}

\author{Stefan Vigerske}
\address{Stefan Vigerske: Institut f\"ur Mathematik, Humboldt-Universit\"at zu Berlin, Unter den Linden 6, 10099 Berlin, Germany.}


\begin{abstract}
We introduce a general method 
to count unlabeled combinatorial
structures and to efficiently generate them at random. 
The approach is based on pointing
unlabeled structures in an ``unbiased'' way that a structure of size $n$ gives rise to $n$ pointed structures. 
We extend P\'olya theory to the corresponding pointing operator, and 
present a random sampling framework based on both the principles of Boltzmann sampling and on P\'olya operators. 
All previously known unlabeled construction principles for 
Boltzmann samplers are special cases of our new results.
Our method is illustrated on several examples: in each case, we
provide enumerative results and efficient random samplers.  
The approach applies to 
unlabeled families of plane and nonplane unrooted
trees, and tree-like structures in general, but
also to families of graphs (such as cacti graphs and outerplanar graphs)
and families of planar maps. 

\emph{This is the extended and revised journal version of a conference
paper with the title ``An unbiased pointing operator for unlabeled structures, with applications to counting and sampling'', which appeared in the Proceedings of ACM-SIAM Symposium on Discrete Algorithms (SODA'07), 7-9 January 2007, New Orleans.}
\end{abstract}

\maketitle

\let\languagename\relax

\maketitle
\section{Introduction}
\label{sect:introduction}
Pointing (or rooting) is an important tool to derive
decompositions of combinatorial structures, with many applications in 
enumerative combinatorics. Such decompositions can for instance 
be used in polynomial-time algorithms that sample structures 
of a combinatorial class uniformly at random.
For a class of \emph{labeled} structures, pointing corresponds to
taking the derivative of the (typically exponential) 
generating function of the class. In other words, each structure of size $n$ gives rise to $n$ pointed (or rooted)
structures. Other important operations on classes of combinatorial structures 
are the disjoint union, the product, and the substitution operation 
-- they correspond
to addition, multiplication, and composition of the associated
generating functions. Together with the usual basic classes 
of combinatorial structures (finite classes, the class of finite sets, the class of finite sequences, and the class of cycles),
this collection of constructions is a powerful device to define a great variety of combinatorial families.

If a class of structures can be described by recursive specifications
involving pointing, 
disjoint unions, products, substitutions, and the basic classes,
then the techniques of analytic combinatorics can be applied to obtain
enumerative results, to study statistical properties of random structures in the
class, and to derive efficient random samplers.
An expository account for this line of research is~\cite{Flajolet}.
Among recent developments in the area of random sampling are
\emph{Boltzmann samplers}~\cite{Boltzmann}, which are 
an attractive alternative to the recursive method of sampling~\cite{NijenhuisWilf,Recmethod}.
Both approaches provide in a systematic way
polynomial-time uniform random generators for decomposable combinatorial classes.
The advantage of Boltzmann samplers over the 
recursive method of sampling 
is that Boltzmann samplers operate in linear time if a small relative tolerance is allowed for the size
of the output, and that they have a small
preprocessing cost, which makes it possible to sample very large structures.

A third general random sampling approach is based on Markov chains. 
This approach does not require recursive decompositions of structures, and  is applicable on a wide range of combinatorial classes. However, Markov chain methods are mostly limited to \emph{approximate} uniformity.
Moreover, it is usually difficult to obtain bounds on the rate of convergence to the uniform distribution~\cite{JeSi}. 

All the results in this paper
concern classes of \emph{unlabeled} combinatorial structures, 
i.e., the structures are considered 
up to isomorphism. 
In the case of the class of all graphs, the labeled and the unlabeled
model do not differ much, which is due to the fact that almost
all graphs do not have a non-trivial automorphism (see e.g.,~\cite[Ch.~9.]{Bollobas} and~\cite{HNBook}). 
However, for many interesting classes of combinatorial structures
(and for most of the classes studied in this paper), 
the difference between the labeled and the unlabeled setting does matter. 

The Markov chain approach can be adapted also to sample
unlabeled structures approximately uniformly at random, 
based on the orbit counting lemma~\cite{Je}. Again, the rate 
of convergence is usually very difficult to analyze, and frequently
these Markov chains do not have a polynomial convergence rate~\cite{GoldbergJerrum}.
For unlabeled structures, the approach faces additional difficulties:
to computationally implement the transitions of the Markov chain,
we have to be able to generate structures with a given symmetry
uniformly at random (this is for instance a difficult task for planar graphs), and we have to be able to generate a symmetry of
a given structure (for the class of all finite graphs, this task is computationally equivalent to the graph isomorphism problem).

To enumerate unlabeled structures, typically
the \emph{ordinary} generating functions are the appropriate tool. 
Disjoint unions and products for unlabeled structures then  still correspond to
addition and multiplication of the associated generating functions as for labeled cases. 
Boltzmann samplers for classes described by 
recursive specifications involving
these operations have recently been developed in~\cite{BoltzmannUnlabeled}.
However, 
the \emph{substitution} operation for unlabeled structures no longer corresponds to 
the composition of generating functions, due to the symmetries an unlabeled
structure might have. This problem can be solved by 
P\'olya theory, which uses the generalization of generating functions to \emph{cycle index sums} to take care of potential symmetries.
P\'olya theory provides a computation rule for the cycle index sum associated to a substitution construction.
The presence of symmetries leads to another problem with the \emph{pointing} (or \emph{rooting}) operator: 
the fundamental property
that a structure of size $n$ gives rise to $n$ pointed structures does not hold. Indeed, if a structure of size $n$
has a non-trivial automorphism, 
then it corresponds to less than $n$ pointed structures 
(because pointing at two vertices in symmetric positions
produces the same pointed structure). Thus, for unlabeled structures, the classical pointing operator does not correspond to the derivative of the ordinary generating function.

In this paper, we introduce an \emph{unbiased} pointing operator for unlabeled structures. The operator is unbiased in the sense that a structure of size $n$ gives rise to $n$ pointed structures. It produces
\emph{cycle-pointed} structures, i.e., combinatorial structures $A$ together with a marked cyclic sequence of atoms of $A$ that is a cycle of an automorphism of $A$. Accordingly, we call our operator \emph{cycle-pointing}. 
The idea is based on 
Parker's lemma in permutation group theory~\cite{Ca} (this will be discussed in Remark~\ref{rem:parker}).
We develop techniques to apply this new
pointing operator to enumeration and random sampling of unlabeled
combinatorial classes. The crucial point is that cycle-pointing is \emph{unbiased}. As a consequence,
performing both tasks of enumeration and uniform random sampling on a combinatorial class
is equivalent to performing these tasks on the associated cycle-pointed class. 

To understand how
we use our operator, it is instructive to look at the class
of \emph{free trees}, i.e., unrooted and nonplane \emph{trees} (equivalently, acyclic connected graphs). 
Building on the work of Cayley and P\'olya, Otter~\cite{Otter} 
determined the exact and asymptotic number of free trees. To this end, he
developed the by-now standard dissimilarity characteristic equation,
which relates the number of free trees with the number of rooted nonplane trees; 
see~\cite{HP73}. The best-known method to sample free trees uniformly at random is due to Wilf~\cite{Wilf}, and uses the concept of the centroid of a tree. 
The method is an example of application of the recursive method of sampling 
and requires a pre-processing step where
a table of quadratic size is computed.

Cycle-pointing provides a new way to count 
and sample free trees.
Both tasks are carried out on cycle-pointed (nonplane) trees.
 The advantage of studying cycle-pointed trees is that 
the pointed cycle provides a starting point
for a recursive decomposition. In the case
of cycle-pointed trees, we can formulate such a decomposition
using standard constructions such as disjoint union, product, and substitution (which 
requires to suitably adapt these constructions to the cycle-pointed framework). 
We want to stress that, despite some superficial similarities,
this method for counting free trees is fundamentally different from the previously existing methods mentioned above, and it proves particulary fruitful in the context of random generation. 
Indeed, the \emph{dissimilarity
characteristic equation}~\cite{Otter} and the \emph{dissymmetry theorem}~\cite{BeLaLe} 
 both  lead 
to generating function equations involving \emph{subtraction}.
However, subtraction yields massive rejection 
when translated into a random generator for the class of structures 
(both for Boltzmann samplers and for the recursive method).
In contrast, the equations produced by the method based on cycle-pointing 
have only positive signs, and the existence of a Boltzmann sampler
for  (cycle-pointed) free trees, with no rejection involved, will follow directly from the general results derived in
this paper. As usual, 
the Boltzmann samplers we obtain have a running time that is linear in
the size of the structure generated, and have small pre-processing cost.

Similarly, we can decompose plane and nonplane trees, 
and more generally all sorts of tree-like structures. 
By the observation that the block decomposition of a graph
has also a tree-like structure, we can apply the method to classes
of graphs where the two-connected components can be explicitly enumerated.
This leads to efficient Boltzmann samplers, for instance, for cacti graphs and outerplanar graphs,
improving on the generators of~\cite{BoKa}.
Further, our strategy is not limited to only tree-like structures, but can also be applied to other classes of structures that allow for a recursive decomposition. To demonstrate this, we sketch how the method can be applied to count and sample certain classes of planar maps.

\subsection*{Outline of the paper}
To formalize our general results on enumeration and sampling 
for classes of unlabeled combinatorial structures in full generality,
we apply the concept of \emph{combinatorial species}; we 
recall this concept in Section~\ref{sect:prelims}. 
In Section~\ref{sec:polya_cycle_pointed} we introduce
\emph{cycle-pointed species} and the cycle-pointing operator.
Section~\ref{sec:appl} is devoted to applications of our
cycle-pointing operator in enumeration. The technique in 
which we use the operator to obtain recursive decomposition
strategies for unlabeled enumeration is very generally applicable; 
we illustrate it by the enumeration of (unrooted) non-plane  and plane trees, cacti graphs, and maps.
Finally, in Section~\ref{sec:sampler}, we present how to
apply the concepts introduced in this paper to obtain highly efficient
random sampling procedures for unlabeled structures. This
is illustrated by applications for sampling of concrete and fundamental classes 
of unlabeled combinatorial structures; several of these 
concrete 
sampling results are either new or improve the state-of-the-art of
sampling efficiency.


\ignore{
Applications:
\begin{enumerate}
\item plane and nonplane unrooted trees
\item all families of plane and nonplane unrooted trees with restriction on the degree of the nodes (binary, ternary, etc...) there we can sell ternary trees saying that they correspond to alcohols (cf Polya and the book by Harary Palmer)
\item cactus graphs (2-connected components are cycles)
\item outerplanar graphs
\item RNA secondary structures (good to sell, they are biologic structures), I am not completely sure (there should be a definition somewhere I hope) they look like cactus graphs but the blocks are connected by "chains" instead of separating vertices
\item Things that can be decomposed recursively
- it would be nice if we can include the discussion of maps
\end{enumerate}
}

\section{Preliminaries}\label{sect:prelims}
In this paper we work with classes of \emph{combinatorial structures}, such as graphs, relational structures, functions, trees, plane trees, maps, words, terms, or permutations. There are many ways to define formally these objects; for example, rooted trees can be coded  as special types of directed graphs, but also as terms. 
Which formal representation of a combinatorial structure is best usually depends on the application.

When we are interested in combinatorial enumeration,
the differences in representation are not essential; in the example
above, we have the same number of rooted trees, no matter how
they are represented. We would thus like to have
a formalism that is sufficiently abstract so that our results apply
to broad classes of combinatorial structures. 
At the same time, we would like
to have a formalism for classes of combinatorial structures that supports fundamental construction operations for combinatorial classes, such as formation of \emph{disjoint unions}, \emph{products}, and \emph{substitution}, and allows to state general results about
enumeration and random sampling.

The theory of combinatorial species is an elegant tool that
fully satisfies the needs mentioned above. We give a brief introduction to species, and refer to Bergeron, Labelle, and Leroux~\cite{BeLaLe} for a broader treatment of the topic. The fundamental and well-known classes of combinatorial structures
that we treat in Section~\ref{sec:appl} provide ample illustration of the concepts we define in this section.



\subsection{Combinatorial Species}
We closely follow the presentation in~\cite{BeLaLe}.
A \emph{species of structures} is a functor $\cA$ from finite sets $U$
to finite sets $\cA[U]$, together with a rule that produces
for each bijection $\sigma: U \rightarrow V$ a function from $\cA[U]$ to $\cA[V]$. Slightly abusing notation, this
function will also be denoted by $\cA[\sigma]$.
The functions $\cA[\sigma]$ must satisfy the following two (functorial) properties:
\begin{itemize}
\item for all bijections $\sigma: U \rightarrow V$ and $\tau: V \rightarrow W$, 
$$ \cA[\tau \circ \sigma] = \cA[\tau] \circ \cA[\sigma] \; ;$$
\item for the identity map $\text{Id}_U: U \rightarrow U$, 
$$\cA[\text{Id}_U]=\text{Id}_{\cA[U]} \; .$$
\end{itemize}
An element $A \in \cA[U]$ is called an \emph{$\cA$-structure on $U$}.
The elements of $U$ are called the \emph{atoms
of $\cA[U]$}, and we refer to $|U|$ as the \emph{size} of $A$. The function $\cA[\sigma]$ is called the \emph{transport} of $\cA$-structures
along $\sigma$. We write $\sigma \bcdot A$ for $\cA[\sigma](A)$.

\begin{definition}
Let $A_1 \in \cA[U]$ and $A_2 \in \cA[V]$ be two $A$-structures.
A bijection $\sigma: U \rightarrow V$ is called an \emph{isomorphism}
from $A_1$ to $A_2$ if $A_2 = \sigma \bcdot A_1 = \cA[\sigma](A_1)$.
An isomorphism from $A_1$ to $A_1$ is called an \emph{automorphism} of $A_1$.
\end{definition}

The advantage of species is that the rule $\cA$ that produces the structures $\cA[U]$ and the transport functions $\cA[\sigma]$ can be described
in any way; the book~\cite{BeLaLe} gives instructive examples
where this description is by axiomatic systems, explicit constructions,
algorithms, combinatorial operations, or functional equations.

\begin{definition}
A species $\cA$ is a \emph{subspecies} of a species $\cB$ (and we write $\cA \subseteq \cB$) if it satisfies the following two conditions:
\begin{itemize}
\item for any finite set $U$, $\cA[U] \subseteq \cB[U]$;
\item for any $\sigma: U \rightarrow V$ we have $\cA[\sigma] = \cB[\sigma] |_{\cB[U]}$.
\end{itemize}
\end{definition}

\subsection{Enumeration} \label{sub:prelim_enum}
For all finite sets $U$, the number of $\cA$-structures on $U$ depends only on the number of elements of $U$ (and not on the elements of $U$). If $\cA$ is a species, we write $a_n$ for $|\cA[\{1,\dots,n\}]|$,
the cardinality of the set of $\cA$-structures on $\{1,\dots,n\}$.
The series 
\begin{equation}
\cA(x):=\sum_{n\geq 0}\frac{1}{n!}a_n x^n
\end{equation}
 is called the \emph{exponential generating series} of the species $\cA$. 
 
In this article we focus on \emph{unlabeled} structures, i.e., we consider structures up to isomorphism. 
We may restrict
ourselves to structures on sets of the form $U=[1..n]:=\{1,\dots,n\}$,
and write $\cA[n]$ for $\cA[\{1,\dots,n\}]$. 
We define the equivalence relation $\sim$ on $\cA[n]$ by setting $A_1 \sim A_2$, for 
$A_1,A_2 \in \cA[n]$, if and only if there is an isomorphism $\sigma: [1..n] \rightarrow [1..n]$ from $A_1$ to $A_2$. We also say that $A_1$ and $A_2$ have the
same \emph{isomorphism type}, and the equivalence classes
of $\cA$-structures on $[1..n]$ with respect to $\sim$ are also called \emph{unlabeled $\cA$-structures of size $n$},
The set of all those equivalence
classes is denoted by $\widetilde{\cA}_n$,
and we write $\widetilde{a}_n$ for 
its cardinality $|\widetilde{\cA}_n|$.
The series
\begin{equation}
\widetilde{\cA}(x):=\sum_{n\geq 0} \widetilde{a}_n x^n
\end{equation}
 is called the \emph{ordinary generating series (OGS)} of $\cA$.
We use the classical notation $[x^n]\widetilde{\cA}(x)$ to denote the  $n$-th 
coefficient $\widetilde{a}_n$ in the power series $\widetilde{\cA}(x)$.

\subsection{Cycle index sums}\label{ssect:cycle-index}
For a species $\cA$ and each  $n \geq 0$, 
a \emph{symmetry of $\cA$ of size $n$} 
is a pair $(A, \sigma)$ where $A$ is from $\cA[n]$ and $\sigma$ is
an automorphism of $A$. We call $A$ the \emph{underlying structure} of the symmetry $(A, \sigma)$.
Notice that the automorphism $\sigma$ can be the identity.
We denote by $\Sym(\cA)$ the species defined by 
$$\Sym(\cA)[n] := \{(A,\sigma) \; | \; A \in \cA[n] \text{ and $\sigma$ is an automorphism of $A$} \} \; .$$ 
The definition of the transport of $\Sym(\cA)$ is obvious
(the definition of symmetries as an auxiliary species is standard; see Section 4.3 in ~\cite{BeLaLe}).
The \emph{weight-monomial} of a symmetry $(A,\sigma)$ of size $n$ is defined as
\begin{equation}
\label{eq:def_weight_unlabeled}
w_{(A,\sigma)}:=\frac{1}{n!}\prod_{i=1}^{n}s_i^{c_i(\sigma)},
\end{equation}
where, for $i\in[1..n]$,  $s_i$ is a 
 formal variable and $c_i(\sigma)$ is the number of cycles of $\sigma$ 
of length $i$.
For simplicity, in the following we will write $c_i$ for $c_i(\sigma)$ if the corresponding automorphism is clear from the context.
The \emph{cycle index sum} of $\cA$, 
denoted by $Z_{\cA}(s_1,s_2,\ldots)$, or 
shortly $Z_{\cA}$, is the formal power series defined
as the sum of the weight-monomials over all the symmetries of $\cA$,
\begin{equation}
Z_{\cA}(s_1,s_2,\ldots):= \sum_{n\geq 0} \left( \sum_{(A,\sigma)\in\Sym(\cA)[n]}w_{(A,\sigma)} \right) .
\end{equation}
Cycle index sums for classes of combinatorial structures have been introduced by P\'olya~\cite{Polya-First}. 
The following fact, which is based
on Burnside's lemma, shows that cycle index sums are a refinement of  ordinary generating series.

\begin{lemma}[P\'olya]
\label{lem:unlabel}
Let $\cA$ be a species of structures. For $n\geq 0$,
 each unlabeled structure $\tilde{A}\in\wt{\cA}_n$ gives rise to $n!$ symmetries, i.e., 
there are $n!$ symmetries $(A,\sigma)$ such that $A \in \tilde{A}$. Hence,
\begin{equation}
\label{eq:spec}
Z_{\cA}(x,x^2,x^3,\ldots)=\wt{\cA}(x).
\end{equation}
\end{lemma}

In the proofs of combinatorial identities it will be convenient to identify species that
are essentially the same. 

\begin{definition} Let $\cA$ and $\cB$ be two species.
An \emph{isomorphism}
from $\cA$ to $\cB$ is a family of bijections $\alpha_U: \cA[U] \rightarrow \cB[U]$ which satisfies the following \emph{naturality condition}:
for any bijection $\sigma: U \rightarrow V$ between two finite sets
$U$ and $V$, and for any $\cA$-structure $A$ on $U$, one must have $$\sigma \bcdot \alpha_U(A) = \alpha_V(\sigma \bcdot A)\; .$$
\end{definition}

It is known (see~\cite{BeLaLe}) 
that when $\cA$ is isomorphic to $\cB$, then $\cA(x)=\cB(x)$, 
$\widetilde \cA(x) = \widetilde \cB(x)$, 
and $Z_\cA(x_1,x_2,\dots) = Z_\cB(x_1,x_2,\dots)$. Hence, for the purposes of combinatorial enumeration we can even identify isomorphic species,
and as in~\cite{BeLaLe} we write $\cA = \cB$ when $\cA$ and $\cB$
are isomorphic species.




\subsection{Basic species and combinatorial constructions}
\label{sub:dictionary_polyaunlabeled}
We now recall a collection of basic species and combinatorial constructions.
We start with the description of some
basic species, and then introduce the three fundamental constructions
of disjoint union, product, and substitution. 

\begin{figure}
\begin{tabular}[t]{|l|l|l|}
  \hline
  \textbf{Basic species} & \textbf{Notation} & \textbf{Cycle index sum} \\
  \hline
  \textbf{Empty species} & {\bf 0} & $\ds Z_{\mathbf{0}} = 0$ \\
  \textbf{Neutral species} & {\bf 1} & $\ds Z_{\mathbf{1}} = 1$ \\
  \textbf{Atomic species} & $X$ & $\ds Z_{X} = s_1$ \\
  \textbf{Sequences with $k$ atoms} & $\Seq^{[k]}$ & $\ds Z_{\Seq^{[k]}} = s_1^k$ \\
  \textbf{Sets with $k$ atoms} & $\Set^{[k]}$ & $\ds Z_{\Set^{[k]}} =[x^k]\exp\Big( \sum_{r\geq 1} \frac{1}{r}x^rs_r\Big) $ \\
  \textbf{Cycles with $k$ atoms} & $\Cyc^{[k]}$ &  $\ds Z_{\Cyc^{[k]}} =\frac{1}{k}\sum_{r|k} \phi(r)s_r^{k/r} $\\
  \textbf{Species of sequences} & $\Seq$ & $\ds Z_{\Seq} = \frac{1}{1-s_1}$ \\
  \textbf{Species of sets} & $\Set$ & $\ds Z_{\Set} =\exp\Big( \sum_{r\geq 1} \frac{1}{r}s_r\Big) $ \\
  \textbf{Species of cycles} & $\Cyc$ &  $\ds Z_{\Cyc} =1+\sum_{r\geq 1} \frac{\phi(r)}{r}\log\left(\frac{1}{1-s_r}\right) $  \\[0.2cm]
\hline
  \hline
  \textbf{Construction} & \textbf{Notation} & \textbf{Cycle index sum}\\
  \hline
  \textbf{Union of two species} & $\cM=\cA + \cB $ &  $\ds Z_{\cM} = Z_{\cA} + Z_{\cB}$ \\
  \textbf{Product of two species} & $\cM=\cA \bcdot \cB $  & $\ds Z_{\cM} = Z_{\cA} \bcdot Z_{\cB}$ \\
  \textbf{Composition of two species} & $\cM=\cA \circ \cB $  & $\ds Z_{\cM} = Z_{\cA} \circ Z_{\cB}$ \\
  \hline
\end{tabular}
\caption{The cycle index sums of basic species and of species composed by $+$, $\bcdot$, and $\circ$.} \label{Basic}
\end{figure}

\subsubsection{Basic Species.}
The most elementary species are 
\begin{itemize}
\item the \emph{empty species} ${\bf 0}$ 
defined by ${\bf 0}[U] := \emptyset$ for all $U$; 
\item the \emph{neutral species} ${\bf 1}$ 
defined by ${\bf 1}[U] := \{\emptyset\}$ if $|U| = 0$ and ${\bf 1}[U] := \emptyset$ otherwise;
\item the \emph{atomic species} $X$
defined by $X[U] := \{U\}$ if $|U| = 1$ and $X[U] := \emptyset$ otherwise;
\item the \emph{species of sets} $\Set$ defined by $\Set[U] := \{U\}$;
\item the species $\Seq$ of sequences defined by
$\Seq[U] := \{(U,<) \; | \; U \text{ is linearly ordered by } <\}$;
\item the species $\Cyc$ of oriented cycles (or cyclic permutations) defined by $$\Cyc[U] := \{(U,\pi) \; | \; \pi \text{ is a cyclic permutation of } U\} \; .$$
\end{itemize}
In each case, the definition of the transport $\cA[\sigma]$ of $\cA$-structures is obvious.

If $\mathfrak A$ is a species, then $\mathfrak A^{[n]}$ denotes
the species defined by $\cA^{[n]}[U] := \cA[U]$ if $|U|=n$ and $\cA[U] := \emptyset$ otherwise.
The transport
$\cA^{[n]}[\sigma]$ for bijections $\sigma: U \rightarrow V$
is defined by $\cA^{[n]}[\sigma] := \cA[\sigma]$ if $|U|=n$ and
$\cA^{[n]}[\sigma] := \text{Id}_{\cA[U]}$ otherwise.
 In particular, $X$ is $\Set^{[1]}$. 
 
\subsubsection{Constructions}\label{sssect:constr}
We now describe the three fundamental constructions that we use to construct species from other species.

\vspace{0.2cm}

\emph{Disjoint union.}
For two species $\cA$ and $\cB$, the species $\cA + \cB$, called the \emph{disjoint union} (or sum) of $\cA$ and $\cB$, is defined by
$(\cA + \cB)[U] = (\{1\} \times \cA[U]) \cup (\{2\} \times \cB[U])$. 
The transport along a bijection $\sigma: U \rightarrow V$ is carried out by setting, for any $(\cA + \cB)$-structure $C$ on $U$, 
\begin{equation*}\label{eq:analytic}
(\cA+\cB)[\sigma]((i,C)) := \left\{
\begin{array}{rl}
\cA[\sigma](C) &  \text{ if } i = 1 \\
\cB[\sigma](C) & \text{ if } i = 2 \; .
\end{array}\right. 
\end{equation*}
Note that when $\cA, \cA', \cB, \cB'$ are species such that
$\cA = \cA'$ (i.e.,  $\cA$ and $\cA'$ are isomorphic) and $\cB = \cB'$, then $\cA + \cB = \cA' + \cB'$.

The disjoint union of a countably infinite sequence $\cA_1,\cA_2,\dots$ of species
is defined analogously, setting $\big (\sum_{i \geq 1} \cA_i \big)[U] := \bigcup_{i \geq 1} (i,\cA_i[U])$ (this species is well-defined provided the set on the right
is finite for each finite set $U$). 


\vspace{0.2cm}

\emph{Product.}
The \emph{cartesian} (often also called \emph{partitional} or \emph{dinary}) 
\emph{product} $\cA\bcdot\cB$ of two species $\cA$ and $\cB$ is the species defined as follows.
An $(\cA \bcdot \cB)$-structure on $U$ is an ordered pair $C = (A,B)$
where $A$ is an $\cA$-structure on a set $U_1 \subseteq U$,
$B$  is a $\cB$-structure on a set $U_2 \subseteq U$,
$U_1 \cap U_2 = \emptyset$, and $U_1 \cup U_2 = U$.
The transport along a bijection $\sigma: U \rightarrow V$ is carried
out by setting, for each $(\cA \bcdot \cB)$-structure $C=(A,B)$ on $U$,

$$ (\cA \bcdot \cB)[\sigma](C) := (\cA[\sigma_1](A),\cB[\sigma_2](B)) $$
where $\sigma_i$ is the restriction $\sigma|_{U_i}$ of $\sigma$ on $U_i$ for $i \in \{1,2\}$.

Again we note that when $\cA, \cA', \cB, \cB'$ are species such that
$\cA = \cA'$ and $\cB = \cB'$, then $\cA \bcdot \cB = \cA' \bcdot \cB'$.

\vspace{0.2cm}

\emph{Substitution.} 
Given two species $\cA$ and $\cB$ such that $\cB[\emptyset] = \emptyset$, the \emph{(partitional) composite} of $\cB$ in $\cA$,
denoted by
$\cA\circ\cB$, is the species $\cC$ obtained as follows.
A $\cC$-structure on $U$ is a triple $(\pi, A,B)$ where
\begin{itemize}
\item $\pi$ is a partition of $U$;
\item $A$ is an $\cA$-structure on the set of classes of $\pi$;
\item $B := (B_p)_{p \in \pi}$ where for each class $p$ of $\pi$, $B_p$ is a $\cB$-structure on $p$.
\end{itemize}
The transport along a bijection $\sigma: U \rightarrow V$ is carried out
by setting, for any $\cC$-structure $C=(\pi,A,(B_p)_{p \in \pi})$ on $U$
$$ \cC[\sigma](C) := (\bar \pi, \bar A, (\bar B_{\bar p})_{\bar p \in \bar \pi})$$ 
where
\begin{itemize}
\item $\bar \pi$ is the partition of $V$ obtained by transport of $\pi$ along $\sigma$;
\item the structure $\bar A$ is obtained from the structure $A$
by $\cA$-transport along the bijection $\bar \sigma: \pi \rightarrow \bar \pi$ induced by $\sigma$ on $\pi$;
\item for each $\bar p = \sigma(p) \in \bar \pi$, the structure $\bar B_{\bar p}$ is obtained from the structure $B_p$ by $\cB$-transport
along $\sigma|_p$.
\end{itemize}
We call the $\cA$-structure $A$ the \emph{core}
of 
$(\pi,A,B)$, and $B_p$ 
in $B=(B_p)_{p \in \pi}$ 
a \emph{component} of $(\pi,A,B)$.
Also for the substitution construction, we have that when $\cA, \cA', \cB, \cB'$ are species such that
$\cA = \cA'$ and $\cB = \cB'$, then $\cA \circ \cB = \cA' \circ \cB'$.


Together with the basic species, these constructions provide an extremely
powerful device for the description of combinatorial families. 
The substitution construction is particularly interesting, as it allows us to express other combinatorial constructions, such as the formation of sequences, sets, and cycles of structures of a species $\cA$, which are specified as 
 $\Seq \circ \cA$, $\Set \circ \cA$, and $\Cyc \circ \cA$, respectively. 

As usual, and to avoid clumsy expressions with many brackets, 
we make the convention that in expressions involving several
of the symbols $+,\bcdot,\circ$, the symbol $\circ$ binds stronger than
the symbol $\bcdot$, and the symbol $\bcdot$ binds stronger
than the symbol $+$.

There are explicit rules to compute
 the cycle index sum for the basic species and for each construction. 
  
 \begin{definition}
 Given two power series
 $f:=f(x_1,x_2,x_3\ldots)$ and $g:=g(x_1,x_2,x_3,\ldots)$ such that
 $g(0,0,\ldots)=0$, 
 the \emph{plethystic composition} of $f$ and $g$, as defined in~\cite{BeLaLe}, is the
 power series 
 \begin{equation}
 f\circ g:=f(g_1,g_2,g_3,\ldots),\ \ \ \mathrm{with}\ g_k=g(x_k,x_{2k},x_{3k},\ldots) \; .
 \end{equation}
 In other words, $f\circ g$ is the series $f$ where each variable $x_k$ is replaced by $g(x_k,x_{2k},x_{3k},\ldots)$.
 \end{definition}

\begin{proposition}[P\'olya, Bergeron et al.~\cite{BeLaLe}]\label{prop:dict}
For each of the basic species $\{ \mathbf{0}, \mathbf{1}$, $X$, $\Seq^{[k]}$, $\Set^{[k]}$, $\Cyc^{[k]}$, $\Seq$, $\Set$, $\Cyc\}$, the 
associated cycle index sum has an explicit expression, as given in Figure~\ref{Basic}.
For each of the fundamental constructions $\wedge\in\{+,\bcdot,\circ\}$, there is an explicit rule to 
compute the cycle index sum of the species $\cA\wedge\cB$, as given in Figure~\ref{Basic}.
\end{proposition}

\begin{remark}\label{remark:OGS}
The ordinary generating series of a species can be obtained
from the cycle index sums for the species. For the sum and product constructions, we obtain
\begin{eqnarray}
\hspace{-1.5cm}\cC=\cA+\cB & \Rightarrow & \wt{\cC}(x)=\wt{\cA}(x)+\wt{\cB}(x),
\end{eqnarray}
\begin{eqnarray}
\hspace{-1.5cm}\cC=\cA\bcdot\cB & \Rightarrow & \wt{\cC}(x)=\wt{\cA}(x)\cdot\wt{\cB}(x).
\end{eqnarray}
For the substitution construction, the computation rule is
\begin{eqnarray}
\hspace{1cm}\cC=\cA\circ\cB & \Rightarrow & \wt{\cC}(x)=Z_{\cA}(\wt{\cB}(x),\wt{\cB}(x^2),\wt{\cB}(x^3),\ldots).
\end{eqnarray}
\end{remark}

\subsection{Recursive Specifications.} \label{sssect:rec}
It is possible to define species via recursive specifications
that involve the fundamental constructions introduced above.

\begin{definition}\label{def:rec-standard}
A \emph{(standard) recursive specification} with variables $x_1,\dots,x_m$ over
the species $\cA_1,\dots,\cA_{\ell}$
is a system $\Psi$ of equations $x_1 = e_1, \dots, x_m = e_m$ 
where each $e_i$ 
is 
\begin{itemize}
\item of the form $a+b$ or $a \bcdot b$ with $a, b \in
\{x_1,\ldots,x_m,\cA_1,\dots,\cA_{\ell}\}$, or 
\item of the form
$a \circ b$ with $a \in \{\cA_1,\dots,\cA_{\ell}\}$ 
and $b \in \{x_1,\dots,x_m, \cA_1,\dots,\cA_{\ell}\}$.
\end{itemize}
\end{definition}

Under a certain condition, 
a recursive specification $\Psi$ with variables $x_1,\dots,x_m$
defines new species $\cX_1,\dots,\cX_m$ as follows.
We first define for each $i \geq 0$ a vector of species
$\cX_1^{(i)},\dots,\cX_m^{(i)}$. For $i=1$, we set
$(\cX_1^{(i)},\dots,\cX_m^{(i)}) = ({\bf 0},\dots,{\bf 0})$.
For $i > 1$ and $1 \leq j \leq m$, the species $\cX^{(i)}_j$ is defined
from $e_j$ by substituting for all $1 \leq k \leq m$ 
the occurrences of $x_k$ in $e_j$ by $\cX_k^{(i-1)}$.
The resulting expression only contains species and
symbols $+$, $\bcdot$, and $\circ$, and hence evaluates to
a species, unless $e_j = a \circ b$ and $b$ is substituted by a species that contains structures of size $0$. 
If this case never occurs, i.e., if $\cB[\emptyset] = \emptyset$  whenever a species $\cB$ is substituted for $b$ in an expression $a \circ b$,
then we call $\Psi$ \emph{admissible}.

Note that the sequence $\cX_k^{(1)},\cX_k^{(2)},\dots$ is \emph{monotone}, that is, $\cX_k^{(i)} \subseteq \cX_k^{(i+1)}$ 
for all $1 \leq k \leq m$ and all $i \geq 1$; this follows from the following basic fact (we omit the proof which is straightforward).

\begin{proposition}
If $\cA,\cA',\cB,\cB'$ are species, 
and $\cA \subseteq \cA'$, $\cB \subseteq \cB'$, then
$\cA + \cB \subseteq \cA' + \cB'$,
$\cA \bcdot \cB \subseteq \cA' \bcdot \cB'$, and $\cA \circ \cB \subseteq \cA' \circ \cB'$.
\end{proposition}

Note that due to this proposition, it is easy to decide for a given
recursive specification whether or not it is admissible
(given also the information which of the species $\cA_1,\dots,\cA_{\ell}$
contains structures of size $0$).

\begin{definition}\label{def:semantics}
Let $\Psi$ be an admissible recursive specification with variables
$x_1,\dots,x_m$ over the species $\cA_1,\dots,\cA_{\ell}$,
and let the species $\cX^{(i)}_j$ be as described
above. If for each $n$, the set $\bigcup_{i \geq 1} \cX^{(i)}[n]$ is finite,
then the species $\cX_1,\dots,\cX_m$ \emph{specified by $\Psi$} 
are defined as follows. We set $\cX_j[n] = \bigcup_{i \geq 1} \cX_j^{(i)}[n]$. By monotonicity, for each $n$ we have that $\cX_j[n]$
equals $\cX_j^{(k)}[n]$ for some $k$,
and so we can define the transport of $\cX_j$ by $\cX_j[\sigma] = \cX_j^{(k)}[\sigma]$.
\end{definition}




\begin{definition}\label{def:dec-standard}
Let $\mathfrak A$ be a class of species.
The class of species that is \emph{decomposable over $\mathfrak A$}
is the smallest class of species $\mathfrak B$ that contains $\mathfrak A$, and that contains all species that can be defined by recursive specifications over species from $\mathfrak B$.
\end{definition}

\subsection{Decomposition of  symmetries}\label{sec:decop_symme}
In this subsection we provide a
 proof of Proposition~\ref{prop:dict}. The proof relies on a
precise description of 
the nature of the automorphisms for the basic species and for the species obtained from one of the constructions $+$, $\bcdot$, or $\circ$.
Even though proofs and details can already be found 
in~\cite{BeLaLe}, we give our own presentation here
since we build on this later on, in particular to define random
generation rules (Section~\ref{sec:sampler}). 

\subsubsection{Automorphisms of some basic species.}\label{sec:auto_basic}
By convention, the neutral species $\bf{1}$ 
has the cycle-index sum $1$ (the structure of size $0$ is assumed to be fixed by the ``empty'' automorphism, of weight $1$).

The cycle index sum of $X$ is $s_1$. The only $X$-structures are
over $U$ with $|U|=1$, and all automorphisms of such structures
consist of a single cycle that has weight $s_1$.

For the species $\Seq$, the only automorphisms are the identity, 
and the identity has weight $s_1^k/k!$. As there are $k!$ sequences of length $k$, the cycle index sum of $\Seq^{[k]}$ is $s_1^k$. 
Since  $\Seq=\sum_{k\geq 0} \Seq^{[k]}$, the cycle index sum for $\Seq$ is
\begin{equation}
Z_{\Seq}=\sum_{k\geq 0}s_1^k=\frac{1}{1-s_1}.
\end{equation}

For the species $\Cyc$, the automorphisms are exactly the `shifts': 
for a cycle $(v_1,\ldots,v_k)$, the \emph{shift} of this cycle by
$m\in[0..k-1]$ maps $v_i$ to $v_{i+m}$ where the indices are
modulo $k$. The automorphisms consist of $k/r$ cycles of length $r$, where
$r$ is the order of $m$ in $\mathbb{Z}/(k\mathbb{Z})$. For each divisor $r$ of $k$, there are $\phi(r)$ elements of order $r$ in $\mathbb{Z}/(k\mathbb{Z})$, where $\phi(.)$ is the Euler totient function.
Hence, for each cycle of length $k$, the sum of the weight-monomials over all the cycles of
size $r$ is $\phi(r) s_{r}^{k/r}/k!$, and the sum of the weight-monomials over all the symmetries  
is $\sum_{r|k}\phi(r) s_{r}^{k/r}/k!$. As there are $(k-1)!$ cycles of length $k$, the cycle index sum of
$\Cyc^{[k]}$ is $1/k\cdot\sum_{r|k}\phi(r) s_{r}^{k/r}$.
For the species $\Cyc=\sum_{k\geq 1}\Cyc^{[k]}$ of all cycles, the sum of the weight-monomials over
all symmetries of size $r$, which we denote by $Z_{{\Cyc},r}$, is thus
\begin{equation}
Z_{{\Cyc},r}=\sum_{k, r|k} \frac{\phi(r)}{k} s_{r}^{k/r}=\frac{\phi(r)}{r}\sum_{m\geq 1} \frac{1}{m}s_r^m=\frac{\phi(r)}{r}\log\left(\frac{1}{1-s_r}\right).
\end{equation}
Therefore, summing $Z_{\Cyc,r}$ over $r\geq 1$, one obtains the expression of $Z_{\Cyc}$
given in Figure~\ref{Basic}.

For the species $\Set$, the automorphisms are all permutations. Hence, the cycle index sum for $\Set$ is simply the exponential generating series for 
all permutations, where each cycle of length $i$ is marked by a variable $s_i$. 
In other words, when $f(x)$ denotes the exponential generating series for permutations, and $f(x;s_1,s_2,s_3,\ldots)$ denotes the same generating series where each variable $s_i$ marks
the number of cycles of length $i$, then the rules for computing exponential generating series
 yield
\begin{equation}
f(x;s_1,s_2,s_3,\ldots)=\exp\left(\sum_{i\geq 1}\frac{1}{i!}x^i (i-1)!s_i\right)=\exp\left(\sum_{i\geq 1}\frac{1}{i}x^is_i\right) \; .
\end{equation}
Hence, $f(1;s_1,s_2,s_3,\ldots)$ is the cycle index sum for $\Set$, and
 the cycle index sum for $\Set^{[k]}$ is $[x^k]f(x;s_1,s_2,s_3,\ldots)$, which corresponds to a restriction to permutations of size $k$. Notice that $Z_{\Set^{[k]}}$ is always a polynomial; for instance, we have
 $$
 Z_{\Set^{[2]}}=\frac{1}{2}(s_1^2+s_2),\ \ Z_{\Set^{[3]}}=\frac{1}{6}(s_1^3+3s_2s_1+2s_3),\ \ Z_{\Set^{[4]}}=\frac{1}{24}(s_1^4+6s_1^2s_2+8s_1s_3+3s_2^2+6s_4).
 $$

\subsubsection{Automorphisms related to the constructions}\label{sec:aut_cons}
We now describe precisely how a symmetry of a species $\cC=\cA\wedge\cB$, with $\wedge\in\{+,\bcdot,\circ\}$, is assembled from symmetries on the species $\cA$ and $\cB$.

\vspace{0.2cm}

\emph{Disjoint union.}
Clearly, each symmetry of $\cC=\cA+\cB$ is either a symmetry of $\cA$ or of $\cB$ depending on whether
the underlying structure is in $\cA$ or in $\cB$. Consequently,
$\Sym(\cA+\cB)$ is the disjoint union of
$\Sym(\cA)$ and $\Sym(\cB)$.
This directly yields the formula $Z_{\cA+\cB}=Z_{\cA}+Z_{\cB}$.

\vspace{0.2cm}

\emph{Product.} 
Consider a product species $\cC=\cA \bcdot \cB$. Then there is the following bijective correspondence between symmetries of $\cC$
and ordered pairs of a symmetry of $\cA$ and a symmetry of $\cB$:
when $((A,B),\sigma)$ is a symmetry of $\cC$ where $U$ are the atoms of $A$ and $V$ are the atoms of $B$, then $((A,B),\sigma)$  is in correspondence
with the symmetry $(A,\sigma|_{U})$ of $\cA$ and the symmetry $(B,\sigma|_{V})$ of $\cB$ (where $\sigma|_{U}$ and $\sigma|_{V}$ denote the restriction of $\sigma$ to $U$ and $V$, respectively).

Hence, $Z_{\cC}=Z_{\cA}\bcdot Z_{\cB}$.
(Indeed, for the species $\cC$ the cycle index sum acts like an exponential 
generating series for the species $\Sym(\cC)$ of symmetries
when taking the refined weights $s_1^{c_1}\cdots s_n^{c_n}/n!$ instead of $x^n/n!$.)
\vspace{0.2cm}

\emph{Substitution.}
For two species $\cA$ and $\cB$ with $\cB[\emptyset]=\emptyset$, 
let $\cC = \cA \circ \cB$.
To understand the automorphisms of $\cC$-structures, consider
a $\cC$-structure $C = (\pi,A,(B_p)_{p \in \pi})$ over $[1..n]$. 
Let $\sigma$ be an automorphism of $C$. 
It is clear that if two atoms $v_1$ and $v_2$ of $C$ are in the same component, then
$\sigma(v_1)$ and $\sigma(v_2)$ have to be on the same component as well, by definition of the transport for $\cC$;
moreover, $\sigma$ induces 
an automorphism $\tau$ of $A$. 

Consider an atom $v$ from a component $B_p$ of $C$, 
and let $k$ be the length of the cycle of $\tau$ containing $v$. 
Note that $\sigma^{k}$ maps $B_p$ to itself, so $\sigma^{k}$ induces an automorphism on $B_p$, 
the resulting symmetry being denoted by $(B_p,\sigma_p)$. 
Consider the cycle $c=(p_1,\ldots,p_k)$ in $\tau$ where $p_1=p$.
Observe that 
the symmetries $(B_{p_i},\sigma_{p_i})$, for $i\in [k]$, 
can be seen as $k$ copies of the same
symmetry of $\cB$, which we denote by $(B_c,\sigma_c)$. For each cycle 
$d=(w_1,\ldots,w_{\ell})$ of $\sigma_c$, 
let $(d_1,\ldots,d_k)$ be the copies of $d$ at $(p_1,\ldots,p_k)$, respectively. 
Then one can merge $d_1,\ldots,d_k$ into
a unique cycle of length $\ell\cdot k$ using a specific operation which we call \emph{composition of cycles}.

\begin{definition}
Let $d=(v_1,\ldots,v_{\ell})$ be a cycle of atoms from $[1..n]$, with $v_1$ the atom of $d$ having the smallest label.
Let $d_1,\ldots,d_k$ be a sequence of $k$ copies of
the cycle. Then the \emph{composed cycle} of $d_1,\ldots,d_k$ 
is the cycle of atoms of length $\ell k$ such that, for $1\leq i<k$ and $1\leq j\leq \ell$, the successor of the atom $v_j$
in $d_i$ is the atom $v_j$ in $d_{i+1}$; and for $1\leq j\leq \ell$,
the successor of the atom $v_j$ in $d_k$ is the atom $v_{(j+1)\text{ mod }\ell}$ in $d_1$.
\end{definition}

\ignore{
\begin{definition}
Consider $k$ cyclically ordered cycles of $\ell$ labeled elements,
and where the overall $\ell\cdot k$ elements have distinct (integer) labels.
Write the $k$ cycles canonically as $c_1=(v_1^{(1)},\ldots$,
$v_{1}^{(\ell)}),\ldots,c_k=(v_k^{(1)},\ldots,v_{k}^{(\ell)})$, where the element with smallest label is in $c_1$, and for each
$i\in[1..k]$ $v_i^{(1)}$ is the element with smallest label in $c_i$. 
Then the \emph{composed cycle} of $L=(c_1,\ldots,c_k)$ is the cycle $C$
made from the $\ell\cdot k$ elements of $L$ 
such that  for $1\leq i<k$ and $1\leq j\leq \ell$, the successor of $v_i^{(j)}$ 
 is $v_{i+1}^{(j)}$; and for $1\leq j\leq \ell$,
the successor of  $v_k^{(j)}$ is $v_1^{(j\!+\!1\ \mathrm{mod}\ \ell)}$.
\end{definition}
}

This definition correctly reflects how each cycle of $\sigma$ is assembled from copies of cycles that are on isomorphic components. 
 Indeed, walking $k$ steps forward on the composed cycle corresponds to 
walking one step forward on one fixed cycle, which corresponds to the fact that
the induced automorphism on each component $B_{v_i}$ is the effect of $\sigma$ iterated $k$ times.

In each symmetry $(A,\sigma)$ where $A$ is an $\cA$-structure over $[1..n]$, the automorphism $\sigma$ induces a partition of $[1..n]$ corresponding to the cycles of $\sigma$.  
We say that $\sigma$ has \emph{type} $t=(c_1,\dots,c_n)$
when $c_i$ is the number of cycles of length $i$ in $\sigma$. 
Note that $n = \sum_{i=1}^n ic_i$ (and such integer sequences $(c_1,\dots,c_n)$ will also
be called \emph{partition sequences (of order $n$)}).

To compute the cycle index sum $Z_{\cA \circ \cB}$,
we first choose a type $t = (c_1,\dots,c_n)$ of a permutation of $[1..n]$.
The number $a(t)$ of symmetries $(A,\sigma)$ of the species $\cA$ where $\sigma$ is of type $t$  equals
$n! \cdot[s_1^{c_1}s_2^{c_2}\ldots s_n^{c_n}](Z_{\cA})$.
To obtain a symmetry $(C,\rho)$ of $\cA \circ \cB$ with $C = (\pi,A,B)$ where $\rho$ induces
on $A$ a permutation $\sigma$ of type $t$, 
we choose a symmetry of $\cA$ of type $t$, and then choose $c_i$ symmetries of $\cB$ for each $i$.
The sum $Z_{\cA\circ\cB}^{(t)}$ of the weight-monomials for all those symmetries is therefore

\begin{equation}\label{eq:aut_cons}
Z_{\cA\circ\cB}^{(t)}=\frac{a(t)}{n!}b_1^{c_1}b_2^{c_2}\cdots b_n^{c_n},\ \ \mathrm{where}\ b_i:=Z_{\cB}(s_i,s_{2i},s_{3i},\ldots).
\end{equation}

Summing over all possible types $t$ of permutations, one obtains
\begin{equation}
Z_{\cA\circ\cB}=Z_{\cA}(b_1,b_2,b_3,\ldots),\ \ \mathrm{where}\ b_i:=Z_{\cB}(s_i,s_{2i},s_{3i},\ldots).
\end{equation}


\section{Cycle-pointed Species}
\label{sec:polya_cycle_pointed}
In this section we introduce cycle-pointed species, and our unbiased pointing operator.

\subsection{Cycle-Pointed Species}
Let $\cA$ be a species. Then the \emph{cycle-pointed species of $\cA$}, denoted by $\cA^{\circ}$, is defined as follows.
For a finite set $U$, the set $\cA^{\circ}[U]$ is defined to be the set of 
all pairs $P=(A,c)$ where $A \in \cA[U]$ and $c = (v_1,\dots,v_{\ell})$ is
a cycle of atoms of $A$ such that there exists \emph{at least one} automorphism of $A$ having $c$ as one of its cycles
(i.e., $(v_1,\dots,v_{\ell})$ is mapped to $(v_2,\dots,v_{\ell},v_1)$).
The cycle $c$ is called the \emph{marked cycle} (or pointed cycle) 
of $P$, and $A$ is called the \emph{underlying
structure} of $P$.
Note that cycle-pointed species are in particular species. Thus, the theory from the previous section also applies to cycle-pointed species.

An automorphism $\sigma$ of $A$ having $c$ as one of its cycles is called a \emph{$c$-automorphism} of $P$, and 
the other cycles of $\sigma$ are called \emph{unmarked}.
By definition, two cycle-pointed structures $P$ and $P'$ are isomorphic if there
exists an isomorphism from the underlying structure of $P$ to
the underlying structure of $P'$ 
that maps the marked cycle of $P$ 
to the marked cycle of $P'$ (i.e., each atom of the marked cycle of $P$ is mapped 
to an atom of the marked cycle of $P'$, and the cyclic order
is preserved).

\begin{definition}
A species $\cP$ is called \emph{cycle-pointed} if  $\cP \subseteq \cA^{\circ}$, for some species $\cA$.
For $\ell\geq 1$, 
$\cP_{(\ell)}$ is the species that consists 
of those structures from $\cP$ whose marked cycle has length $\ell$.  
\end{definition}

We define $\cA^{\cst}$ to be the subspecies of $\cA^{\circ}$ 
where in all structures the marked cycle has length greater than $1$, i.e., $\cA^{\cst}=\sum_{\ell\geq 2}(\cA^{\circ})_{(\ell)}$. (All cycle-pointed species that will be considered in the applications -- except for maps -- are either of the form $\cA^{\circ}$ or $\cA^{\cst}$.)

\begin{definition}
Let $\cP$ be a cycle-pointed species. 
We define $\cP^{\dag}$ to be the species obtained from $\cP$ by
removing the marked cycle from all $\cP$-structures; that is,
the set of $\cP^{\dag}$-structures on $[1..n]$ is $\{ A \; | \; (A,c) \in \cP_n\ \mathrm{for\ some\ cycle\ }c\ \mathrm{of\ atoms\ in\ }A  \}$. 
\end{definition}

Clearly, for any species $\cA$ we have that $(\cA^{\circ})^\dag = \cA$.

\subsection{Cycle Index Sums}
In order to develop P\'olya theory for cycle-pointed species,
we introduce the terminology of \emph{c-symmetry}
and \emph{rooted c-symmetry}. 
Given a cycle-pointed species $\cP$, a \emph{c-symmetry} on $\cP$ is a pair $(P,\sigma)$ where $P=(A,c)$ is a cycle-pointed structure in $\cP$ and $\sigma$ is a c-automorphism of $A$. 
A \emph{rooted c-symmetry} is a triple $(P,\sigma,v)$,
where $(P,\sigma)$ is a c-symmetry, and $v$ is one of the atoms of the marked cycle of $P$;
this atom $v$ is called the \emph{root} of the rooted c-symmetry.
The species of rooted c-symmetries
of $\cP$ is denoted by $\RSym(\cP)$.

The \emph{weight-monomial} of a rooted c-symmetry of size $n$ is defined as 
\begin{equation}
\label{eq:def_weight_cycle_pointed}
w_{(P,\sigma,v)}:=\frac{1}{n!}t_{\ell}\prod_{i=1}^{n}s_i^{n_i(\sigma)},
\end{equation}
where $t_{\ell}$ and the $s_i$'s are 
formal variables, $\ell$ is the length of the marked cycle, and for $i\in[n]$, $n_i(\sigma)$ is the number 
of unmarked cycles of $\sigma$ of length $i$, i.e., $n_i(\sigma)=c_i(\sigma)$ if $i\neq\ell$ and $n_{\ell}(\sigma)=c_{\ell}(\sigma)-1$.
For simplicity, in the following we will write $n_i$ for $n_i(\sigma)$ if the corresponding automorphism is clear from the context.
We define the \emph{pointed cycle index sum} $\bar Z_{\cP}(s_1,t_1;s_2,t_2;\ldots)$ of $\cP$, denoted by
$\bar Z_{\cP}$, as 
the sum of the weight-monomials over all rooted c-symmetries of $\cP$, 
\begin{equation}
\bar Z_{\cP}(s_1,t_1;s_2,t_2;\dots):=\sum_{(P,\sigma,v)\in\RSym(\cP)}w_{(P,\sigma,v)}.\\
\end{equation}

The following lemma is the counterpart of Lemma~\ref{lem:unlabel} for cycle-pointed species;
it shows that pointed cycle index sums refine ordinary generating functions.
Recall that for a species $\cP$, the set $\wt{\cP}$ denotes the unlabeled $\cP$-structures, see Section~\ref{sub:prelim_enum}.

\begin{lemma}
\label{lem:cpointedcounting}
Let $\cP$ be a cycle-pointed species.
For $n\geq 0$, each unlabeled structure $\tilde{P}\in\wt{\cP}_n$ gives rise to exactly $n!$ rooted c-symmetries, i.e., there are $n!$ rooted c-symmetries $(P,\sigma,v)$
such that $P \in \tilde{P}$. As a consequence,
\begin{equation}
\label{eq:PolyaPoint}
\wt{\cP}(x) = \bar Z_{\cP}(x,x;x^2,x^2;\ldots).
\end{equation}
\end{lemma}
\begin{proof}

Lemma~\ref{lem:unlabel} implies that $\tilde{P}_n$ gives rise to 
$n!$ symmetries. 
Now we establish a bijection between these symmetries and 
rooted c-symmetries from $\tilde{P}_n$.
Fix a c-symmetry $(P_0, \sigma_0)$ where $P_0$ 
is from $\tilde P_n$ (by definition of cycle-pointed species, such
a c-symmetry exists). Now, 
let $(P_1,\sigma_1)$ be a symmetry of $\cP$ where $P_1$ is also from
$\tilde{P}_n$. The marked cycle $(v_1,\ldots,v_{\ell})$ of $P_1$ is preserved by 
$\sigma_1$, so that all its elements are \emph{shifted} by the same
value $r \in [1..\ell]$ modulo $\ell$, i.e., $\sigma_1$ maps $v_i$ 
to $v_{(i+r) \mod \ell}$. 
Since both $P_0$ and $P_1$ are from $\tilde P_n$, 
there is an isomorphism $\sigma$ from
$P_1$ to $P_0$. Moreover, the permutation 
$\tau:=\sigma^{-1} \sigma_0 \sigma$ is a c-symmetry of 
$P_1$. 
Observe that the permutation 
$\tau^{-r+1}\sigma_1$ is an automorphism of $P_1$ that 
moves an atom of the marked cycle $r$ steps forward 
(because of $\sigma_1$) 
and then $r-1$ steps backward (because of $\tau^{-r+1}$). Hence,
$\tau^{-r+1}\sigma_1$ is a c-symmetry of $P_1$. 
The desired bijection maps 
$(P_1,\sigma_1)$ to the 
rooted c-symmetry $(P_1,\tau^{-r+1}\sigma_1,v)$ where $v$ is the atom of
the marked cycle having the $r$-th smallest label. 
This correspondence can be inverted easily, and hence we have found a bijection
between the symmetries and 
the rooted c-symmetries for structures having $\tilde P_n$ as unlabeled structure.
\end{proof}

Observe that a rooted c-symmetry of $\cAp$
is obtained from a symmetry 
$(A,\sigma)$ of $\cA$ by choosing an atom $v$ of $A$ 
and marking the cycle
of $\sigma$ containing $v$.
Therefore, each symmetry $(A,\sigma)$ of size $n$ of the species $\cA$ yields $n$ rooted c-symmetries on $\cAp$, and hence
\begin{equation}
\bar Z_{\cA^{\circ}}(s_1,t_1;s_2,t_2;\ldots)|_{\{t_1=s_1,t_2=s_2,\ldots\}}=\frac{d}{dt}Z_{\cA}(ts_1,t^2s_2,t^3s_3,\ldots)|_{\{t=1\}}.
\end{equation}
Further, a rooted c-symmetry of $\cAp$ can equivalently be obtained by marking a cycle of atoms 
that corresponds to a cycle of $\sigma$ and choosing an atom
of the cycle as the root of the rooted c-symmetry.
Together, these observations yield the equality
\begin{equation} \label{eq:ZcUpl}
\bar Z_{\cA^{\circ}_{(\ell)}}(s_1,t_1;s_2,t_2;\ldots)|_{\{t_1=s_1,t_2=s_2,\ldots\}}=\ell\ t_{\ell}\frac{\partial}{\partial s_{\ell}}Z_{\cA}(s_1,s_2,\ldots)\ \ \text{ for all }\ \ell\geq 1.
\end{equation}
For $\ell=1$, which corresponds
to structures of $\cA$ with a unique distinguished vertex (the \emph{root}), 
we recover the well-known equation 
relating the cycle index sum of a species and of the associated 
\emph{rooted} species;
see~\cite[Sec.~1.4.]{BeLaLe} and~\cite{HP73}.

\begin{figure}
\begin{center}
\includegraphics[width=12cm]{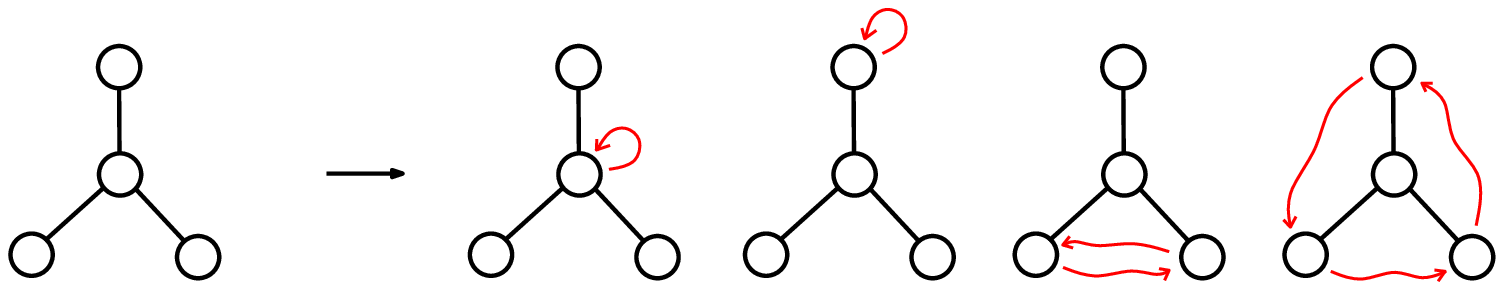}
\end{center}
\caption{An unlabeled  
nonplane tree of size $4$ yields 4 unlabeled cycle-pointed trees.}
\label{fig:unbiased}
\end{figure}

As stated below and illustrated in Figure~\ref{fig:unbiased},
 pointing a cycle of symmetry instead of a single atom (as the classical pointing
 operator does) yields an unbiased pointing operator in the unlabeled setting.
 
\begin{theorem}[unbiased pointing]
\label{theo:npointed}
Let $\cA$ be a species.
Then, for $n\geq 0$,
 each unlabeled structure of $\wt{\cA}_n$ gives rise to exactly $n$ 
unlabeled structures in $\wt{\cA^{\circ}_n}$; that is,
for each  $A \in \cA[n]$
there are exactly $n$ non-isomorphic structures $(A,c)$
in $\cA^{\circ}[n]$.
Hence, the ordinary generating series 
of $\cA^{\circ}$ satisfies
\begin{equation} \label{eq:Up}
\wt{\cA^{\circ}}(x)=x\frac{\mathrm{d}}{\mathrm{d}x}\wt{\cA}(x).
\end{equation}
\end{theorem}

\begin{proof}
Given $\tilde{A}\in\wt{\cA}_n$, let $S$ be the set of unlabeled pointed structures of $\wt{\cA^{\circ}}$
 whose underlying unpointed structure is $\tilde{A}$. The proof of the lemma reduces
to proving that $S$ has cardinality $n$.
Let $\Sym(\tilde{A})$ be the set of symmetries for the structure 
$\tilde{A}$, and
let $\RSym(S)$ be the set of rooted c-symmetries for
structures from $S$.
Lemma~\ref{lem:unlabel} shows that $\tilde{A}$ has $n!$
symmetries and Lemma~\ref{lem:cpointedcounting} shows that
each structure of $\wt{\cA^{\circ}_n}$ has $n!$ rooted c-symmetries.
Hence $|\Sym(\tilde{A})|=n!$ and $|\RSym(S)|=|S|n!$.
In addition, we have seen that a symmetry $(A,\sigma)$ of size $n$ on $\cA$ has $n$ rooted c-symmetries on $\cAp$.
Hence, $|\RSym(S)|=n |\Sym(\tilde{A})|$. Thus, 
we obtain $|S|n!=nn!$, and therefore $|S|=n$.
\end{proof}

\begin{remark}\label{rem:parker}
Theorem~\ref{theo:npointed} is equivalent to a result known as Parker's lemma~\cite[Section 2.8]{Ca}. For a subgroup $G$
of the symmetric group $\mathfrak{S}_n$, say that a cycle $c$ in $g\in G$
is equivalent to a cycle $c'$ in $g'\in G$ if there exists $h\in G$
that maps the elements of $c$ to the elements of $c'$ and preserves
the cyclic order, i.e., 
for each element $x\in c$, the successor of $x$ in $c$ is mapped
by $h$ to the successor of $h(x)$ in $c'$. Let $a_k$
be the number of inequivalent cycles of length $k$. Then Parker's lemma states that $\sum_{k=1}^na_k=n$.

If this lemma is applied to the automorphism group of a fixed structure $A$ of size $n$,
then $a_k$ is the number of unlabeled cycle-pointed structures arising
from $A$ and such that the marked cycle has length $k$.
So Parker's lemma states that there are $n$ unlabeled cycle-pointed structures arising from $A$, i.e., it implies Theorem~\ref{theo:npointed} 
(conversely, each permutation group is the automorphism group of a structure, so Parker's lemma can be deduced from Theorem~\ref{theo:npointed}).  
\end{remark}

\begin{remark}
The classical pointing operator, which selects a single atom in a structure, 
yields an equation similar to~(\ref{eq:Up}) for \emph{exponential} generating series, which is useful 
for \emph{labeled} enumeration. Given a species $\cA$, let $\cA^{\bullet}$ be the species
of structures from $\cA$ where an atom is distinguished. Then
\begin{equation}\label{eq:}
\cA^{\bullet}(x)=x\frac{\mathrm{d}}{\mathrm{d}x}\cA(x).
\end{equation}
An important contribution of this article is to define a pointing operator $\cA\mapsto\cA^{\circ}$ that yields  
 the same equation, Equation~\eqref{eq:Up}, in the \emph{unlabeled} case.
\end{remark}

\ignore{
Let us mention how we discovered cycle-pointing independently.
Random generation was our motivation to find an unbiased pointing
operator $\cA\mapsto\cA^{\circ}$ 
in the unlabeled setting (so that the uniform distribution on $\wt{\cA_n^{\circ}}$
induces  the uniform distribution on $\wt{\cA_n}$). We guessed how to define
the operator from the counting series. At the level of ordinary generating
series, unbiased pointing means that $\wt{\cA^{\circ}}(x)=x\tfrac{\mathrm{d}}{\mathrm{d}x}\wt{\cA}(x)$, i.e., each monom $x^n$ in $\wt{A}(x)$ is multiplied by $n$.
This is not of great help for guessing, but if instead of the ordinary series
one works with the finer cycle index sum then asking for each monomial
$s_1^{c_1}\cdots s_k^{c_k}$ to be multiplied by $n=\sum_iic_i$ (the size
of the automorphism) means that the derivation operator has to be
$\Delta:=\sum_{\ell\geq 1}s_{\ell}\frac{\partial}{\partial s_{\ell}}$.   
This indicates that we have to point at cycles in automorphisms (with the pointed
cycle of arbitrary length $\ell\geq 1$). 
}


\subsection{Basic Cycle-pointed Species and Constructions}
\label{sub:dictionary_polyacycle_pointed}

\begin{figure}
\begin{tabular}{|l|l|l|}
  \hline
\multicolumn{3}{|l|}{  \textbf{Basic cycle-pointed species}: $\cAp$, $\cA^{\cst}$,$^{\phantom{1^{\phantom{1^{\phantom{1}}}}}}$} \\[0.1cm]
\multicolumn{3}{|c|}{ where $\cA\in\{{\bf 0}, {\bf 1}, X, \Seq^{[k]}, \Set^{[k]}, \Cyc^{[k]}, \Seq, \Set, \Cyc\}$}\\[0.2cm]
 \multicolumn{3}{|l|}{ \textbf{Pointed cycle-index sum}:}\\
 \multicolumn{3}{|c|}{$\ds \bar Z_{\cAp}=\sum_{\ell\geq 1}\ell\ t_{\ell}\frac{\partial}{\partial s_{\ell}}Z_{\cA}$}\\
 \multicolumn{3}{|c|}{$\ds \bar Z_{\cA^{\cst}}=\sum_{\ell\geq 2}\ell\ t_{\ell}\frac{\partial}{\partial s_{\ell}}Z_{\cA}$}\\
  \hline
  \hline
  \textbf{Construction} & \textbf{Notation} & \textbf{Pointed cycle index sum}\\
  \hline
  \textbf{Pointed union} & $\cR=\cP + \cQ $ &  $\bar Z_{\cR} = \bar Z_{\cP} + \bar Z_{\cQ}^{\phantom{1}}$ \\[0.1cm]
  \textbf{Pointed product} & $\cR=\cP\star\cB $  & $\bar Z_{\cR}=\bar Z_{\cP} \bcdot Z_{\cB}$ \\[0.1cm]
  \textbf{Pointed substitution} & $\cR=\cP \circledcirc \cB $  & $\bar Z_{\cR}=\bar Z_{\cP} \circledcirc Z_{\cB}$ \\[0.1cm]
  \hline
\end{tabular}
\caption{The rules to calculate the pointed cycle index sums of 
basic cycle-pointed species and of cycle-pointed species composed by $+$, $\star$, and $\circledcirc$.} \label{BasicPointed}

\end{figure}

\subsubsection{Basic cycle-pointed species}
Species of the form $\cA^{\circ}$ or $\cA^{\cst}$
where $\cA$ is a basic species as introduced in Subsection~\ref{sub:dictionary_polyaunlabeled}
will be called \emph{basic cycle-pointed species}.
The derivation rule in Equation~(\ref{eq:ZcUpl}) allows us to compute the pointed cycle index sum
of basic cycle-pointed species, as indicated in Figure~\ref{BasicPointed} (upper part).

\vspace{0.2cm}

\subsubsection{Pointed constructions}
It is clear that the disjoint union of two cycle-pointed species
as defined in Section~\ref{sssect:constr} is again a cycle-pointed
species. We now adapt the constructions of 
product and substitution to obtain a pointed product and a pointed substitution operation that produce cycle-pointed species.

\vspace{0.2cm}

\emph{Pointed product.} 
Let $\cA,\cB$ be species, and let $\cP \subseteq \cA^\circ$ be
a cycle-pointed species. 
Then the pointed product $\cP \star \cB$ of $\cP$ and $\cB$ is 
the subspecies of $(\cA\bcdot\cB)^{\circ}$ of all those 
structures $((A,B),c)$ in $(\cA\bcdot\cB)^{\circ}$ where the pointed cycle $c$ is from $A$, and $(A,c) \in\cP$.

\vspace{0.2cm}


\emph{Pointed substitution.}
Let $E=((\pi,A,(B_p)_{p\in\pi}),c)$ be a structure in $(\cA\circ\cB)^{\circ}$. The study of automorphisms of structures in $\cA\circ\cB$ performed in Section~\ref{sec:aut_cons} shows that
\begin{itemize}
\item $c$ is the cycle composed from cycles $c_1,\ldots,c_k$ on different components $B_{p_1},\ldots,B_{p_k}$; 
\item the structure $(A,c'=(p_1,...,p_k))$ is cycle-pointed.
\end{itemize}
We call $(A,c')\in\cA^{\circ}$ the \emph{core-structure} of $E$.

We can now define a substitution construction for cycle-pointed species. Let $\cP\subseteq\cA^{\circ}$  and let $\cB$ be a species such that $\cB[\emptyset]=\emptyset$.
Then $\cP\circledcirc\cB$ is defined as the subspecies of structures from $(\cA\circ\cB)^{\circ}$ whose core-structure is in $\cP$.

\begin{figure}
\begin{center}
\includegraphics[width=11cm]{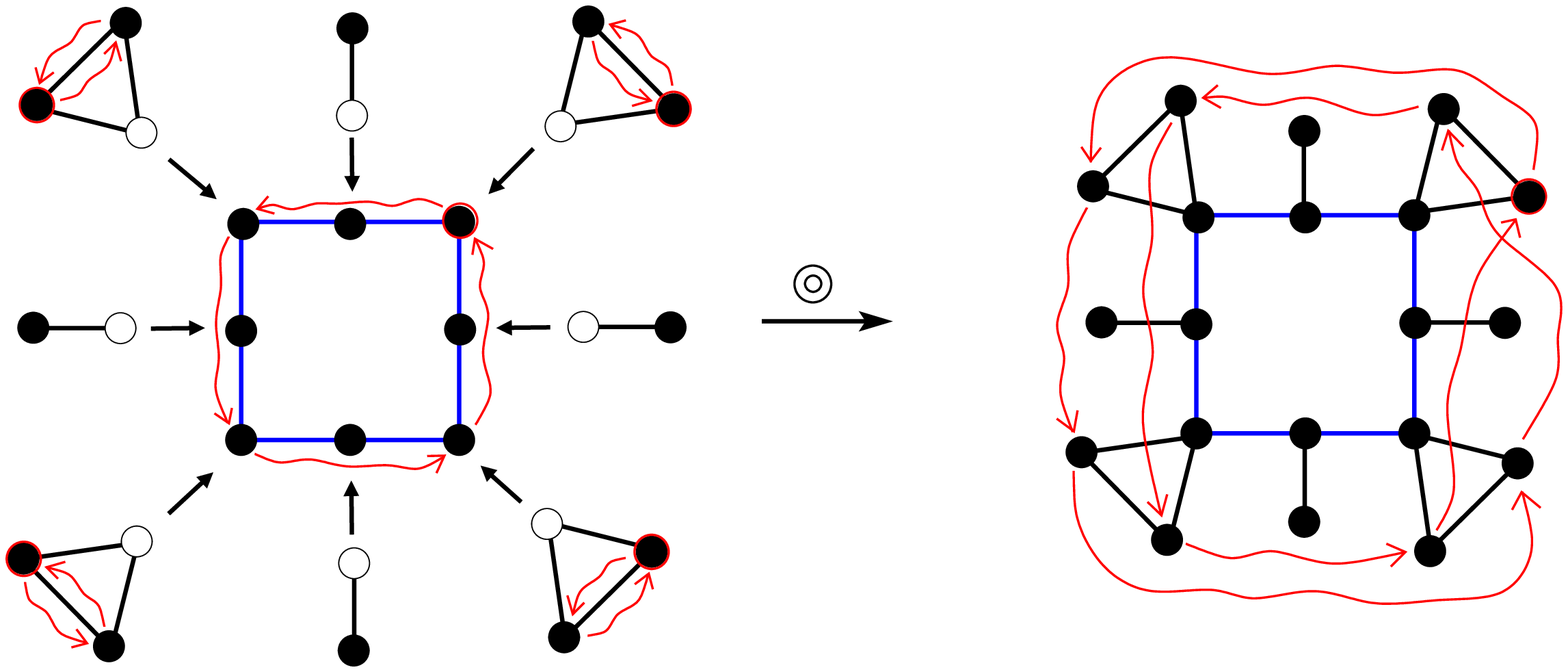}
\end{center}
\caption{A cycle-pointed graph obtained from a substitution.}
\label{fig:composition}
\end{figure}

As in the previous section, 
we make the convention that in expressions involving several
of the symbols $+,\bcdot,\star, \circ,\circledcirc$, the symbols $\circ,\circledcirc$ bind stronger than
the symbols $\bcdot,\star$, and the symbols $\bcdot,\star$ bind stronger
than the symbol $+$.


In a similar way to the labeled framework~\cite{BeLaLe}, 
our pointing operator behaves well with the three constructions $+$, $\bcdot$, and $\circ$:

\begin{proposition}\label{prop:point_rules}
The cycle-pointing operator obeys the following rules: 
\begin{eqnarray}
\left(\cA+\cB\right)^{\circ}&=&\cAp+\cB^{\circ},\label{eq:pointunion}\\
\left(\cA \bcdot \cB\right)^{\circ}&=&\cAp \star \cB\ +\ \cB^{\circ} \star \cA,\label{eq:pointproduct}\\
\left( \cA \circ \cB\right)^{\circ}&=&\cAp\circledcirc \cB \; .\label{eq:pointcomp}
\end{eqnarray}
\end{proposition}
\begin{proof}
It is easy to see that $(\cA+\cB)^{\circ}$ is isomorphic to $\cA^{\circ} + \cB^{\circ}$. For the product, 
note that the pointed cycle $c$ of a structures from 
$(\cA \bcdot \cB)^{\circ}$ has to be entirely on $A$ or entirely on $B$. 
The species that contains all structures $((A,B),c)$ where $c$ is on $A$ is isomorphic to $\cAp \star \cB$, 
and the species that contains all structures $((A,B),c)$ where
$c$ is on $B$ is isomorphic to $\cB^{\circ} \star \cA$. 
Hence, $(\cA\bcdot\cB)^{\circ} = \cAp\star\cB+\cB^{\circ}\star\cA$.
For the expression for the substitution operation, we in fact have not only isomorphism, but even equality of species.
This is clear from the fact that \emph{all} core-structures of structures
from $(\cA\circ\cB)^{\circ}$ are from $\cA^\circ$.
\end{proof}

As in the unpointed case, there are explicit rules to compute
 the pointed cycle index sums for each basic species and for each construction. To this end we need the following
 notion of composition for power series. 
 
 \begin{definition}
 Let $f$ and $g$ be two power series of the form
 $\bar{f}:=f(x_1,y_1;x_2,y_2;\ldots)$ and $g:=g(x_1,x_2,\ldots)$ such that
 $g(0,0,\ldots)=0$. 
 Then the \emph{pointed plethystic composition} of $f$ with $g$
 is the power series 
 \begin{align}
 \bar{f} \circledcirc g := f(g_1,\bar{h}_1;g_2,\bar{h}_2;\ldots), 
 \end{align}
with $g_k=g(x_k,x_{2k},x_{3k},\ldots)$  
and
$\bar{h}_k=\bar{h}(x_k,y_k;x_{2k},y_{2k};\ldots)$ for $\bar{h} := \sum_{\ell \geq 1} \ell t_{\ell} \frac{\partial}{\partial s_{\ell}} g$.
\end{definition}

The following proposition is the counterpart of Proposition~\ref{prop:dict} for cycle-pointed species.

\begin{proposition}[computation rules for pointed cycle index sums]\label{prop:comp_cyc}
For each basic species $\cA$,  the pointed cycle index sum 
of the pointed species $\cA^{\circ}$ and $\cA^{\cst}$ is given by the explicit expression
given in Figure~\ref{BasicPointed} (upper part) in terms of the
cycle index sum of $\cA$.
For each of the fundamental pointed constructions $+$, $\star$, and $\circledcirc$ there is an explicit rule, 
given in Figure~\ref{BasicPointed} (lower part), to 
compute the pointed cycle index sum of the resulting species.
\end{proposition}
\begin{proof}
The statement is clear for the cycle-pointed atomic species and the pointed union.
Let $\cA, \cB$ be species, and let $\cP \subseteq \cA^{\circ}$ be
a cycle-pointed species.
For the cycle-pointed product $\cQ=\cP\star\cB$ notice that, similarly as for a partitional product for species, a rooted c-symmetry on $\cQ$ decomposes into a rooted c-symmetry on $\cP$
and a symmetry on $\cB$, since the automorphism has to act separately on the two component structures.
Therefore, $\RSym(\cQ)$ can be considered as a partitional product of $\RSym(\cP)$ and $\Sym(\cB)$,
which yields $\bar Z_{\cQ}=\bar Z_{\cP}\bcdot Z_{\cB}$.

For the substitution construction, 
the proof is similar. 
Let $\cP\subseteq \cA^{\circ}$ be a cycle-pointed species and 
let $\cB$ be a species such that $\cB[\emptyset]=\emptyset$. 
Let $\cQ=\cP \circledcirc \cB\subseteq(\cA\circ\cB)^{\circ}$.
Consider a  rooted c-symmetry $(Q,\sigma,v)$.  
As we have seen in
Section~\ref{sec:decop_symme}, 
the core structure $A$ is endowed with an induced automorphism $\sigma'$.
In addition, the automorphism is naturally rooted at the atom $v' \in A$ where the $\cB^{\circ}$-component that contains the root $v$ is substituted. We denote the cycle of $\sigma'$ 
that contains $v'$ by $c'$. 
Now we have that $(A,c')$ is cycle-pointed and the automorphism $\sigma'$ rooted at $v'$
is a rooted c-symmetry on $P=(A,c')$. 
In addition, the components  
substituted at each atom of a cycle $c=(u_1,\ldots,u_k)$ of $\sigma'$  are 
isomorphic copies of a same symmetry on $\cB$. 
The components that are substituted at the atoms
of the marked cycle $c'$ are naturally rooted at the isomorphic representant of $v$. 
Finally, this decomposition is reversable: one can go back to the original composed symmetry
using the composition of cycle operation. 

To express these observations in an equation, we
define 
the \emph{type of the rooted c-symmetry} $(P,\rho,v)$ with $P=(A,c')$ to be
the sequence $(\ell; n_1,n_2,\dots,n_k)$ where $\ell$ 
is the length of $c'$, and $n_i$ is the number of unmarked cycles of length $i$ in $\rho$. Note that the size of $P$ is $\ell+\sum_ii n_i$.
The number $a(t)$ of rooted c-symmetries of type $t=(\ell;n_1,n_2,\dots,n_n)$ on $\cP$ is $n![t^{\ell}s_1^{n_1} \cdots s_n^{n_n}]\bar Z_{\cP}$. 
The \emph{core type} of a rooted c-symmetry on $\cP \circledcirc \cB$ is defined as the 
type of the rooted c-symmetry induced on the core structure.

Let $\bar Z_{\cR}^{(t)}$ be the pointed cycle index sum of $\cP$ restricted to the rooted c-symmetries
with core-type $t=(\ell;n_1,n_2,\dots,n_n)$. From the above discussion, we have
\begin{equation}\label{eq:comp_cyc_subs}
\bar Z_{\cR}^{(t)}=\frac{a(t)}{n!}q_{\ell}b_1^{n_1}b_2^{n_2}\ldots b_n^{n_n},\ \mathrm{where}\ b_i:=Z_{\cB}(s_1,s_2,\ldots),\ q_{\ell}:=\bar Z_{\cB^{\circ}}(s_{\ell},t_{\ell};s_{2\ell},t_{2\ell};\ldots).
\end{equation}
Summing over all possible types of rooted $c$-symmetries $t$, we obtain\\[.2cm]
\phantom{1}\hspace{0.7cm}$
\ds\bar Z_{\cR}=\bar Z_{\cP}(b_1,q_1;b_2,q_2;\ldots),\ \mathrm{where}\ b_i:=Z_{\cB}(s_1,s_2,\ldots),\ q_{\ell}:=\bar Z_{\cB^{\circ}}(s_{\ell},t_{\ell};s_{2\ell},t_{2\ell};\ldots).
$
\end{proof}

In the following we introduce recursive specifications that involve
pointed constructions. Cycle-pointed specifications are like standard recursive specifications (Definition~\ref{def:rec-standard}),
 but with two sorts of variables (where one is reserved for
 cycle-pointed species) and where we are allowed to use
 additionally the pointed constructions.


\begin{definition}\label{def:dec_cycle}
A \emph{recursive cycle-pointed specification}
with variables $x_1,\dots,x_m,y_1,\dots,y_{m'}$
over the species $\cA_1,\dots,\cA_\ell$ and over cycle-pointed species
$\cB_1,\dots,\cB_k$
is a system $\Psi$ of equations $x_1=e_1,\dots,x_m=e_m,y_1=f_1,\dots,y_{m'}=f_{m'}$ where each $e_i$ is 
\begin{itemize}
\item of the form $a+b$ or $a \bcdot b$ with $a, b \in
\{x_1,\ldots,x_m,\cA_1,\dots,\cA_\ell\}$, or 
\item of the form
$a \circ b$ with $a \in \{\cA_1,\dots,\cA_\ell\}$ 
and $b \in \{x_1,\dots,x_m, \cA_1,\dots,\cA_\ell\}$,
\end{itemize}
and each $f_i$ is 
\begin{itemize}
\item of the form $a+b$ with $a, b \in
\{y_1,\ldots,y_{m'},\cB_1,\dots,\cB_k\}$, or
\item of the form $a \star b$ with $a \in
\{y_1,\ldots,y_{m'},\cB_1,\dots,\cB_k\}$ and $b \in
\{x_1,\ldots,x_{m},\cA_1,\dots,\cA_\ell\}$, or
\item of the form $a \circledcirc b$ 
with $a \in \{\cB_1,\dots,\cB_k\}$ 
and $b \in \{x_1,\dots,x_m, \cA_1,\dots,\cA_\ell\}$.
\end{itemize}
\end{definition}

To define the species $\cX_1,\dots,\cX_m,\cY_1,\dots,\cY_{m'}$
that are given by a recursive cycle-pointed specification $\Psi$
with variables $x_1,\dots,x_{m},y_1,\dots,y_{m'}$ over the species
$\cA_1,\dots,\cA_\ell,\cB_1,\dots,\cB_k$ where $\cB_1,\dots,\cB_k$ are pointed, we again (as in Section~\ref{sssect:rec}) consider sequences of species $\cX_j^{(i)}$ and $\cY_j^{(i)}$ for $i \geq 1$. For $i=1$,
we define $\cX_j^{(i)} = \bf 0$ for all $1 \leq j\leq m$,
and $\cY_j^{(i)} = \bf 0^\circ$ for all $1 \leq j\leq {m'}$.
For $i > 1$ the species $\cX^{(i)}_j$ and $\cY^{(i)}_j$ are 
obtained by evaluating the corresponding expressions for $x_j$ and $y_j$, respectively (as in Section~\ref{sssect:rec}).
We say that $\Psi$ is \emph{admissible} if in expressions of the
form $a \circ b$ or $a \circledcirc b$ the species substituted for $b$ never 
contain structures of size $0$. 

Note that also the new pointed constructions are monotone, so
in case that for each $n$ the sets
$\bigcup_{i \geq 1}  \cX^{(i)}[n]$ and $\bigcup_{i \geq 1}  \cY^{(i)}[n]$ are finite
it is straightforward (and analogous to Definition~\ref{def:semantics}) to define the species $\cX_1,\dots,\cX_m,\cY_1,\dots,\cY_{m'}$ specified by admissible recursive specifications $\Psi$ over 
$\cA_1,\dots,\cA_\ell,\cB_1,\dots,\cB_k$.


\begin{definition}
Let $\mathfrak A$ be a class of species.
The class of species that is \emph{cycle-pointing decomposable
over $\mathfrak A$} 
is the smallest class of species $\mathfrak B$ that contains $\mathfrak A$, contains all species that can be specified by 
cycle-pointed recursive specifications over pointed and
unpointed species from $\mathfrak B$,
and contains all species obtained from species of the form $\cA^{\circ}$ 
in $\mathfrak B$ by applying the unpointing operation.
\end{definition} 
Plenty of examples of species that are decomposable over
simple basic species can be found in Section~\ref{sec:appl}.

\begin{proposition}
If a species $\cA$ is decomposable (in the sense of Definition~\ref{def:dec-standard}), then the pointed species $\cA^{\circ}$
is cycle-pointing-decomposable.
\end{proposition}
\begin{proof}
Follows directly from Proposition~\ref{prop:point_rules}.
\end{proof}

\begin{remark}\label{remark:OGS_point}
The ordinary generating series inherit 
simple computation rules from the ones 
for pointed cycle index sums. As expected, for the sum and product constructions, one gets
\begin{eqnarray}
\hspace{-1.5cm}\cR=\cP+\cQ & \Rightarrow & \wt{\cR}(x)=\wt{\cP}(x)+\wt{\cQ}(x),
\end{eqnarray}
\begin{eqnarray}
\hspace{-1.5cm}\cQ=\cP\star\cB & \Rightarrow & \wt{\cQ}(x)=\wt{\cP}(x)\cdot\wt{\cB}(x).
\end{eqnarray}
For the substitution construction, the computation rule is:
\begin{eqnarray}
\hspace{1cm}\cQ=\cP\circledcirc \cB & \Rightarrow & \wt{\cQ}(x)=\bar Z_{\cP}(\wt{\cB}(x),\wt{\cB^{\circ}}(x);\wt{\cB}(x^2),\wt{\cB^{\circ}}(x^2);\ldots),
\end{eqnarray}
where $\wt{\cB^{\circ}}(x)=x\frac{\mathrm{d}}{\mathrm{d}x}\wt{\cB
}(x)$.
Hence, to compute the ordinary generating series of a decomposable cycle-pointed species, the only place where the cycle index sum or pointed cycle index sum is needed (as a refinement
of ordinary generating series) is for the species that is the first argument of a substitution or pointed 
substitution construction.
\end{remark}

\begin{remark} As an exercise, the reader can check just by standard algebraic manipulations that the computation
rules for cycle-index sums 
are consistent with Proposition~\ref{prop:point_rules}. For instance, proving
$\bar Z_{(\cA\circ\cB)^{\circ}}=\bar Z_{\cA^{\circ}\circledcirc \cB}$ is equivalent (by the 
computation rules) to proving the equality
$\bar Z_{(\cA\circ\cB)^{\circ}}=\bar Z_{\cA^{\circ}}\circledcirc Z_{\cB}$,
which reduces to checking the following identity on power series:
\begin{equation}\Delta (f \circ g)=(\Delta f)\circledcirc g \; ,\end{equation}
where $\Delta$ is the operator that associates to a power series $f(x_1,x_2,x_3,\ldots)$
the power series 
$$\Delta f(x_1,y_1;x_2,y_2;\ldots):=\sum_{\ell\geq 1}\ell\ y_{\ell}\frac{\partial}{\partial x_{\ell}}f(x_1,x_2,x_3,\ldots).$$
Similarly, to prove $\bar Z_{(\cA+\cB)^{\circ}}=\bar Z_{\cA^{\circ}+\cB^{\circ}}$ and $\bar Z_{(\cA\bcdot\cB)^{\circ}}=\bar Z_{\cA^{\circ}\star\cB+\cB^{\circ}\star\cA}$, one has to check the 
identities $\Delta(f+g)=\Delta f+\Delta g$, and $\Delta(f\cdot g)=\Delta f\cdot g+f\cdot\Delta g$,
respectively.
\end{remark}

\section{Application to Enumeration}
\label{sec:appl}
In this section we demonstrate that cycle-pointing provides a new way of counting many classes of combinatorial structures in the unlabeled setting. 
Typically, species satisfying a ``tree-like''
decomposition are amenable
to our method. This includes of course species of trees, but also species
of graphs (provided that the species is closed under taking 2-connected components, and that the sub-species of 2-connected graphs is tractable), and species of planar maps.

The general scheme to enumerate unlabeled structures of a 
species $\cA$,
i.e., to obtain the coefficients $|\wt{\cA_n}|$, is as follows. First,
we observe that the task is equivalent to the task to
enumerate unlabeled \emph{cycle-pointed}
structures from $\cA^{\circ}$, because 
$|\wt{\cA^{\circ}_n}|=n|\wt{\cA_n}|$.
Enumeration for $\cA^{\circ}$ turns out to be easier since
the marked cycle usually provides a starting point for a recursive decomposition.

For a cycle-pointed species of trees (and more generally for 
species satisfying tree-like decompositions), the first step of the decomposition scheme is 
to distinguish whether the 
marked cycle has length $1$ or greater than $1$.
The general equation is
\begin{equation}\label{eq:count_cycl}
\cA^{\circ}=X^{\circ}\star\cA'+\cA^{\cst},
\end{equation}
where $\cA'$ is the \emph{derived species} of $\cA$, consisting of structures
from $\cA$ where one atom is marked with a special label, say $*$, as defined in
\cite{BeLaLe}\footnote{Note that the derived species $\cA'$ is not cycle-pointed. However, it can be identified with $\cA^{\circ}_{(1)}$. Indeed, by adding a new label and a pointing loop on the
marked atom, one obtains a bijective correspondence between $\cA^{\circ}_{(1)}$ and $X^{\circ}\star\cA'$.}.
Then each of the two species $\cA'$ and $\cA^{\cst}$ has to be decomposed.
For derived structures (the species $\cA'$) we follow the classical root decomposition. For symmetric cycle-pointed structures (the species $\cA^{\cst}$) our decomposition strategy is different, and leads us to introduce the notion of \emph{center of symmetry}.

\subsection{Trees}
We first illustrate our decomposition method for trees, which are 
 defined as connected acyclic graphs (i.e., unless mentioned otherwise, trees are \emph{unrooted}),
and we start with the formal definition of the \emph{center of symmetry}.
Let $T$ be a symmetric cycle-pointed tree. 
A path of $T$ connecting two consecutive atoms of the marked cycle
  is called a \emph{connecting path} (thus the number of connecting
  paths is the size of the marked cycle).
  
  \begin{claim}[center of symmetry]
  Given a symmetric cycle-pointed tree $T$, 
  all connecting paths of $T$
  share the same middle $v_c$, called the \emph{central point} 
  for the marked cycle of $T$. 
  The central point $v_c$  is the middle of an edge $e$ if these paths
  have odd length, and is a vertex $v$ if these paths have even length.
  In the first (second) case, the edge $e$ (the vertex $v$, resp.) is called
  the \emph{center of symmetry} of $T$. 
  \end{claim}  
  \begin{proof}
  We prove here that all connecting paths share the same middle.
  Let $U$ be the subgraph of $T$ formed by the union of all connecting paths.
  Observe that $U$ is connected, so $U$ is a subtree of $T$.
  In addition, $U$  is globally fixed by any c-automorphism of $T$
  (indeed, the property of being on a connecting path is invariant under
  the action of a c-automorphism), and it contains
  the atoms of the marked cycle of $T$. Hence $U$ is the underlying structure of a cycle-pointed tree $(U,c)$.
  Consider the classical center of $U$, obtained by pruning the leaves (at each step,
  all leaves are simultaneously deleted) until the resulting tree is reduced to an edge
  or a vertex~\cite{BeLaLe}. The central point $v_c$ of $U$ is defined as follows:
  if the center of $U$ is a vertex $v$, then $v_c:=v$, if the center of $U$ is an edge
  $e$, then $v_c$ is the middle of $e$.
  Let $\sigma$ be a c-automorphism of $(U,c)$ and let $\langle\sigma\rangle$ be the group
  of automorphisms generated by $\sigma$.
  It is well known that the central point is fixed by any automorphism on the tree,
  hence $v_c$ is equidistant from all atoms of the marked cycle,
  as the group $\langle\sigma\rangle$ acts transitively on the vertices 
  of the marked cycle.   In addition, $v_c$ is on at least one connecting 
  path of $T$ (because $U$ is the union of these connecting paths). Hence, $v_c$ has to be on all connecting paths, as the group $\langle\sigma\rangle$ acts transitively on the connecting paths.
  Thus, $v_c$ has to be the middle of all connecting paths simultaneously. 
    \end{proof}
  
\begin{figure}
  \begin{center}
  \includegraphics[width=12cm]{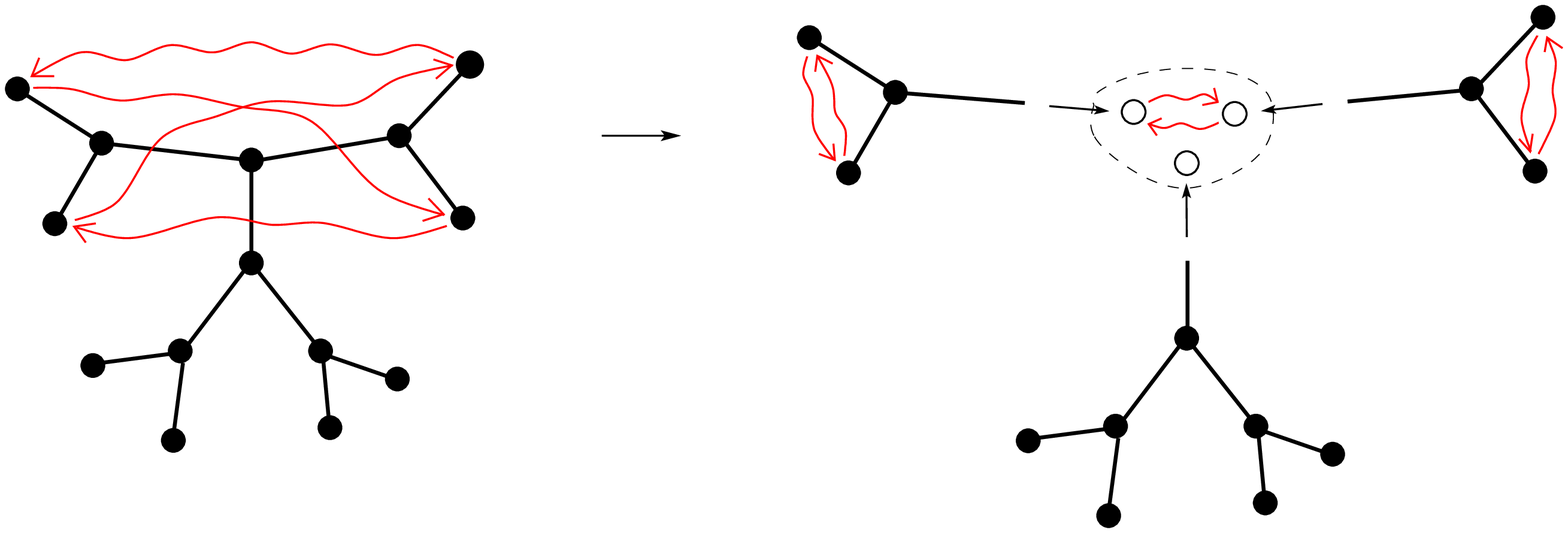}
  \end{center}
  \caption{Decomposition of a nonplane tree at its center of symmetry (in case  
  the center of symmetry is a vertex).}
  \label{fig:trees}
  \end{figure}

  \begin{remark}
  Notice that the center of symmetry  might not
  coincide with the classical center of the tree, as shown in Figure~\ref{fig:trees}.
  However, in the case of \emph{plane trees}, the two notions of center coincide.
  \end{remark}
  
\subsubsection{Nonplane trees}
Let $\cF$ be the species of \emph{free trees}, i.e., unrooted nonplane trees (equivalently, acyclic connected
graphs), where the vertices are taken as atoms.
Let $\cF\ \!'$ be the derived species of $\cF$ (also the species of derived nonplane trees). 
Rooted nonplane trees can be decomposed
at the root. Since the root does not count as an atom and since the children of the root node are unordered, we classically have 
\begin{equation}\label{eq:first_trees}
\cF\ \!'=\Set\circ(X \bcdot \cF\ \!').
\end{equation}

In contrast, the decomposition of symmetric cycle-pointed trees does not start at atoms 
of the marked cycle, but at the center of symmetry,
which is either an edge or a vertex (see Figure~\ref{fig:trees} for an illustration of the decomposition). In order to write down the decomposition, we introduce the species $\cL$ consisting of a single one-edge graph. Note that $\cL\simeq\Set^{[2]}$, and that  $\cL^{\cst}$ consists of the link graph carrying a marked cycle of length 2 that exchanges the two extremities of the edge.

\begin{claim}\label{cl:sym_trees}
The species $\cFs$ of symmetric cycle-pointed free trees satisfies
\begin{equation}\label{eq:second_trees}
\cFs=\cL^{\cst}\circledcirc\cR+X \bcdot \Set^{\cst}\circledcirc\cR,
\end{equation}
where $\cR:=X\bcdot\cF$ is the species of all pointed trees.
\end{claim}
\begin{proof}
Consider a tree produced from the species $\cL^{\cst}\circledcirc\cR+X \bcdot \Set^{\cst}\circledcirc\cR$ 
(for an example of a tree produced from
 $X \bcdot \Set^{\cst}\circledcirc\cR$, see the transition between the right and the
left drawing in Figure~\ref{fig:trees}).
Clearly, such a tree is  
 free and cycle-pointed and it is symmetric because the marked cycle of the 
core-structure -- an edge $e$ in the first case, a cycle-pointed set 
attached to a vertex $v$ 
in the second case -- already has length greater than 1. 
Hence $\cFs\supseteq\cL^{\cst}\circledcirc\cR+X \bcdot \Set^{\cst}\circledcirc\cR^{\circ}$.
Notice also that in the first (second) case, $e$ ($v$, respectively) 
 is the center of symmetry
of the resulting tree. Indeed each connecting path connects vertices on two
different subtrees attached at the center of symmetry, which, by symmetry,
stands in the middle of such a path. 

Conversely, for each symmetric cycle-pointed free tree $T$, we color blue its center
of symmetry, which plays the role of a \emph{core-structure} for $T$. 
Partition $\cFs$ as $\cFvs+\cFes$, where $\cFvs$ ($\cFes$, respectively)
gathers the trees in $\cFs$ whose center of symmetry is a vertex
(an edge, respectively).
Define also $\cMv$ ($\cMe$) as the species of free trees with a distinguished vertex
(edge, resp.) that is colored blue.
Clearly $\cMv=X \bcdot (\Set \circ \cR)$ and $\cMe=\cL\circ\cR$.
 From Proposition~\ref{prop:point_rules}, we obtain $\cMvs = X^{\circ}\star (\Set \circ \cR) +X \bcdot \Set^{\circ}\circledcirc\cR$ and $\cMep= \cL^{\circ}\circledcirc\cR$. Observe that $\cFvs$ ($\cFes$) 
 contains the structures of $\cMvs$ 
 (of $\cMes$) where the 
 blue vertex (edge, resp.) is the center of symmetry. It is clear that the 
 structures of $X^{\circ}\star (\Set \circ \cR)$ have their marked cycle of length 1,
 so they are not in $\cFvs$.
 Concerning the structures of $X\bcdot\Set^{\circ}_{(1)}\circledcirc\cR$,
 the atoms of the marked cycle are on a same subtree attached at the blue vertex,
 so that  the blue vertex is not the center of symmetry. Hence the structures of
 $X\bcdot\Set^{\circ}_{(1)}\circledcirc\cR$ are not in  $\cFv^{\cst}$.
 Similarly, the structures of $\cL^{\circ}_{(1)}\circledcirc\cR$ are not
 in $\cFes$. Therefore we obtain the second inclusion 
 $\cFs\subseteq\cL^{\cst}\circledcirc\cR+X \bcdot \Set^{\cst}\circledcirc\cR$.
 \end{proof}



\begin{proposition}[decomposing and counting free trees]\label{prop:gram_trees}
The species $\cF^{\circ}$ of cycle-pointed free trees has
the following cycle-pointed recursive specification 
over the species $\Set$, $\mathcal{L}^{\cst}$, $X$:
\begin{equation}
\left\{
\begin{array}{rcl}
\cFp&=&X^{\circ}\star\cF\ \!'+\cFs,\\
\cF\ \!'&=&\Set\circ \cR,\ \ \cR=X\bcdot\cF\ \!',\\
\cFs&=&\mathcal{L}^{\cst}\circledcirc\cR+X \bcdot \Set^{\cst}\circledcirc\cR,\\
\cR^{\circ}&=&X^{\circ}\star (\Set\circ\cR)+X\bcdot\Set^{\circ}\circledcirc\cR, \label{eq:gram_trees}
\end{array}
\right.
\end{equation}

The ordinary generating function $f(x):=\wt{\cF}(x)$ of free trees satisfies the equations
\begin{eqnarray}
xf'(x) &=& r(x)+x^2r'(x^2)+\left( \sum_{\ell\geq 2}x^\ell r'(x^\ell)\right)r(x)\label{eq:first_f}\\
&=& xr'(x)(1-r(x))+x^2r'(x^2),\label{eq:second_f}
\end{eqnarray}
where $r(x)$ is specified by  $\ds r(x)=x\exp\left(\sum_{i\geq 1}\frac{1}{i}r(x^i)\right)$.
\end{proposition}
\begin{proof}
The first three lines of the grammar are Equations~(\ref{eq:count_cycl}),~(\ref{eq:first_trees}), and~(\ref{eq:second_trees}), 
respectively. The fourth line\footnote{The fourth line of Equation~(\ref{eq:gram_trees}) is not needed for enumeration, but it is necessary to make the grammar completely recursive, and, as such, will be necessary for 
writing down a random generator in Section~\ref{sec:sampler}.} 
is obtained from the second line (i.e., $\cR=X\bcdot (\Set \circ \cR)$) using the derivation rules of Proposition~\ref{prop:point_rules}.

Concerning the OGSs, let $r(x):=\wt{\cR}(x)$ be the OGS of the species $\cR$.
Note that $\wt{\cF^{\circ}}(x)=xf'(x)$ and $\wt{\cR^{\circ}}(x)=xr'(x)$ by 
Theorem~\ref{theo:npointed}. 
By the computation rules for OGSs (Remark~\ref{remark:OGS} and Remark~\ref{remark:OGS_point}), 
the second line of the grammar, i.e., $\cR=X\bcdot (\Set \circ \cR)$, yields 
$r(x)=x\exp(\sum_{i\geq 1}r(x^i)/i)$;
and the third
line of the grammar yields 
$$\wt{\cF^{\cst}}(x)=\oZ_{\cL^{\cst}}(r(x),xr'(x);r(x^2),x^2r'(x^2);\ldots)+x\oZ_{\Set^{\cst}}(r(x),xr'(x);r(x^2),x^2r'(x^2);\ldots).$$ 
Applying the derivation rule~(\ref{eq:ZcUpl})  to $\cL = \Set^{[2]}$ and  
$\Set$, we get the expressions  
$\oZ_{\cL^{\cst}}\!=\!t_2$ 
and $\oZ_{\Set^{\cst}}\!=\!(\sum_{\ell\geq 2}t_{\ell})\bcdot Z_{\Set}$. 
Hence $\wt{\cF^{\cst}}(x)\!=\!x^2r'(x^2)+\big(\sum_{\ell\geq 2}x^{\ell}r'(x^{\ell})\big)\cdot r(x)$.
Finally, the first line of the grammar yields $\wt{\cF^{\circ}}(x)=r(x)+\wt{\cF^{\cst}}(x)$,
which gives Expression~(\ref{eq:first_f}) of $xf'(x)$. 
Using $xr'(x)=r(x)(1+\sum_{\ell\geq 1}x^{\ell}r'(x^{\ell}))$, this expression simplifies to Expression~(\ref{eq:second_f}) of $xf'(x)$.
\end{proof}

\begin{remark}
The expression $xf'(x)=xr'(x)(1-r(x))+x^2r'(x^2)$ clearly agrees with Otter's formula~\cite{Otter}: 
\begin{equation}
f(x)=r(x)-\frac{1}{2}(r^2(x)-r(x^2)),
\end{equation}
which can be obtained either from Otter's dissimilarity equation or from 
the dissymmetry theorem~\cite{BeLaLe}.
The new result of our method is to yield an expression for $xf'(x)$ 
-- Equation~(\ref{eq:first_f}) -- that  has only positive signs, as it reflects a positive decomposition grammar. This is crucial to obtain random generators without rejection in Section~\ref{sec:sampler}.
\end{remark}

All the arguments we have used for free trees can be adapted to decompose and enumerate species $\cFo$  
of trees where the degrees of the vertices lie in a finite integer set $\Omega$ that contains $1$. 
It is helpful to define the auxiliary species $\cRo$ 
that consists of  trees from $\cFo$ rooted at a leaf that does not count
as an atom. By decomposing trees at the root, we note that $\cRo$ has the recursive specification
$$
\cRo=X\bcdot\Set_{\Omega-1}\circ\cRo,\ \ \mathrm{with}\ \ \Set_{\Omega-1}:=\cup_{k\in\Omega}\Set^{[k-1]} \; .
$$
The species $\cRo$ serves as elementary rooted species 
to express the pointed species arising from $\cFo$. 

\begin{proposition}[decomposing and counting degree-constrained trees]
For any finite set $\Omega$ of positive integers containing $1$,  let $\cFo$ be 
the species of nonplane trees whose vertex degrees are in $\Omega$. Then the species $\cFop$ has the following cycle-pointed recursive specification, where $\Set_{\Omega}:=\cup_{k\in\Omega}\Set^{[k]}$ and $\Set_{\Omega-1}:=\cup_{k\in\Omega}\Set^{[k-1]}$:
\begin{equation}
\left\{
\begin{array}{rcl}
\cFop&=&X^{\circ}\star\cFo\ \!'+\cFos,\\
\cFo\ \!'&=&\Seto\circ\cRo,\ \ \cRo=X\bcdot\Set_{\Omega-1}\circ\cRo,\\
\cFos&=&\mathcal{L}^{\cst}\circledcirc\cRo+X \bcdot \Setos\circledcirc\cRo,\\
\cRop&=&X^{\circ}\star\Seto\circ\cRo+X\bcdot\Setop\circledcirc\cRo. \label{eq:gram_degree_nonplane}
\end{array}
\right.
\end{equation}

The ordinary generating function $f_{\Omega}(x):=\wt{\cFo}(x)$ satisfies the
equation
\begin{equation}
\label{eq:ex_trees}
x\fomeg '(x) = xZ_{\Set_{\Omega}}(r_{\Omega}(x),r_{\Omega}(x^2),\ldots)+x^2\romeg'(x^2)+x \oZ_{\Seto^{\cst}}(\romeg(x),x\romeg'(x);\romeg(x^2),x^2\romeg'(x^2);\ldots),
\end{equation}
where $\romeg(x)$ is specified by  $\ds \romeg(x)=x\bcdot Z_{\Set_{\Omega-1}}(\romeg(x),\romeg(x^2),\romeg(x^3),\ldots)$. The power series $Z_{\Set_{\Omega-1}}$, $Z_{\Seto}$ and $\oZ_{\Seto^{\cst}}$
appearing in the equation
are polynomials that can be computed explicitly.

\end{proposition}

\begin{example}\emph{Unrooted nonplane binary trees.}
Trees whose vertex degrees are in $\Omega:=\{1,3\}$ are called 
unrooted nonplane  binary trees (note that rooting such a tree at a leaf, one obtains
 a  rooted nonplane binary tree, i.e., each internal node has two unordered children).
 In that case, the elementary cycle index sums required in Equation~\eqref{eq:ex_trees}
 are
 $$
 Z_{\Set_{\Omega-1}}=1+\frac1{2}s_1^2+\frac1{2}s_2,\ \  Z_{\Seto}=s_1+\frac1{6}s_1^3+\frac1{2}s_1s_2+\frac1{3}s_3,\ \ \oZ_{\Seto^{\cst}}=t_2s_1+t_3.
 $$
 Let $f(x)$ be the OGS of unrooted nonplane binary trees and $r(x)$ the OGS of
 rooted nonplane binary trees (rooted at a leaf that does not count as an atom).
 Firstly, from the expression of $Z_{\Set_{\Omega-1}}$ one obtains
 $$r(x)=x\bcdot(1+\frac1{2}r(x)^2+\frac1{2}r(x^2)).$$
Then, Equation~\eqref{eq:ex_trees} yields
 $$
 xf'(x)=x\bcdot (r(x)+\frac1{6}r(x)^3+\frac1{2}r(x)r(x^2)+\frac1{3}r(x^3))+x^2r'(x^2)
 +x\bcdot(x^2r'(x^2)r(x)+x^3r'(x^3)),
 $$
 i.e.,
 $$
 f'(x)=r(x)+\frac1{6}r(x)^3+\frac1{2}r(x)r(x^2)+\frac1{3}r(x^3)+xr'(x^2)\bcdot(1+xr(x))+x^3r'(x^3).
 $$
 From this equation one can extract the counting coefficients of unrooted nonplane
 binary trees with respect to  the number of vertices (after extracting first the coefficients of $r(x)$):
 $$
 xf'(x)=2\cdot\mathbf{1}\cdot x^2 +4\cdot\mathbf{1}\cdot x^4+6\cdot\mathbf{1}\cdot x^6+8\cdot\mathbf{1}\cdot x^8+10\cdot\mathbf{2}\cdot x^{10}+12\cdot\mathbf{2}\cdot x^{12}+14\cdot\mathbf{4}\cdot x^{14}+16\cdot\mathbf{6}\cdot x^{16}+\ldots $$
Hence the first counting coefficients with respect to the number of internal nodes (starting with $0$ internal nodes) are $1$, $1$, $1$, $1$, $2$, $2$, $4$,  $6$. Pushing further one gets
 $1$, $1$, $1$, $1$, $2$, $2$, $4$,  $6$, $11$, $18$, $37$, $66$, $135$, $265$, $552$,  $1132$, which coincides with Sequence A000672 in~\cite{Sloane} (the number of 
 trivalent trees with $n$ nodes).
\end{example}

\subsubsection{Plane trees.}

A \emph{plane tree} is a tree endowed with an explicit \emph{embedding} in the plane.
Hence, a plane tree is a tree where the cyclic order around each vertex matters.
Let $\cE$ be the species of plane trees, where again the atoms are the vertices. 
As usual the startegy to count plane trees is to decompose $\cE^{\circ}$, distinguishing whether the marked cycle has length $1$ or larger than $1$:
\begin{equation}
\cEp=X^{\circ}\star\cE'+\cEs.
\end{equation}

The species $\cE'$ is decomposed with the help of another species of plane trees: denote by $\cA$ the species of plane trees rooted at a leaf which does not count as an atom. Decomposing $\cA$ at the root, we get
\begin{equation}
\cA=X\bcdot\Seq\circ\cA.
\end{equation}
Again the species $\cA$ serves as elementary rooted species to express species of pointed plane trees:

\begin{proposition}[decomposing and counting plane trees]
The species $\cEp$ of cycle-pointed plane trees has the following
cycle-pointed recursive specification.
\begin{equation}
\left\{
\begin{array}{rcl}
\cEp&=&X^{\circ}\star\cE\ \!'+\cEs,\\
\cE\ \!'&=&\Cyc\circ\cA,\ \ \cA=X\bcdot\Seq\circ\cA,\\
\cEs&=&\mathcal{L}^{\cst}\circledcirc\cA+X \bcdot \Cyc^{\cst}\circledcirc\cA,\\
\cA^{\circ}&=&X^{\circ}\star\Seq\circ\cA+X\bcdot\Seq^{\circ}\circledcirc\cA, \label{eq:gram_plane}
\end{array}
\right.
\end{equation}
The ordinary generating function $\wt{\cE}(x)$ of plane trees satisfies the 
equation:
\begin{eqnarray}
\label{eq:ex_trees_plane}
e'(x) &=& 1+\sum_{\ell\geq 1}\frac{\phi(\ell)}{\ell}\log\frac1{1-a(x^{\ell})}+xa'(x^2)+\sum_{\ell\geq 2}\phi(\ell)\frac{x^\ell a'(x^\ell)}{1-a(x^{\ell})},
\end{eqnarray}
where $a(x)$ is the series of Catalan numbers: $\ds a(x)=\frac1{2}\big(\ \!1-\sqrt{1-4x}\ \!\big)=\sum_{n\geq 0}\frac{1}{n+1}\binom{2n}{n}x^{n+1}$.

By coefficient extraction, one gets the following formula for the number $e_n$
of plane trees with $n+1$ vertices (entry A002995 in~\cite{Sloane}):
\begin{equation}
e_n=\frac1{n+1}\Big[   \frac1{2n}\binom{2n}{n}+\frac{n+1}{2n}\sum_{k|n,k\neq n}
\phi(n/k)\binom{2k}{k}+\mathbf{1}_{\left\{\substack{n\ \mathrm{odd}\\ n=2n'\!+\!1}\right\}}\cdot \binom{2n'}{n'}      \Big].
\end{equation}
\end{proposition}
\begin{proof}
The grammar is obtained by arguments similar to those used to derive the
grammar~\eqref{eq:gram_trees} for free trees.
The only difference is that the cyclic order of the neighbors around each vertex
matters, so a $\Set$ construction in the grammar for free trees typically has
to be replaced by a $\Cyc$ construction in the grammar for plane trees.  
\end{proof}

All the arguments apply similarly for species of plane trees where the degrees of
vertices are constrained. As a counterpart to Proposition~\ref{prop:gram_trees}, we obtain:  

\begin{proposition}[decomposing and counting degree-constrained plane trees]\label{prop:gram_plane_dgree}
For any finite  set $\Omega$ of positive integers containing $1$, let $\cEo$ be the species of free trees where the degrees of vertices are constrained to lie in $\Omega$. Then the
cycle-pointed species $\cEop$ is decomposable, it satisfies the following decomposition grammar, where $\Cyco:=\cup_{k\in\Omega}\Cyc^{[k]}$ and $\Seq_{\Omega-1}:=\cup_{k\in\Omega}\Seq^{[k-1]}$:
\begin{equation}
\left\{
\begin{array}{rcl}
\cEop&=&X^{\circ}\star\cEo\ \!'+\cEos,\\
\cEo\ \!'&=&\Cyco\circ\cAo,\ \ \cAo=X\bcdot\Seq_{\Omega-1}\circ\cAo,\\
\cEos&=&\mathcal{L}^{\cst}\circledcirc\cAo+X \bcdot \Cycos\circledcirc\cAo,\\
\cAop&=&X^{\circ}\star\Seq_{\Omega-1}\circ\cAo+X\bcdot\Seqominusone\circledcirc\cAo. \label{eq:gram_degree_plane}
\end{array}
\right.
\end{equation}

The ordinary generating function $e_{\Omega}(x):=\wt{\cEo}(x)$ satisfies the
equation\footnote{To obtain this equation we use the formula
$\ds Z_{\Cyc^{[k]}}=\frac1{k}\sum_{r|k}\phi(r)s_r^{k/r}$.}:
\begin{equation}
\label{eq:ex_trees_degree_plane}
\eomeg\ \!'(x) = \sum_{\substack{k\in\Omega\\r|k}}\frac{\phi(r)}{k}a_{\Omega}(x^r)^{k/r}+xa_{\Omega}\ \!'(x^2)+\sum_{\substack{k\in\Omega\\r|k,r>1}}\phi(r)x^ra_{\Omega}\ \!'(x^r)a_{\Omega}(x^r)^{k/r-1},
\end{equation}
where $a_{\Omega}(x)$ is specified by  $\ds a_{\Omega}(x)=x\sum_{k\in\Omega}a_{\Omega}(x)^{k-1}$.
\end{proposition}

\begin{example}\emph{$d$-regular plane trees.}
For $d\geq 3$, 
a $d$-regular plane tree is a plane tree such that each internal node has degree $d$,
which corresponds to the case $\Omega=\{1,d\}$ in 
Proposition~\ref{prop:gram_plane_dgree}.
It is easily shown that such a tree with $n$ internal nodes has $m=n(d-2)+2$ leaves.
Let $\cE_{[d]}$ be the species of $d$-regular plane trees, where the atoms are the \emph{leaves} (it proves here more convenient to take leaves as atoms and to write
the counting coefficients according to the number of internal vertices). Let $\cA_{[d]}:=\cE_{[d]}\ \!\!\!'$ be the corresponding derived species, which 
satisfies $\cA_{[d]}=X+(\cA_{[d]})^{d-1}$. The decomposition for the cycle-pointed species
$\cE_{[d]}\ \!\!\!^{\circ}$ is
$$
\cE_{[d]}\ \!\!\!^{\circ}=X^{\circ}\star\cA_{[d]}+\cL^{\cst}\circledcirc\cA_{[d]}+(\Cyc^{[d]})^{\cst}\circledcirc\cA_{[d]}.
$$
Hence, the OGS $e_{[d]}(x):=\wt{\cE_{[d]}}(x)$ satisfies:
$$
xe_{[d]}\ \!\!\!'(x)=xa_{[d]}(x)+x^2a_{[d]}\ \!\!\!'(x^2)+\sum_{r|d,r>1}\phi(r)x^ra_{[d]}\ \!\!\!'(x^r)a_{[d]}(x^r)^{d/r-1},\ 
\mathrm{where}\ a_{[d]}(x)=x+a_{[d]}(x)^{d-1}.$$
The coefficients of each of the summand series (such as
$a_{[d]}\ \!\!\!'(x)a_{[d]}(x)^{k-1}$) have a closed formula, which can be found for
instance using the univariate Lagrange inversion formula. From these formulas, 
we obtain the following expression for the number 
$e_{n,[d]}$ of $d$-regular plane trees with $n$ internal nodes:

\begin{equation}\label{eq:dplane}
e_{n,[d]}=\frac{1}{m}\!\left[ \frac1{m-1}\binom{n\!+\!m\!-\!2}{n}+\mathbf{1}_{\left\{\substack{2|m,\\ (d\!-\!2)|(m\!-\!2)/2}\right\}}\cdot\binom{n'\!+\!m_2\!-\!1}{n'}+\!\!\!\sum_{\substack{r>1,r|d,r|m \\ (d\!-\!2)|(m-d)/r}}\!\!\!\phi(r)\binom{n_r\!+\!m_r\!-\!1}{n_r}\right]
\end{equation}
where $$m=n(d-2)+2,\ \ \ n'\!=\!\frac{m-2}{2(d-2)}\!=\!\frac{n}{2},\ \ \ n_r\!=\!\frac{m-d}{r(d-2)}\!=\!\frac{n-1}{r},\ \ \ \ m_r=m/r. $$
One can extend this formula to any degree distribution on vertices, by adding variables marking the degree of each vertex and
applying the multivariate Lagrange inversion formula.  
A general enumeration formula is given in~\cite{BBLL00} using the dissymmetry theorem.  
\end{example}
\begin{remark}
As the only  automorphisms for plane trees are rotations, a simple application of Burnside's lemma is enough to get~\eqref{eq:dplane}. (An adaptation
of Burnside's lemma to unrooted plane graphs is given in~\cite{Li81}.)
In contrast, free trees require more involved counting techniques using 
cycle index sums. Currently, these techniques are  Otter's dissimilarity equation, the dissymmetry theorem (these two methods being closely related), and now cycle-pointing. 
\end{remark}



\begin{figure}[!ht]
\begin{center}
\includegraphics[width=12cm]{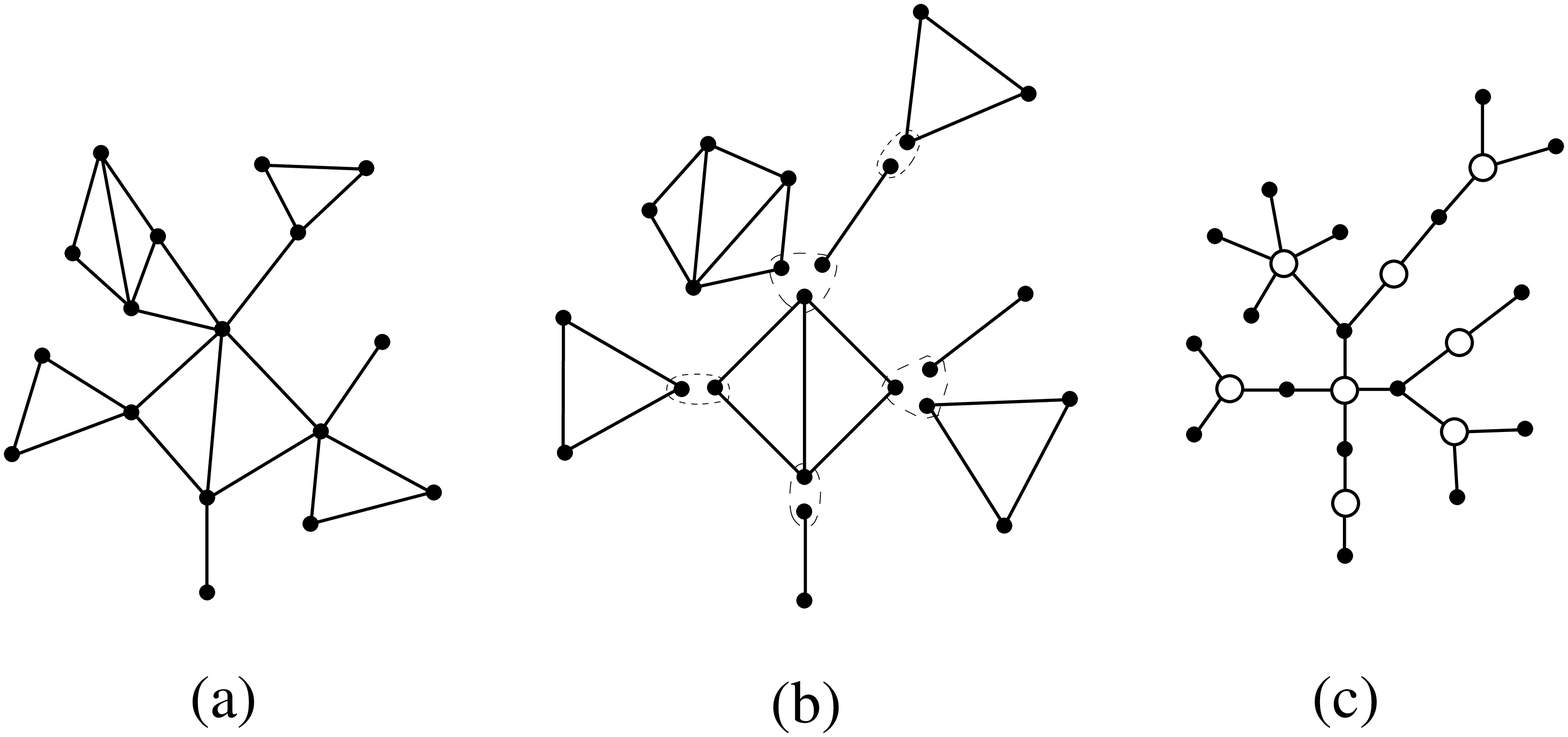}
\end{center}
\caption{(a) A connected (outerplanar) graph, (b) its block-decomposition, (c) the 
associated Bv-tree.}
\label{fig:block_decomp}
\end{figure}

\subsection{Graphs}
\label{sec:app_graph}
We extend here the decomposition principles which we have developed for trees to the 
more general case of a species of connected graphs, by taking advantage
of a well-known ``tree-like'' decomposition of a connected graph into
2-connected components. (A 2-connected graph is a graph that has at least two vertices and has no separating vertex.) 
Given a connected graph $G$, a maximal
2-connected subgraph of $G$ is called a \emph{block} of $G$. The set of vertices of $G$
is denoted $\mathfrak{V}(G)$ and its set of blocks is denoted $\mathfrak{B}(G)$.
The \emph{Bv-tree} of $G$ is the bicolored graph with vertex-set $\mathfrak{V}(G)\cup\mathfrak{B}(G)$ and edges corresponding to the adjacencies between the blocks and the 
vertices of $G$, see Figure~\ref{fig:block_decomp}. 
It can be shown that the Bv-tree of $G$ is indeed a 
tree, see~\cite[p.10]{HP73} and~\cite{MoTh} for details.




\begin{proposition}\label{prop:gramm_graph}
Let $\cG$ be a species of connected graphs that satisfy the following stability property: ``a connected graph is in $\cG$ iff all its blocks are in $\cG$''. Let $\cB$ be the subspecies of graphs in $\cG$ that are 2-connected. Then $\cG$ admits a decomposition grammar from the species of 2-connected structures $\cB'$, $\cBs$, and $(\cB')^{\circ}$: 
\begin{equation}
\label{eq:dec_graphs}
\left\{
\begin{array}{rcl}
\cGp&=&X\bcdot\cG'+\cGs.\\
\cG'&=&\Set\circ \cK, \quad   \cK=\cB'\circ\cH, \quad  \cH=X\bcdot\cG',\ \\\
\cGs&=&\cBs\circledcirc\cH+X\bcdot\Set^{\cst}\circledcirc\cK,\\
\cHp&=&X^{\circ}\star\Set\circ\cK+X\bcdot\Set^{\circ}\circledcirc\cK,\ \ \cKp=(\cB')^{\circ}\circledcirc\cH.

\end{array}
\right.
\end{equation}
Hence, if the species of 2-connected structures $\cB^{\cst}$ and $\cB'$ are
decomposable (the latter implies that $(\cB')^{\circ}$ is decomposable), 
then the cycle-pointed species $\cG^{\circ}$ is decomposable as well. More generally, if $Z_{\cB'}$ and $\oZ_{\cB^{\cst}}$ are both solutions
of an equation system involving the operations $\{+,\star,\circ,\circledcirc\}$ and
basic cycle-index sums, 
then $Z_{\cG^{\circ}}$ is also a solution of such an equation system.
\end{proposition}
\begin{proof}
The first line of the grammar is obtained as usual by distinguishing whether the 
marked cycle has length 1 or greater than 1. 
The second line easily follows from the block decomposition,
as shown for instance in~\cite{gimeneznoy}.
To wit, the marked vertex
of a graph in $\cG'$ is incident to a collection of blocks, and a connected graph is 
possibly attached at each non-marked vertex of these blocks.
Let us prove the third line in a similar way as for free trees (Claim~\ref{cl:sym_trees}).
Consider a graph $G$ in $\cG'$, and let $T$ be the Bv-tree of $G$. Clearly the Bv-tree
of a graph has less structure than the graph itself, so any automorphism of $G$
induces an automorphism on $T$. In particular $T$ is a symmetric cycle-pointed tree,
hence it has a center of symmetry that either corresponds to a block or to a vertex of $G$. The species of graphs in $\cG^{\cst}$ whose center of symmetry
in the Bv-tree is a vertex (a block) is denoted $\cGsv$ ($\cGsB$, resp.). 
Let $\cGv$ ($\cGB$) be the species of graphs in $\cG$ with a marked vertex (block, resp.) that is colored blue. Then clearly $\cGv=X\bcdot\cG'$ and $\cGB=\cB\circ(X\bcdot\cG')$. Hence, following the notations introduced
in the grammar, $\cGv=\cH=X\bcdot\Set\circ\cK$  and $\cGB=\cB\circ\cH$.
Note that the structures in $\cGsv$ ($\cGsB$) are 
the graphs in $\cGv\ \!\!^{\cst}$ ($\cGB\ \!\!^{\cst}$) such 
that the center of symmetry of the associated Bv-tree is the blue vertex (block, resp.).
It is easy to check, in a similar way as for free trees, that this property holds only for the graphs of $\cGv\ \!\!^{\cst}$ that
are in $X\bcdot\Set^{\cst}\circledcirc\cK$ and only for the graphs of $\cGB\ \!\!^{\cst}$ that are in
$\cB^{\cst}\circledcirc\cH$.
Finally, the 4th line, which is necessary to have only species of 2-connected structures as terminal species, is obtained from the second line by applying the derivation rules (Proposition~\ref{prop:point_rules}).
\end{proof}

\begin{remark}
Trees are exactly connected graphs where each block
is an edge. In other words, the species $\cF$ of free trees is 
the species $\cG$ of connected graphs formed from
the species $\cB=\cL$ (the one-element species that consists of the link graph).
One easily checks that, in that case, the grammar~\eqref{eq:dec_graphs} for $\cG$ 
is equivalent to the grammar~\eqref{eq:gram_trees} 
for free trees.
\end{remark}

\subsubsection{Cacti graphs}\label{sec:cacti}
Cacti graphs form an important class of graphs that have several
algorithmic applications. 
 They consist of cycles attached together
in a tree-like fashion; in other words, the species of cacti graphs 
arises from the species of 2-connected structures as $\cB=\cL+\cP$ where $\cL$ is the species of the link graph  
and $\cP$ is the speices of polygons with at least 3 edges (i.e., $\cB$
is  the species of polygons where one allows the degenerated 2-sided polygon).

Thanks to the grammar~\eqref{eq:dec_graphs}, the unlabeled enumeration of 
connected cacti graphs reduces to the calculation of the cycle-index sums
for the species of 2-connected structures $\cB'$ and $\cBs$ (the cycle-index sum $\oZ_{(\cB')^{\circ}}$
is also required, but it can directly be deduced from  $Z_{\cB'}$ by differentiation).
Since the 2-connected cacti graphs are polygons, 
the possible automorphisms are from the dihedral group.
In addition, the presence of a marked vertex (for $\cB'$) or cycle (for $\cBs$) restricts the symmetries.
For instance, if a structure in $\cB'$ has a marked (unlabeled) vertex $v$, then the automorphisms have to fix $v$; there are only two such symmetries for each polygon, the identity and the unique reflection
whose axis passes by $v$. Accordingly, we have two terms in the expression of
$Z_{\cB'}$ below, the first one for the identity, and the second 
one for reflections (where one 
distinguishes whether the polygon has odd or even length).
$$Z_{\cB'}=\frac1{2}\frac{s_1}{1-s_1}+\frac1{2}\frac{s_1+s_2}{1-s_2} $$

For $\cBs$, all symmetries must be nontrivial and have 
to respect the marked cycle. These symmetries are of two types: rotation and 
reflection, which yields the two main terms in the expression of $\oZ_{\cBs}$ below.
$$
\oZ_{\cBs}=\sum_{r\geq 2}\frac{\phi(r)}{2}\frac{t_r}{1-s_r}+\frac{t_2(s_1^2+2s_1+1)}{2(1-s_2)^2}
$$

The expressions for $Z_{\cB'}$ and $\oZ_{\cBs}$ can be used to enumerate unlabeled cacti graphs.
We just have to translate (using the computation rules in Remark~\ref{remark:OGS} and Remark~\ref{remark:OGS_point}) the grammar~\eqref{eq:dec_graphs} -- applied to the species of cacti graphs -- 
into an equation system satisfied by the corresponding ordinary generating functions.  

\begin{proposition}[enumeration of unlabeled cacti graphs]
The ordinary generating function $c(x)=\sum_nc_nx^n$ of unlabeled cacti graphs counted with respect to
the number of vertices satisfies
\begin{eqnarray*}
xc'(x)&=&H(x)+I(x),\\
H(x)&=&x\exp\Big(\sum_{r\geq 1}\frac{K(x^r)}{r}\Big),\ \ \mathrm{with}\ K(x)=\frac1{2}\frac{H(x)}{1-H(x)}+\frac1{2}\frac{H(x)+H(x^2)}{1-H(x^2)},\\
I(x)&=&\sum_{r\geq 2}\frac{\phi(r)}{2}\frac{x^rH'(x^r)}{1-H(x^r)}+\frac{x^2H'(x^2)(H(x)^2+2H(x)+1)}{2(1-H(x^2))^2}+H(x)\sum_{r\geq 2}x^rK'(x^r),
\end{eqnarray*}
from which one can extract\footnote{The calculations have been done with the help
of the computer algebra system Maple.} the counting coefficients $c_n$ (after firstly extracting the coefficients of $H(x)$):
$$xc'(x)=1\cdot\mathbf{1}\cdot x+2\cdot\mathbf{1}\cdot x^2 +3\cdot \mathbf{2}\cdot x^3+4\cdot\mathbf{4}\cdot x^4+5\cdot\mathbf{9}\cdot x^5+6\cdot\mathbf{23}\cdot x^6+7\cdot\mathbf{63}\cdot x^7+\ldots.$$
\end{proposition}

\subsubsection{Outerplanar graphs}\label{sec:outerplanar}
Outerplanar graphs are graphs that can be drawn in the plane so that 
all vertices are incident to the outer face. They form a fundamental subspecies of the species of
planar graphs, which already captures some difficulties of the species 
of all planar graphs; for example, the convergence rate of sampling procedures using the Markov Chain approach is not known.
However, outerplanar graphs are easier to tackle with the decomposition approach. For enumeration, we use the well-known property that 
2-connected outerplanar graphs, except for the one-edge graph, 
have a unique hamiltonian cycle.
Hence, the species $\cB$ of 2-connected outerplanar graphs can be identified
with the species of dissections of a polygon (allowing a degenerated 2-sided dissection).
This time, to obtain the cycle index sums $Z_{\cB'}$ and $\oZ_{\cBs}$, 
we have to count not polygons (as for cacti graphs) but \emph{dissections} of a polygon under 
 the action of the dihedral group. 
We only sketch the method here (the principles for counting such dissections are well known, going back to earlier articles of Read~\cite{Read}, see also~\cite{BFKV} for more
detailed calculations). 

\begin{figure}
\begin{center}
\includegraphics[height=3cm]{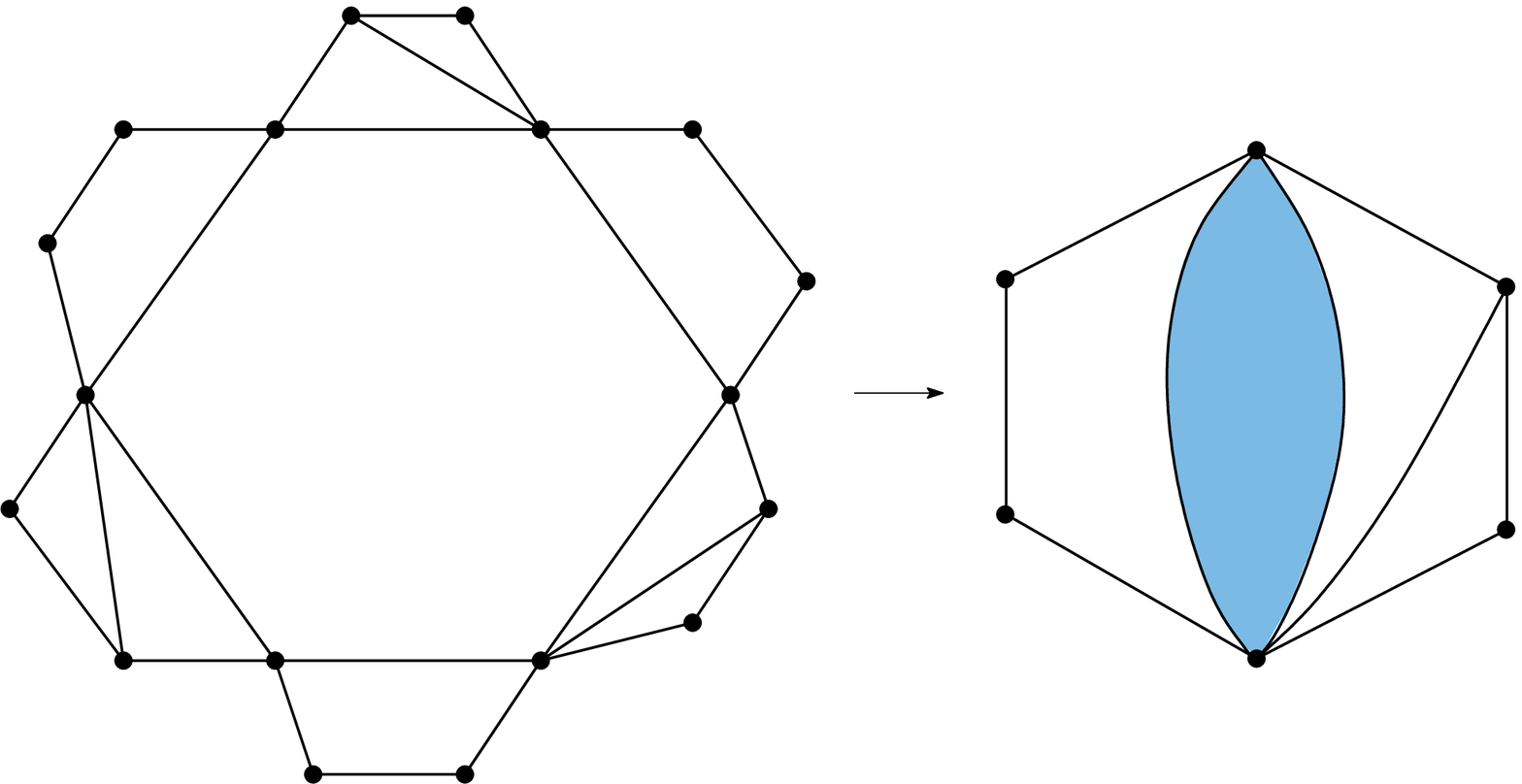}\hspace{1cm}\includegraphics[height=2.5cm]{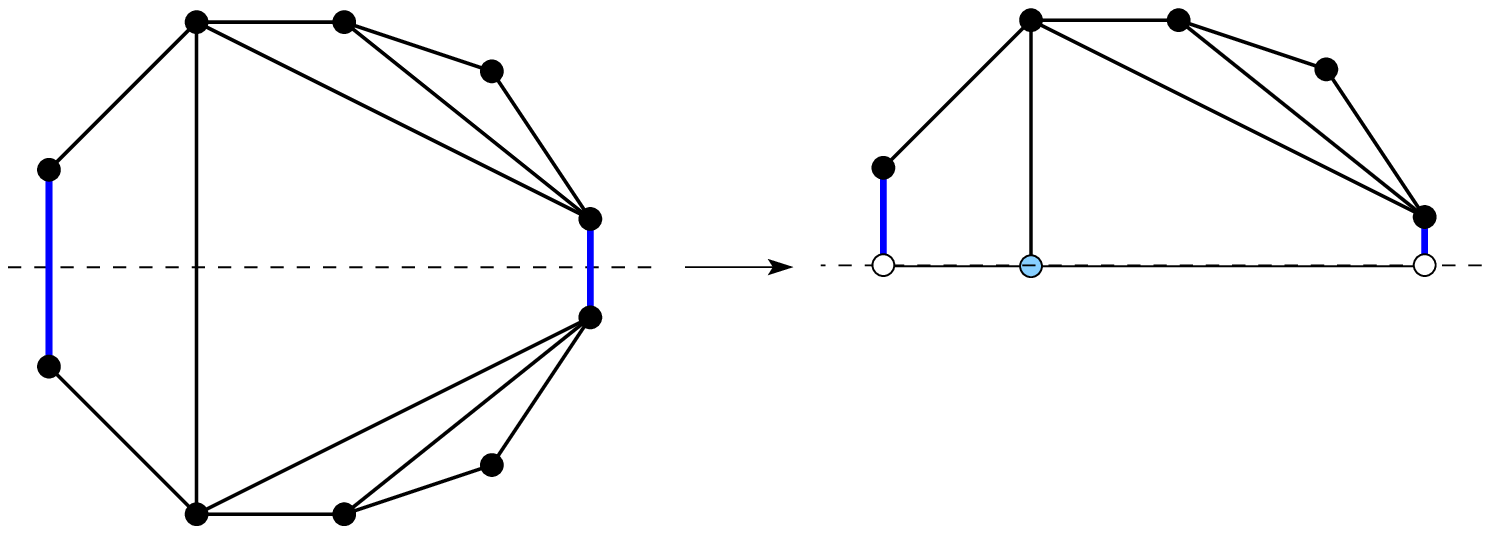}
\end{center}
\caption{Left (Right, resp.): the quotient of a dissection under a rotation
(reflection, resp.).}
\label{fig:quotient}
\end{figure}

For each type of symmetry (rotation or reflection), one considers the ``quotient 
dissection'', as shown in Figure~\ref{fig:quotient}. Notice that a dissection fixed by 
a rotation has either a central edge (only for the rotation of order two)
or a central face.
In case of a central edge $e$, it turns out to be more convenient to ``double'' $e$, so as to always have a central face before taking the quotient.
Thus, the quotient dissection 
has a marked face (the quotient of the central face)
that might have degree one (only for rotations of order at least three) or two (only for rotations of 
order at least two). 
Concerning quotient dissections
under a reflection, there are two special vertices $v_1$ and $v_2$ on the boundary (the intersections of the
original polygon with the reflection-axis), and there might be some other special vertices, all of degree three, on the boundary path from $v_1$ to $v_2$; see Figure~\ref{fig:quotient}.

\begin{figure}
\begin{center}
\includegraphics[height=3cm]{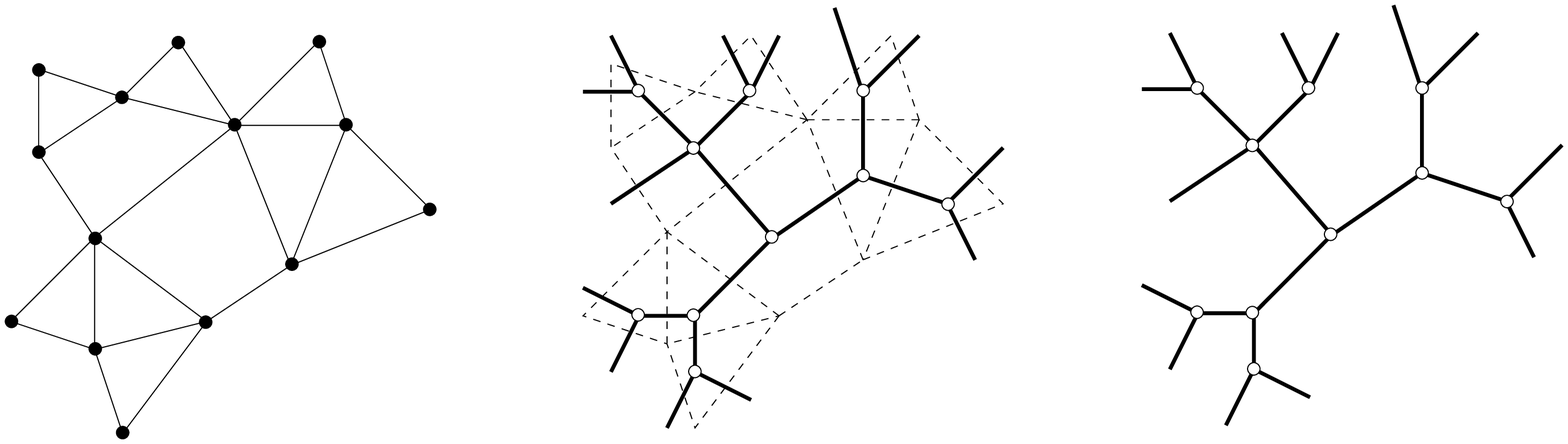}
\end{center}
\caption{The dual tree of a dissection.}
\label{fig:duality}
\end{figure}

 The second ingredient is to take the dual
of such quotient dissections in order to obtain plane trees, which are easier to decompose and to count. Notice that if the rotation is the identity rotation, then the associated
plane tree is in the species $\cF$ of plane trees with no vertex of degree two.
Notice also that each leaf of the tree corresponds to a vertex of the dissection; 
see Figure~\ref{fig:duality} for an example. The generating function of $\cF$ with respect to the number of leaves
satisfies  
\begin{equation}
F(x)=x+\frac{F(x)^2}{1-F(x)}.
\end{equation}

To calculate  $Z_{\cB'}$ and
$Z_{\cBs}$, one computes separately the contributions
of rotations and reflections to  $Z_{\cB'}$ and to $Z_{\cBs}$. In each case, using 
duality, the contribution is easily expressed in terms of the series $F(x)$.
All calculations done, one finds:
$$
Z_{\cB'}=\frac1{2}\big(F(s_1)+(s_1+s_2)P(s_2)\big),\ \ \ \ \oZ_{\cBs}=\frac1{2}\sum_{r\geq 2}\phi(r)t_rG(s_r)+\frac1{2}t_2\big(1+s_1^2Q(s_2)+2s_1R(s_2)+S(s_2)\big),
$$
where
\begin{center}
$\ds
G(x)=\frac{F'(x)}{1-F(x)},\ \ P(x)=\frac1{x}\frac{F(x)}{1-2F(x)},
$
\\
$\ds
Q(x)=\frac{\mathrm{d}}{\mathrm{d}x}P(x),\ \ R(x)=\frac{\mathrm{d}}{\mathrm{d}x}(xP(x)),\ \ S(x)=\frac{\mathrm{d}}{\mathrm{d}x}(x^2P(x)).
$
\end{center}
\vspace{.2cm}

Similarly as for cacti graphs, the expressions for $Z_{\cB'}$ and $\oZ_{\cBs}$  make it possible to 
enumerate unlabeled connected outerplanar graphs. 
Translating the grammar~\eqref{eq:dec_graphs}
into an equation system on the corresponding generating functions, we obtain the following.
 
\begin{proposition}[Enumeration of unlabeled connected outerplanar graphs]
The ordinary generating function $o(x)=\sum_no_nx^n$ of unlabeled connected outerplanar graphs counted with respect to
the number of vertices satisfies the system:
\begin{eqnarray*}
xo'(x)&=&H(x)+I(x),\\
H(x)&=&x\exp\Big(\sum_{r\geq 1}\frac{K(x^r)}{r}\Big),\ \ \mathrm{with}\ K(x)=\frac1{2}F(H(x))+\frac1{2}(H(x)+H(x^2))\cdot P(H(x^2)),\\
I(x)&=&\frac1{2}\sum_{r\geq 2}\phi(r)x^rH'(x^r)G(H(x^r))+H(x)\sum_{r\geq 2}x^rK'(x^r)\\
&&+\frac1{2}x^2H'(x^2)\big(1+H(x)^2Q(H(x^2))+2H(x)R(H(x^2))+S(H(x^2))\big),
\end{eqnarray*}
where the series $F$, $G$, $P$, $Q$, $R$, $S$ are defined above.
One extracts from this system (extracting firstly the coefficients in $F$, $Q$, $R$, $S$,
then in $H$ and $K$, then in $I$) the counting coefficients $o_n$:
$$xo'(x)=1\cdot\mathbf{1}\cdot x+2\cdot\mathbf{1}\cdot x^2 +3\cdot \mathbf{2}\cdot x^3+4\cdot\mathbf{5}\cdot x^4+5\cdot\mathbf{13}\cdot x^5+6\cdot\mathbf{46}\cdot x^6+7\cdot\mathbf{172}\cdot x^7+\ldots.$$
\end{proposition}

\subsection{Maps}
A map is a planar graph embedded on a sphere up to isotopic deformation, i.e., it is a planar graph together with a cyclic order of the neighbors around each vertex. There is a huge literature on maps 
 since the pioneering work of Tutte~\cite{Census}. As we show next, the decomposition grammar~(\ref{eq:dec_graphs}) for maps is actually 
simpler than for graphs, and it allows us to enumerate 
(unrooted unlabeled) 2-connected maps in terms of not necessarily connected maps.
To write down the grammar, it turns out to be more convenient to take half-edges as atoms instead of vertices. 
Denote by $\cM$ the species of maps 
 -- so $\cR:=\cZ\bcdot\cM'$ is the species of \emph{rooted} maps (maps with a marked
 half-edge) -- 
 and by $\cB$ the species of 2-connected maps (the loop-map is  considered as 2-connected).

\begin{proposition}
The species of rooted maps and symmetric cycle-pointed maps (in each length $\ell\geq 2$ of the marked cycle) have the
following recursive specification over
the corresponding species 
of rooted and cycle-pointed \emph{2-connected} maps.
\begin{equation}
\label{eq:maps}
\left\{
\begin{array}{rcl}
\cR$\!\!\!$&=&$\!\!\!$\cK\bcdot\Seq\circ\cK,\ \  \cK=X\bcdot\cB'\circ\cH,\ \ \cH=X\bcdot(1+\cR) \; ,\\
\cM^{\circ}_{(\ell)}$\!\!\!$&=&$\!\!\!$\cB^{\circ}_{(\ell)}\circledcirc\cH+\Cyc^{\circ}_{(\ell)}\circledcirc\cK\ \mathrm{for}\ \ell\geq 2 \; .
\end{array}
\right.
\end{equation}
\end{proposition}
\begin{proof}
The arguments are similar to the proof for graphs 
(Proposition~\ref{prop:gramm_graph}).  
The only difference is that one takes the embedding into
account, hence corners (which are in one-to-one correspondence with half-edges for a given map) play for maps a similar role as vertices do for graphs, and 
a $\Set$ construction typically becomes a $\Cyc$ construction here. 
Let us comment here on the decomposition for symmetric cycle-pointed maps (the one for rooted maps is well known, see~\cite{Census}). One has
\begin{equation}\label{eq:maps_2conn}
\cMs=\cBs\circledcirc\cH+\Cyc^{\cst}\circledcirc\cK,
\end{equation}
where the first (second) term takes account of the maps whose center of symmetry
 -- for the associated block-decomposition tree -- is a block (vertex, respectively).
Further simplification is possible, since a rooted map 
has only the identity as automorphism. Hence, the species of rooted maps $\cH$ and $\cK$ satisfy $\cH^{\circ}=\cH^{\circ}_{(1)}$ and 
$\cK^{\circ}=\cK^{\circ}_{(1)}$. Thus, Equation~\eqref{eq:maps_2conn} can be ``sliced'' into 
 a collection of equations, one for each length $\ell\geq 2$  of the marked cycle.
\end{proof}

An important property of any map automorphism  
 -- as shown by Liskovets~\cite{Li81} -- is that 
all its
cycles  have the same length $\ell$, which is also the order of the automorphism.
Hence, the number of half-edges of a cycle-pointed map with a marked cycle of length $\ell$ is divisible by $\ell$.
For $\ell\geq 1$, 
 denote by $\Mell(x)$ ($\Bell(y)$) the series counting unlabeled cycle-pointed maps
(2-connected maps, respectively),
according to the number of half-edges, divided by $\ell$.
In particular, $R(x):=M_{(1)}(x)$ and $S(x):=B_{(1)}(x)$ are the series
counting rooted maps and rooted 2-connected maps,  respectively.
We clearly have 
$$
\oZ_{\cM^{\circ}_{(\ell)}}=\frac{t_{\ell}}{s_{\ell}}\Mell(s_{\ell}),\ \ \ \ \oZ_{\cB^{\circ}_{(\ell)}}=\frac{t_{\ell}}{s_{\ell}}\Bell(s_{\ell})  \; .
$$
Given this simplification, the grammar~\eqref{eq:maps} is translated into the 
following system relating the series counting species of maps and species of 2-connected maps:



\begin{eqnarray}
R(x)&=&\frac{K(x)}{1-K(x)},\ \ \ \ H(x)=x(1+R(x)),\ \ \ \ K(x)=\frac{x}{H(x)}S(H(x))\label{eq:map_root},\\
\Mell(x)&=&\frac{xH'(x)}{H(x)}\Bell(H(x))+\phi(\ell)\frac{xK'(x)}{1-K(x)}\ \ \mathrm{for}\ \ell\geq 2.\label{eq:map_kroot}
\end{eqnarray}

In the case of maps, the decomposition grammar is used in the other direction, i.e.,
one obtains the enumeration of (unrooted) 2-connected maps from maps. 
Indeed, unconstrained 
maps are easier to count, by a method of quotient~\cite{Li81} similar to the one
we have used for counting dissections in Section~\ref{sec:outerplanar}.

Let us first review (from Tutte~\cite{Census}) how one obtains an expression for the series $S(y)$ counting rooted 2-connected maps from an expression
for the series $R(x)$ counting rooted maps.
One starts from the following expression of $R(x)$:
\begin{equation}
R(x)=\frac{\beta(2-9\beta)}{(1-3\beta)^2},\ \ \ \mathrm{with}\ \ \beta=\beta(x)\ \ \mathrm{specified\ by}\ \ \beta=x^2+3\beta^2.
\end{equation}
Next, notice that the change of variable $y=H(x)=x(1+R(x))$ between rooted maps and rooted 2-connected maps is such that $y^2$ is also rational in $\beta$:
$$
y^2=x^2(1+R(x))^2=\beta\frac{(1-4\beta)^2}{(1-3\beta)^3}.
$$
Equivalently:
$$
y^2=\eta(1-\eta)^2,\ \ \mathrm{where}\ \ \eta:=\frac{\beta}{1-3\beta},
$$
so the dependence between $\eta$ and $\beta$ is invertible: $\beta=\eta/(1+3\eta)$. Notice also from~\eqref{eq:map_root} that
$$
S(y)=\frac{H(x)}{x}K(x)=(1+R(x))K(x)=R(x)=\frac{\beta(2-9\beta)}{(1-3\beta)^2}.
$$
 Replacing $\beta$ by $\eta/(1+3\eta)$, one gets
 $$
 S(y)=\eta(2-3\eta),\ \ \mathrm{with}\ \eta=\eta(y)\ \mathrm{specified\ by}\ \eta=\frac{y^2}{(1-\eta)^2},
 $$
 which can also be written as
 \begin{equation}
 S(y)=\oS(y^2),\ \ \mathrm{with}\ \oS(y)=\oeta(2-3\oeta),\ \mathrm{and}\ \oeta:=\oeta(y)\ \mathrm{specified\ by}\ \oeta=\frac{y}{(1-\oeta)^2}.
 \end{equation}
 
 In a similar way, if two series $f(x)=g(y)$ are related by the change of variables $y=H(x)$ and if $f(x)$ is rational in $\beta(x)$, then $g(y)$ is rational in $\eta(y)$
 (replacing $\beta$ by $\eta/(1+3\eta)$). For instance, for $\ell\geq 3$, it has been shown by Liskovets using the quotient method (see~\cite{Fu06} for the reformulation on series) that 
 $$
 \Mell(x)=\phi(\ell)\frac{6\beta(x)}{1-6\beta(x)}.
 $$
 Since $H(x)/x=1+R(x)$ and $K(x)=R(x)/(1+R(x))$ 
 are rational in $\beta$, as well as $xH'(x)$ and $xK'(x)$
 (noticing that $x\mathrm{d}f/\mathrm{d}x=2x^2\mathrm{d}f/\mathrm{d}x^2=2\beta(1-3\beta)\cdot(\mathrm{d}f/\mathrm{d}\beta)/(\mathrm{d}x^2/\mathrm{d}\beta)$),
 one finds from~\eqref{eq:map_kroot} a rational expression in $\beta$ for the series $\Bell(H(x))$. Replacing $\beta$ by $\eta/(1+3\eta)$ in that expression, one finally gets:
 $$
  \Bell(y)=\phi(\ell)\frac{2\eta(y)}{1-3\eta(y)}\ \ \mathrm{for}\ \ell\geq 3,
  $$
  which can also be written as
 \begin{equation}
 \Bell(y)=\phi(\ell)G(y^{2})\ \ \mathrm{for}\ \ell\geq 3,\ \mathrm{with}\ G(y)=\frac{2\oeta(y)}{1-3\oeta(y)}.
 \end{equation}
  
In a similar way, starting from the expression (given in~\cite{Fu06})
$$
M_{(2)}(x)=-1+\frac{1-2\beta(x)}{(1-6\beta(x))(1-3\beta(x))}+x\cdot\frac{2}{(1-6\beta(x))(1-3\beta(x))},
$$ 
one obtains the following expression for $B_{(2)}(y)$:
$$
B_{(2)}(y)=\frac{\eta(y)(3-\eta(y))}{1-3\eta(y)}+y\frac{2}{1-3\eta(y)},
$$
which can also be written as 
\begin{equation}
B_{(2)}(y)=P(y^2)+yQ(y^2), \mathrm{with}\ P(y)=\frac{\oeta(y)(3-\oeta(y))}{1-3\oeta(y)}\  \mathrm{and}\ Q(y)=\frac{2}{1-3\oeta(y)}.
\end{equation}
 
 \begin{proposition}[counting unrooted 2-connected maps, recover~\cite{LiWa83}]
 The number $t_n$ of (unrooted unlabeled) 
 2-connected maps with $n$ edges satisfies:
 \begin{equation}
 t_n=\frac1{2n}\Big( s_n+u_n +\frac1{2}\sum_{k |n,k\neq n}\phi(n/k)\cdot(9k^2-9k+1)s_k\Big), 
 \end{equation}
 where $\ds s_n=\frac{2(3n-3)!}{n!(2n-1)!}$, $\ds u_n=\frac{n(n+1)}{2}s_{(n+1)/2}$ if $n$ is odd, and $\ds u_n=\frac{(3n-4)n}{8}s_{n/2}$ if $n$ is even.
 \end{proposition}
 \begin{proof}
 Cycle-pointing ensures that the generating function $\wt{\cB^{\circ}}(y)=\sum_n2nt_ny^{2n}$
 satisfies
 $$
\wt{\cB^{\circ}}(y)=S(y)+B_{(2)}(y^2)+\sum_{\ell\geq 3}\Bell(y^{\ell})=S(y)+\Big(B_{(2)}(y^2)-G(y^4)\Big)+\sum_{\ell\geq 2}\phi(\ell)G(y^{2\ell}).
 $$
 Extracting the coefficient $[y^{2n}]$ in this equation yields
 \begin{equation}\label{eq:2ntn}
 2nt_n=s_n+u_{n}+\sum_{\ell\geq 2,\ell | n}\phi(\ell)v_{n/\ell}=s_n+u_{n}+\sum_{k | n, k\neq n}\phi(n/k)v_{k},
 \end{equation}
 where $s_n=[y^{n}]\oS(y)$, $v_n=[y^{n}]G(y)$, $u_n=[y^{n}](B_{(2)}(y)-G(y^2))$, which 
 is $[y^{n/2}]P(y)-v_{n/2}$ if $n$ is even and $y^{(n-1)/2}Q(y)$ if $n$ is odd. 
Notice that the series $\oS(y)$, $G(y)$, $P(y)$, and $Q(y)$ are rational in the 
simple series $\oeta(y)=y/(1-\oeta(y))^2$.
Hence the  Lagrange inversion formula~\cite[Section 3.1]{BeLaLe} allows us to extract exact formulas for the 
coefficients $s_n$, $v_n$, and $u_n$. 
Substituting these exact expressions in~\eqref{eq:2ntn},
one obtains the announced formula for $t_n$.
  \end{proof}
  
  The enumeration of unrooted 2-connected maps has first been done by Liskovets and
  Walsh~\cite{LiWa83} using the quotient method in a quite involved way.   More recently, the 
  counting formula has been recovered in~\cite{Fu06} using a method of extraction
  at a center of symmetry on quadrangulations. What we do here is 
  equivalent to~\cite{Fu06}, but the cycle-pointed framework allows us to write the
  equations on generating functions in a more systematic way. 
    
\subsection{Asymptotic enumeration}
Cycle-pointing makes it possible to easily obtain an asymptotic 
estimate for the coefficients counting the number of unlabeled structures from a species,
provided that the singular
behavior of the OGS counting the associated species of 
\emph{rooted} unlabeled structures is known. 

We illustrate the method on free trees. Let $R(x)$ be the OGS of rooted unlabeled nonplane trees, which is specified by $R(x)=x\exp(\sum_{i\geq 1}R(x^i)/i)$.
It is well known that
$R(x)$ has a dominant singularity $\rho<1$ of the square-root type~\cite[VII.5]{Flajolet}. That is, in the slit complex neighborhood $D_\epsilon := \{x \; | \; x-\rho\notin\mathbb{R}_+\ \mathrm{and}\ |x-\rho|<\epsilon\}$ we have the expansion
\begin{equation}\label{eq:R}
R(x)=1-a\ \!X+o(X),\ \ \mathrm{where}\ X=\sqrt{1-x/\rho}, 
\end{equation}
which yields -- using transfer theorems of analytic combinatorics~\cite[VI]{Flajolet} --
 the asymptotic estimate 
\begin{equation}\label{eq:asympt_Rn}
R_n \sim c\, n^{-3/2} \, \rho^{-n}, \ \ \mathrm{where}\ c=\frac{a}{2\sqrt{\pi}}\approx 0.43922,\ \ \rho\approx 0.33832,
\end{equation}
for the number of rooted unlabeled nonplane trees with $n$ vertices.

To obtain a similar estimate for \emph{free trees}, we consider the OGS
$P(x)$ of cycle-pointed nonplane trees and start from
the expression of $P(x)$ obtained in Proposition~\ref{prop:gram_trees}:
$$
P(x)=x^2R'(x^2)+\big(1+\sum_{\ell\geq 2}x^{\ell}R'(x^{\ell})\big)R(x).
$$
Notice that, since $\rho<1$, the series $A:=x^2R'(x^2)$ and 
 $B:=1+\sum_{\ell\geq 2}x^{\ell}R'(x^{\ell})$ are analytic at $x=\rho$,
and the value at $x=\rho$ of $B$ is the positive constant 
\begin{equation}
b:=1+\sum_{\ell\geq 2}\rho^{\ell}R'(\rho^{\ell}).
\end{equation}
Therefore, from the singular expansion~\eqref{eq:R} of $R(x)$, we obtain 
\begin{equation}\label{eq:sing_P}
P(x)=P(\rho)-ab\ \!X+o(X) \; .
\end{equation}
Let us simplify further the positive constant $b=B(\rho)$. 
First, by deriving the equation that specifies $R(x)$, one obtains 
$xR'(x)=R(x) \big(1+\sum_{i\geq 1}x^i R'(x^i)\big)$, which yields 
$$
B(x)R(x)=xR'(x)(1-R(x)) \; .
$$
By deriving the singular expansion of $R(x)$, one obtains 
$$
R'(x)= \frac{a}{2\rho}X^{-1}+o(X^{-1}) \; ,
$$
hence $xR'(x)(1-R(x))$ converges to $a^2/2$ as $x\to\rho$, i.e., $b=a^2/2$. 

\begin{proposition}[asymptotic enumeration of free trees]\label{prop:asympt_trees}
The number $F_n$ of unlabeled free trees with $n$ 
vertices satisfies
\begin{equation}\label{eq:asympt_Fn}
F_n\sim (2\pi c^3)n^{-5/2}\rho^{-n},
\end{equation}
where $c$ is the constant and $\rho^{-1}$ is the growth ratio 
in the estimate~\eqref{eq:asympt_Rn} for \emph{rooted} nonplane trees ($R_n\sim c\ \!n^{-3/2}\rho^{-n}$). 
\end{proposition}
\begin{proof}
From the singular expansion~\eqref{eq:sing_P} of $P(x)$ we obtain (again by the transfer theorems of singularity analysis) 
$$
x^n[P(x)]\sim \frac{ab}{2\sqrt{\pi}}n^{-3/2}\rho^{-n} \; .
$$
Using $c=a/(2\sqrt{\pi})$ and $b=a^2/2$, we have
$ab/(2\sqrt{\pi})=2\pi c^3$. Finally, Theorem~\ref{theo:npointed} (unbiased pointing) yields 
$F_n=\frac{1}{n}[x^n]P(x)$, so $F_n\sim (2\pi c^3)n^{-5/2}\rho^{-n}$.
\end{proof}

It is also possible to get the estimate of $F_n$ from Otter's dissimilarity
equation (or from the dissymmetry theorem). 
However, we find that cycle-pointing
provides a more transparent explanation why the asymptotic estimate
of the coefficients $F_n$ counting unlabeled structures from an 
unrooted ``tree-like'' species $\cF$ is of 
the universal type $cn^{-5/2}\rho^{-n}$. The argument is very simple:
\begin{itemize}
\item
The OGS $P(x)$ of the cycle-pointed species $\cFp$ is positively expressed
in terms of the OGS of the rooted species, which has a square-root dominant singularity. Therefore, $P(x)$ inherits the same singularity and singularity type (square-root). 
\item
Transfer theorems of singularity analysis ensure that a square-root
dominant singularity yields an asymptotic estimate in $cn^{-3/2}\rho^{-n}$ for the coefficients $[x^n]P(x)$.
Since $F_n=\frac1{n}[x^n]P(x)$, one gets $F_n\sim c\ \!n^{-5/2} \, \rho^{-n}$. 
\end{itemize}

This strategy applies to all species of trees encountered in this section,
 as well as to cacti graphs and connected outerplanar graphs 
(in all cases one starts from the singular
expansion of the OGS counting the corresponding \emph{rooted} species).

\section{Application to Random Generation}
\label{sec:sampler}
Recently, so-called 
 Boltzmann samplers have been introduced by Duchon et al~\cite{Boltzmann} as a general
method to efficiently (typically in linear time) generate uniformly
at random 
combinatorial structures that admit a decomposition.
In contrast to the more costly \emph{recursive method of
sampling}~\cite{NijenhuisWilf}, which is based on counting coefficients of the recursive decomposition,
Boltzmann samplers are primarily based on
generating functions. Until now Boltzmann samplers were 
developed in the labeled setting~\cite{Boltzmann} 
and partially in the unlabeled setting~\cite{BoltzmannUnlabeled}.

In this section we provide a more complete method in the unlabeled setting.
In order to deal with the substitution construction and the cycle-pointing
operator (which are not covered in~\cite{BoltzmannUnlabeled}),
 we have to describe samplers not solely based on  
ordinary generating functions, but on cycle index sums -- also known as P\'olya 
operators. Therefore we call these 
random generators \emph{P\'olya-Boltzmann samplers}.

With these refined samplers we are able to design in a systematic way (via specific generation rules)  
a P\'olya-Boltzmann sampler for species that admit a recursive decomposition,
thereby allowing in the decomposition all operators that have been described in this article. When
specialized suitably, a 
P\'olya-Boltzmann sampler reduces to an ordinary Boltzmann sampler, hence it provides a uniform random sampler for species of unlabeled structures.
In particular, we obtain highly efficient random generators for the species in Section~\ref{sec:appl}:
for trees, cacti graphs, outerplanar graphs, etc.

\subsection{Ordinary Boltzmann Samplers}
Let $\cA$ be a species of structures, and let 
$\wt{\cA}(x)$ be the ordinary generating series for $\cA$. 
A real number $x>0$ is said to be \emph{admissible} 
iff the sum defining $\wt{\cA}(x)$ converges 
($x$ within the disk of convergence of the series).
Given a fixed admissible value $x>0$, 
an \emph{ordinary Boltzmann sampler}
for unlabeled structures from ${\cA}$ is a random generator 
$\Gamma \wt{\cA}(x)$ that draws each structure
$\gamma\in\wt{\cA}$ with probability 
\begin{equation}\mathbb{P}_x(\gamma)=\frac{x^{|\gamma|}}{\wt{\cA}(x)} \; .
\end{equation} 
Notice that this distribution has the fundamental property to be 
\emph{uniform}, i.e., any 
two unlabeled structures of the species with the same size have the same probability.

\subsubsection{Automatic rules to design Boltzmann samplers}
As described in \cite{Boltzmann}, there are simple rules to assemble Boltzmann samplers for the two classical constructions Sum and Product ($\Bern(p)$ stands for a Bernoulli law, returning ``true'' with probability $p$ and ``false'' with probability $1-p$).

\vspace{0.2cm}

\noindent\begin{tabular}{rrl}
$\cC=\cA+\cB$.& $\Gamma \wt{\cC}(x)$:&{\bf if} $\Bern (\wt{\cA}(x)/\wt{\cC}(x))$ {\bf return} $\Gamma\wt{\cA}(x)$ {\bf else return} $\Gamma \wt{\cB}(x)$\\
$\cC=\cA\ \! \bcdot\ \!\cB$.& $\Gamma \wt{\cC}(x)$:&{\bf return} $(\Gamma \wt{\cA}(x),\Gamma \wt{\cB}(x))$ \{independent\ calls\} \\
\end{tabular}

\vspace{0.2cm}

These rules can be used recursively. For instance, 
the species $\cT$ of 
rooted binary trees satisfies
\begin{equation}
\cT=X+\cT\bcdot\cT,
\end{equation} 
which translates to the following Boltzmann sampler:

\vspace{0.2cm}
\noindent\begin{tabular}{ll}
$\Gamma \wt{\cT}(x)$:& {\bf if} $\Bern\left(\frac{x}{\wt{T}(x)}\right)$ {\bf return} leaf\ {\bf else return} $(\Gamma \wt{\cT}(x),\mathrm{node},\Gamma \wt{\cT}(x))$.
\end{tabular}
\vspace{0.2cm}

\subsubsection{The complexity model}\label{sec:comp_model}
Typically, when $\cF$ has a recursive specification 
over the species $\cA_1, \dots, \cA_l$, then 
our sampling procedure $\Gamma\wt{\cF}(x)$ will require that we can
evaluate the ordinary generating functions for $\cA_1,\dots,\cA_l$
at real values $x$.
Indeed, for a species $\cF$ defined as $\cA_1 + \cA_2$,
each Bernoulli choice requires to draw a uniform value in $[0,1]$ and 
compare it with a ratio of the form $\wt{\cA_1}(x)/\wt{\cF}(x)$. 
In the following we work with the complexity model where we
assume that there exists an \emph{oracle} that 
provides at unit cost the exact values of
these generating functions at $x$, and that a random number in $[0,1]$ can be generated and compared with a fixed value such as $\wt{\cA}(x)/\wt{\cC}(x)$ in constant time as well. We will
refer to this complexity model as the \emph{real-arithmetic complexity
model} in the following.

The model is justified since in many applications
we obtain expressions for the ordinary generating series
that allow a rapid numeric evaluation of those series at given values,
for example with the Newton method.
Then, in practice, 
one works at a fixed precision, say $N$ bits (typically $N=64$, correspondingly roughly to 20 decimal digits).

Let us mention that the Boltzmann samplers for the constructions Multiset and Cycle---as given in~\cite{BoltzmannUnlabeled} and
recovered in a more general framework here---require typically the
values of the generating functions not only at $x$, but at all powers $x^i$. Since combinatorial species often
have exponential growth rate (which is the case for all examples presented here), the dominant
singularity $\rho$ satisfies $\rho<1$, hence $x\leq\rho<1$.
Therefore the values $\wt{\cA}(x^i)$ decrease exponentially fast 
with $i$. When working at fixed precision of $N$ bits, one can thus 
 discard the powers greater than $k=N/\log_2(1/\rho)$ and assume that the oracle provides the evaluations of 
the generating functions at $x, x^2,\ldots,x^k$. For a more detailed study and  implementation
of the evaluation procedures we refer to the recent article by Pivoteau, Salvy, 
and Soria~\cite{PiSaSo}.

\begin{proposition}[Duchon et al.~\cite{Boltzmann}]\label{prop:duch1}
Let $\cF$ be a species that can be decomposed recursively from $\{\mathbf{1},X\}$ 
in terms of the constructions $\{+,\bcdot\}$ (this is meant analogous to but more restricted than Definition~\ref{def:dec-standard}). 
Then one can obtain in a systematic way (from the recursive specification) an ordinary Boltzmann sampler $\Gamma\wt{\cF}(x)$ for $\cF$. In addition, in the real-arithmetic complexity model, 
$\Gamma\wt{\cF}(x)$ operates in linear time in the 
size of the output.
\end{proposition}
 
This result was recently extended in~\cite{BoltzmannUnlabeled} to other constructions, 
such as the Multiset and Cycle constructions (and their counterpart with fixed number of components).
In this section we extend this result to the substitution construction,
and to the cycle-pointed constructions.
It follows that any species that is cycle-pointed decomposable over species where we already have a
P\'olya-Boltzman sampler also has a P\'olya-Boltzman sampler.

\begin{remark} 
Note that in the general results on sampling we consider
species \emph{up to isomorphism}. Theoretically,
this is a necessary assumption, since isomorphisms between species
might in artificial examples be non-effective. However,
in all the presented examples and applications in this article,
the isomorphisms between species are straightforward and
efficiently computable. The effectiveness of isomorphisms between
the output species of the sampling procedures and the 
actual species are actually more related to the question how combinatorial structures are
represented on a computer, and in particular they do not
concern the complexity of the sampling task itself.
\end{remark}

\subsubsection{Targeting Boltzmann samplers}
Boltzmann samplers often lead to very efficient  \emph{exact-size} and \emph{approximate-size}
random samplers. In order to draw unlabeled structures uniformly at random from a species $\cA$ at (in case of exact-size sampling) or around (in case of approximate-size sampling)
a \emph{target-size} $n$, 
one simply repeats calling the Boltzmann sampler $\Gamma\wt{\cA}(x)$ -- with a suitably chosen value of $x$ -- 
until the size of the output is $n$ (exact-size sampling) or is in $[n(1-\epsilon),n(1+\epsilon)]$ 
(approximate-size sampling), 
where $\epsilon$ is a \emph{tolerance-parameter} fixed by the user.
It turns out that for a wide class of species, 
which covers the species encountered in Section~\ref{sec:appl}
(trees, cacti graphs, outerplanar graphs),
this method works very well, as proved in~\cite{Boltzmann}.

\begin{proposition}[Duchon et al~\cite{Boltzmann}]
\label{lem:boltzmann}
Let $\cF$ be a species such that 
asymptotically $$|\wt{\cF}_n| \sim c \; n^{-3/2} \; \rho^{-n}$$ 
for some positive constant $c$,  and where $\rho$ is the radius of convergence of $\wt{\cF}(x)$, assuming that $\wt{\cF}(\rho)$ is convergent\footnote{The asymptotic behaviour $c\rho^{-n}n^{-3/2}$ is called \emph{universal}~\cite{Burris}, as it
 is widely encountered in combinatorics.}. 
Also suppose that there is
a Boltzmann sampler $\Gamma\wt{\cF}(\rho)$ at $x=\rho$ such that the cost of generating a structure 
is linearly bounded by the size of the structure all along the generation process.
Then $\Gamma\wt{\cF}(\rho)$ yields an exact-size (approximate-size, resp.) 
sampler for unlabeled structures from $\cF$ with expected complexity $O(n^2)$ ($O(n/\epsilon)$, resp.),
where $n$ is a target-size and $\epsilon$ is a tolerance-ratio.
\end{proposition}
The exact-size and approximate-size samplers are obtained by running $\Gamma \wt{\cF}(\rho)$
until the size of the output is in the target domain $\Omega_n$ (that is, $\Omega_n=\{n\}$ for exact-size sampling, and 
$\Omega_n=[n(1-\epsilon),n(1+\epsilon)]$ 
for approximate-size sampling). To obtain the stated complexity, 
it is necessary that the generation of too large structures 
is aborted as soon as the size of the generated object  
gets larger than $\mathrm{Max}(\Omega_n)$.

\subsection{P\'olya-Boltzmann Samplers for classical species}
\label{sec:extendedsampler}
Let $\cA$ be a species.
Recall that a \emph{symmetry} on a species $\cA$ is a pair $(A,\sigma)$ where $A\in\cA$ and $\sigma$
is an automorphism of $A$. 
A symmetry has a weight-monomial $w_{(A,\sigma)}$, 
as defined in~\eqref{eq:def_weight_unlabeled}; 
and the cycle index sum $Z_{\cA}(s_1,s_2,\ldots)$ 
is the sum of the weight-monomials over all the 
symmetries on $\cA$.  Similarly as for the one-variable case, a
vector $(s_i)_{i\geq 1}$ of nonnegative real values is said to be 
\emph{admissible} if the sum defining $Z_{\cA}(s_1,s_2,\ldots)$ 
converges.
Given an admissible vector $(s_i)_{i\geq 1}$, 
a \emph{P\'olya-Boltzmann sampler}
is a procedure $\Gamma Z_{\cA}(s_1,s_2,\ldots)$ that randomly samples symmetries
on $\cA$ such that  each symmetry $(A,\sigma)$
is drawn with probability 
\begin{equation}\mathbb{P}_{(s_1,s_2,\ldots)}(A,\sigma)=\frac{w_{(A,\sigma)}}{Z_{\cA}(s_1,s_2,\ldots)},\end{equation} 
where the weight-monomial $w_{(A,\sigma)}$ is evaluated at $(s_1,s_2,\ldots)$. This probability
distribution is called the \emph{P\'olya-Boltzmann distribution} for $\cA$ at $(s_i)_{i\geq 1}$. 
The following simple lemma ensures that 
P\'olya-Boltzmann samplers are a refinement of ordinary Boltzmann samplers, 
in the same way as cycle index sums are a refinement of ordinary generating functions.

\begin{lemma}[P\'olya-Boltzmann samplers extend ordinary Boltzmann samplers]
Consider a species $\cA$ having a P\'olya-Boltzmann sampler $\Gamma Z_{\cA}(s_i)_{i\geq 1}$. 
Then, for any  value $x$ admissible for $\wt{\cA}(x)$,
the sampler $\Gamma Z_{\cA}(x,x^2,x^3,\ldots)$ 
is an ordinary Boltzmann sampler for unlabeled structures from
$\cA$ at $x$.
\end{lemma}
\begin{proof}
The generator $\Gamma Z_{\cA}(x,x^2,\ldots)$ gives weight 
 $x^n/(n!Z_{\cA}(x,x^2,\ldots))$ to each symmetry of size $n$.
Since $Z_{\cA}(x,x^2,\ldots)=\wt{\cA}(x)$
by Lemma~\ref{lem:unlabel}, this weight simplifies to $x^n/(n!\wt{\cA}(x))$. In addition, we have seen in Lemma~\ref{lem:unlabel} that 
each unlabeled structure $\gamma \in\tilde{\cA}_n$ gives rise to
$n!$ symmetries. Hence, each unlabeled structure of size $n$ has weight $x^n/\wt{\cA}(x)$ when
calling $\Gamma Z_{\cA}(x,x^2,\ldots)$, i.e., $\Gamma Z_{\cA}(x,x^2,\ldots)$ is an ordinary 
Boltzmann sampler for unlabeled structures from ${\cA}$. 
\end{proof}

In the next two sections, we describe P\'olya-Boltzmann samplers for unlabeled structures from basic species and
for the constructions $\{+,\bcdot,\circ\}$. 
Note that the output of a P\'olya-Boltzmann sampler for a species $\cA$
consists of a species from ${\cA}[n]$ together with an automorphism on that structure.
In all the random generators to be described (as well as in the procedures for cycle-pointed species), the resulting structure $S$ is made well-labeled by applying a procedure \textsc{DistributeLabels}
that substitutes $[1,\dots,|S|]$ for the atoms of $S$ uniformly at random
(i.e., according to a permutation of size $|S|$ taken uniformly at random).

\begin{figure}\small
\begin{flushleft}
\begin{tabular}{|l|}
\hline
  \\[-.2cm]
\hspace{-.1cm}{\bf (1) Sequence.}\\[.1cm]
\hspace{.5cm}{\bf Algorithm } $~~\Gamma Z_{\Seq}(s_1,s_2,\ldots)$,~with $s_1< 1$:\\
\hspace{1.5cm} $k\leftarrow \Geom(s_1)$;\\
\hspace{.5cm}{\bf return} a sequence of $k$ atoms (endowed with the identity-automorphism).\\[.2cm]
\hline
\\[-.2cm]
\hspace{-.1cm}{\bf (2) Set.}\\[.1cm]
Define the probability distribution relative to $(s_i)_{i\geq 1}$:\\
\hspace{1cm}$\ds\Pr(K\leq k)=\frac1{Z_{\Set}(s_1,s_2,\ldots)}\prod_{j\leq k}\exp\left(\tfrac{1}{j}s_j\right)$.\\
Let $\textsc{Max\_Index}(s_1,s_2,\ldots)$ be a  generator for this
distribution.\\[.2cm] 
\hspace{.5cm}{\bf Algorithm } $~~\Gamma Z_{\Set}(s_1,s_2,\ldots)$~:\\
\hspace{1.5cm}$J\leftarrow \textsc{Max\_Index}(s_1,s_2,\ldots)$;\\
\hspace{1.5cm}{\bf for} $j$ {\bf from} $1$ {\bf to} $J-1$ {\bf do} $\ds k_j \leftarrow \Pois\left( \tfrac{s_j}{j} \right)$ {\bf end for}\\
\hspace{1.5cm} $\ds k_J \leftarrow \Pois_{\geq 1}\left( \tfrac{s_{J}}{J}\right)$ \{Poisson conditioned to output a 
strictly positive integer\}\\
\hspace{.5cm}{\bf return} a collection of cycles of atoms where there are $k_j$ cycles in each length $j>0$.\\[.2cm]
\hline
\\[-.2cm]
\hspace{-.1cm}{\bf (3) Cycle.}\\[.1cm]
Given $(s_i)_{\geq 1}$ such that $Z_{\Cyc}(s_1,s_2,\ldots)$ converges, consider the probability distribution\\
\hspace{1cm}$\ds\Pr(R=r)=\frac1{Z_{\Cyc}(s_1,s_2,\ldots)}\frac{\varphi(r)}{r} \log\left(1/(1-s_r)\right)$ for $r\geq 1$.\\
Let $\textsc{ReplicOrder}(s_1,s_2,\ldots)$ be a generator of this distribution.\\[.2cm]
\hspace{.5cm}{\bf Algorithm } $~~\Gamma Z_{\Cyc}(s_1,s_2,\ldots)$ \\
\hspace{1.5cm} $r\longleftarrow$ $\textsc{ReplicOrder}(s_1,s_2,\ldots)$; \\
\hspace{1.5cm} $j\longleftarrow$ $\Loga \left( s_r \right)$; \\
\hspace{1.5cm} Draw an integer $b\in[1..r-1]$ that is relatively prime to $r$  uniformly at random ; \\
\hspace{.5cm}{\bf return} the cycle of length $j\times r$ endowed with the automorphism:\\
\hspace{1.5cm}``each atom is mapped to the atom that is $j\times b$ units further on the cycle''.\\[.2cm]
\hline
\hline
 \\[-.2cm]
\hspace{-.1cm}{\bf (1') Sequence of size $k$.}\\[.1cm]
\hspace{.5cm}{\bf Algorithm } $~~\Gamma Z_{\Seq^{[k]}}(s_1,s_2,\ldots)$~:\\
\hspace{.5cm}{\bf return} a sequence of $k$ atoms (endowed with the identity-automorphism).\\[.2cm]
\hline
\\[-.2cm]
\hspace{-.1cm}{\bf (2') Set of size $k$.}\\[.1cm]
\hspace{.5cm}{\bf Algorithm } $~~\Gamma Z_{\Set^{[k]}}(s_1,s_2,\ldots)$~:\\
\hspace{1cm} Draw a partition sequence $\pi$ of order $k$ such that
$\ds \Pr(\pi)=\frac{s_1^{n_1}s_2^{n_2}\ldots s_k^{n_k}\cdot 
[s_1^{n_1}...s_k^{n_k}]Z_{\Set^{[k]}}}{Z_{\Set^{[k]}}(s_1,s_2,\ldots)}$\\[.3cm]
\hspace{.5cm}{\bf return} a collection of $n_1$ cycles of length $1$, $n_2$ cycles of length $2$, $\ldots$, $n_k$ cycles of length $k$.\\[.2cm]
\hline
\\[-.2cm]
\hspace{-.1cm}{\bf (3') Cycle of size $k$.}\\[.1cm]
\hspace{.5cm}{\bf Algorithm } $~~\Gamma Z_{\Cyc^{[k]}}(s_1,s_2,\ldots)$~:\\
\hspace{1.5cm}Draw a divisor $r$ of $k$ with distribution $\ds\Pr(r)=\frac{\phi(r)s_r^{k/r}}{kZ_{\Cyc^{[k]}}(s_1,s_2,\ldots)}$\\
\hspace{1.5cm}Draw an integer $b\in[1..r-1]$ that is relatively to $r$  uniformly at random; \\
\hspace{.5cm}{\bf return} the cycle of length $k$ endowed with the automorphism:\\
\hspace{1.5cm}``each atom is mapped to the atom that is $kb/r$ units further on the cycle''.\\[.2cm]
\hline
 \end{tabular}

\end{flushleft}
\caption{\label{fig:basic}
P\'olya-Boltzmann samplers for basic species. In all these random samplers, the output structure is made well-labeled using the procedure \textsc{DistributeLabels}.}
\end{figure}

We also assume to have
generators for classical distributions: 
\begin{itemize}
\item $\Geom(p)$ returns an integer under the geometric law of parameter $p\in[0,1]$:
$\Pr(k)=p^{k-1}(1-p)$;
\item $\Pois(\lambda)$ returns an integer under the Poisson law of parameter $\lambda$:
$\Pr(k)=e^{-\lambda}\lambda^k/k!$; and
\item $\Loga(\lambda)$ returns an integer under the distribution
$\Pr(k)=(\log(1/(1-\lambda)))^{-1}\lambda^k/k$; this distribution we call
\emph{Loga law of parameter $\lambda$}.
\end{itemize}

The generators for those distributions 
(more generally, any generator for an explicit distribution on integers) 
can be easily obtained from the ``inversion
method''~\cite[\S2.1]{Devroye86b} and~\cite[\S4.1]{Knuth98}).

\subsubsection{P\'olya-Boltzmann samplers for basic species}
At first, let us describe P\'olya-Boltzmann samplers for the basic species $\Seq$, $\Set$, $\Cyc$,
and their counterparts $\Seqk$, $\Setk$, $\Cyck$. 
In each case, the design of the sampler is guided by the expressions of the cycle index sums for the basic species as given
in Figure~\ref{Basic}.

\begin{proposition}\label{prop:basic_samp}
The random generators shown in Figure~\ref{fig:basic} are P\'olya-Boltzmann samplers for the corresponding species.
\end{proposition}
\begin{proof}
For $\Seq$, the proof is easy. Since $Z_{\Set}=\sum_{k\geq 1}s_1^k$, the 
probability of a sequence to have size $k$ must be $s_1^k/Z_{\Set}$,
i.e., the size distribution is a geometric law of parameter $s_1$.

For $\Set$, observe that the sum of weight-monomials over all 
symmetries of type $(n_1,n_2,\dots,n_k)$
in $Z_{\Set}$ is 
$$\frac{s_1^{n_1}}{n_1!}\frac{(s_2/2)^{n_2}}{n_2!}\cdots\frac{(s_k/k)^{n_k}}{n_k!}.$$
Therefore, a P\'olya-Boltzmann sampler has 
to 
draw a collection of cycles such that the number $n_i$ 
of cycles of length $i$ follows
a  Poisson law of parameter $s_i/i$ for $i\geq 1$, and the $n_i$'s are independent.
This is precisely what the algorithm $\Gamma Z_{\Set}$ in Figure~\ref{fig:basic} does,
upon choosing a priori the size of the largest cycle to be drawn.

For $\Cyc$, the argument is similar. As we have seen in Section~\ref{sec:auto_basic},
the sum of the weight-monomials over all the symmetries of order $r$ is 
$$
Z_{\Cyc}^{(r)}:=\frac{\phi(r)}{r}\log\Big(\frac{1}{1-s_r}\Big).
$$
Therefore the order of the automorphism has to be chosen with probability $Z_{\Cyc}^{(r)}/Z_{\Cyc}$.
In addition, for each fixed order $r$, the probability of the cycle being $r\times k$ has to
be $s_r^k/k/\log(1/(1-s_r))$, i.e., the size (divided by $r$) has to follow a Loga law
of parameter $s_r$. Finally, for all automorphisms of size $r$ and size $r\times k$,
all possible `rotation angles' (there are $\phi(r)$ possibilities) have to be equiprobable.
This is exactly what the generator $\Gamma Z_{\Cyc}$ does.

The proof that the generators for the species $\Seq^{[k]}$, $\Set^{[k]}$, and $\Cyc^{[k]}$ are Polya-Boltzmann samplers
follows similar arguments upon restricting to structures of size $k$.  
\end{proof}

\subsubsection{P\'olya-Boltzmann samplers for combinatorial constructions}

As shown in Figure~\ref{fig:sampling_rules}, P\'olya-Boltzmann 
samplers  make 
 it possible to have a simple sampling rule 
 for each of the standard constructions; that is, sampling rules not only for sum and product, but also for the substitution construction.

 \begin{figure}
\begin{tabular}{|l|}
\hline
\\[-.2cm]
{\bf Sum:} $\cC=\cA+\cB$.\\
Given $(s_i)_{i\geq 1}$ such that $Z_a:=Z_{\cA}(s_1,s_2,\ldots)$ and $Z_b:=Z_{\cB}(s_1,s_2,\ldots)$ converge:\\[.2cm]
\hspace{.5cm}{\bf Algorithm} $\Gamma Z_{\cC}(s_1,s_2,\ldots)$~:\\[.1cm]
\hspace{.5cm}{\bf if} $\ds\Bern\big(Z_a/(Z_a+Z_b)\big)$ {\bf then return} $\Gamma Z_{\cA}(s_1,s_2,\ldots)$\\[.1cm]
\hspace{.5cm}{\bf else return}  $\Gamma Z_{\cB}(s_1,s_2,\ldots)$ {\bf end if}\\[.2cm]
\hline
\\[-.2cm]
 {\bf Product:} $\cC=\cA\bcdot\cB$.\\
Given $(s_i)_{i\geq 1}$ such that $Z_{\cA}(s_1,s_2,\ldots)$ and $Z_{\cB}(s_1,s_2,\ldots)$ converge:\\[.2cm]
\hspace{.5cm}{\bf Algorithm} $\Gamma Z_{\cC}(s_1,s_2,\ldots)$~:\\[.1cm]
\hspace{.5cm}{\bf return} $(\Gamma Z_{\cA}(s_1,s_2,\ldots), \Gamma Z_{\cB}(s_1,s_2,\ldots))$ \{independent calls\}\\[.2cm]
\hline
\\[-.2cm]
{\bf Substitution:} $\cC=\cA\circ\cB$.\\
Given $(s_i)_{i\geq 1}$ such that $b_i:=Z_{\cB}(s_i,s_{2i},s_{3i},\ldots)$ converges 
for each $i\geq 1$,\\ 
and $Z_{\cA}(b_1,b_2,b_3,\ldots)$ 
converges:\\[.2cm]
\hspace{.5cm}{\bf Algorithm} $\Gamma Z_{\cC}(s_1,s_2,\ldots)$~:\\
\hspace{1.5cm}Compute $(A,\sigma_A) \leftarrow \Gamma Z_{\cA}(b_1,b_2,b_3,\ldots)$;\\
\hspace{1.5cm}{\bf for each} {cycle $C=(u_1,\ldots,u_k)$ of $\sigma_A$} ($u_1$ has smallest label in $C$)  
{\bf do}\\
\hspace{2cm}Compute $(B,\sigma_B) \leftarrow \Gamma Z_{\cB}(s_k,s_{2k}\ldots)$;\\
\hspace{2cm}Replace each atom of $C$ by a copy of $B$;\\
\hspace{2cm}{\bf for each} cycle $D$ of $\sigma_B$ {\bf do}\\
\hspace{2.5cm}Let $E_D$ be the cycle composed of the copies of $D$ at $u_1,\ldots,u_k$\\
\hspace{2cm}{\bf end for}\\
\hspace{1.5cm}{\bf end for}\\
\hspace{.5cm}{\bf return} the resulting structure and the automorphism consisting of\\
\hspace{1.6cm} the composed cycles $E_D$\\[.2cm]
\hline
\end{tabular}
\caption{The rules to specify a P\'olya-Boltzmann sampler for a species that has a recursive specification. In all these random samplers, the finally
returned structure is made well-labeled using the procedure \textsc{DistributeLabels}.}
\label{fig:sampling_rules}
\end{figure}

\begin{proposition}\label{prop:cons_samp}
Let $\cC=\cA\wedge\cB$, with $\wedge\in\{+,\bcdot,\circ\}$. 
When there are P\'olya-Boltzmann samplers for $\cA$ and $\cB$, 
then there is also a P\'olya-Boltzmann sampler $\Gamma Z_{\cC}(s_1,s_2,\ldots)$ for $\cC$ that can be constructed from the samples for $\cA$ and $\cB$, as given in Figure~\ref{fig:sampling_rules}.
\end{proposition}
\begin{proof}
For $\cC=\cA+\cB$, the proof is easy. Note that $\Sym(\cA+\cB)=\Sym(\cA)+\Sym(\cB)$. As
shown in~\cite{Boltzmann} 
(for the standard weight $x^n/n!$), a disjoint union yields 
a Bernoulli switch on Boltzmann samplers, with probability 
corresponding to the ratio of the series for $\cA$ divided by the series for $\cC$ 
(this argument works as well here, where we take the refined weight $s_1^{n_1}s_2^{n_2}\ldots s_k^{n_k}/n!$).
Therefore the probability of the Bernoulli switch has to be $Z_{\cA}/Z_{\cC}$.

For product, $\cC=\cA\bcdot\cB$, $\Sym(\cC)$ is like a partitional product of $\Sym(\cA)$ and $\Sym(\cB)$.
Therefore, a Boltzmann sampler classically consists of two independent calls to Boltzmann samplers,
as shown in~\cite{Boltzmann} (again, for the standard weight $x^n/n!$).
All the arguments work the same way for the refined weight $s_1^{n_1}s_2^{n_2}\ldots s_k^{n_k}/n!$.
Therefore, one has to call independently a P\'olya-Boltzmann sampler for $\cA$ 
and a P\'olya-Boltzmann
sampler for $\cB$.

For substitution, $\cC=\cA\circ\cB$, recall (from Section~\ref{sec:aut_cons}, 
Equation~\eqref{eq:aut_cons}) 
that for each partition sequence $\pi$, the sum of the weight-monomials over all the symmetries
on $\cC$ of type $\pi=(n_1,\dots,n_n)$ satisfies the expression
$$
Z_{\cC}^{(\pi)}=\frac{a_{\pi}}{n!}b_1^{n_1}b_2^{n_2}\ldots b_k^{n_n},\ \ \mathrm{where}\ 
a_{\pi} = n![s_1^{n_1}\cdots s_n^{n_n}]Z_{\cA}(s_1,s_2,\ldots) \ \mathrm{and} \ 
b_i=Z_{\cB}(s_i,s_{2i},\ldots).
$$
Hence, a P\'olya-Boltzmann sampler for $\cC$ must draw the core structure following 
the P\'olya-Boltzmann distribution for $\cA$ with parameters $(b_1,b_2,\ldots)$.
In addition, as discussed in Section~\ref{sec:aut_cons}, once the type $\pi$ of the core symmetry is fixed, 
the structures substituted at the cycles of the core-automorphism form a partitional product of the form
$$
\Sym(\cB)^{n_1}\bcdot\{\Sym(\cB)\ \mathrm{duplicated}\}^{n_2}\bcdot(\cdots)\bcdot\{\Sym(\cB)\ \mathrm{replicated}\ k\ \mathrm{times}\}^{n_k}.
$$
Recall that a partitional product yields independent Boltzmann samplers. Hence, 
once the core-automorphism $(A,\sigma_A)$ is drawn, the symmetries in $\cB$ that are substituted
at each cycle of $\sigma_A$ must be independent calls of a P\'olya-Boltzmann sampler for $\cB$,
and the parameters of the sampler must be $(s_i,s_{2i},\ldots)$ if the replication order is $i$
($i$ is also the length of the cycle of the core structure where the substitution occurs).
This is precisely what the generator $\Gamma Z_{\cC}$ does.
\end{proof}

\subsubsection{P\'olya-Boltzmann samplers for decomposable species}

The random generation rules shown in 
Figure~\ref{fig:basic} for basic species and Figure~\ref{fig:sampling_rules}
for general constructions can be combined to design a P\'olya-Boltzmann sampler for species that are decomposable over 
those basic species. 


\begin{definition}
A P\'olya-Boltzmann sampler $\Gamma Z_{\cA}(s_1,s_2,\ldots)$ 
is called \emph{linear} if, in the real-arithmetic complexity
model described above, the number of computation steps for generating a structure from $\cA$ is 
linearly bounded by the size of the structures all along the generation process.
\end{definition}

\begin{theorem}\label{thm:dec-sampling}
Let $\cA$ be a species that is decomposable over species $\cA_1,\dots,\cA_l$  (Definition~\ref{def:dec-standard}), each having a linear P\'olya-Boltzmann sampler. Then also the species $\cA$ has a linear
P\'olya-Boltzmann sampler $\Gamma Z_{\cA}(s_1,s_2,\ldots)$.
\end{theorem}
\begin{proof}
Given Proposition~\ref{prop:cons_samp}, the P\'olya-Boltzmann sampler for $\cA$ is straightforward by `unrolling' the recursive specifications that show that $\cA$ is decomposable; so we are
left with the task to show that we also have a \emph{linear}
P\'olya-Boltzmann sampler for $\cA$.

Our argument is based on the following concept.
The \emph{decomposition tree} for structures $A$ from $\cA$ is defined as follows. The leaves of this tree are structures from
the species $\cA_1,\dots,\cA_l$. The other vertices of this tree are structures from the species $\cX^{(i)}_j$ that appear in the definition of recursive specifications (Definition~\ref{def:semantics}).
Suppose that $\cA$ is specified by variable $x_s$ in the recursive specification, and that $n$ is the size of $A$. The root of
the decomposition tree is $A$ itself, which is a structure from 
$\cX^{(n)}_s$. To define the children of a vertex $A'$ from 
$\cX^{(i)}_j$ in the decomposition tree, 
consider the species $\cX^{(i-1)}_k$ for which $x_k$ appears in the expression $e_j$ in the recursive specification.
The children are the structures of those species from which
$A$ has been constructed.
The \emph{size} of a decomposition tree is the number of its internal 
(i.e., non-leaf) vertices.

Let $A$ be a structure from $\cA$, and let $A_1, \dots, A_t$
be the leafs of the decomposition tree for $A$.
Note that the cost of the sampling procedure for producing $A$
is linearly bounded by the size of the decomposition tree for $A$ plus the total cost for sampling $A_1,\dots,A_t$.
The size of $A$ is distributed over all structures $A_1,\dots,A_t$ 
from terminal classes, consequently $|A|=|A_1|+\dots+|A_t|$. 
Since the samplers for the terminal classes are linear,
we conclude that the total cost for sampling $A_1,\dots,A_t$ 
is linear in $|A|$.

We finally show that the size
of the decomposition tree for $A$ is
linear in the size $n$ of $A$. More specifically, 
when $m$ is the number of variables  in the recursive specification of $\cA$, then the size of the decomposition tree is bounded by $2mn$.
Otherwise, there would be a branch in the decomposition tree
that contains two vertices $B_1$ from $\cX^{(n_1)}_j$ and $B_2$ from $\cX^{(n_2)}_j$
(corresponding to the same variable $x_j$) such that $B_1$ and
$B_2$ have the same size $n$. But then, by looping through the corresponding cycle in the decomposition grammar, 
we can produce infinitely many distinct structures of size $n$
from the recursive specification, contradicting the fact that
$\cX_j[n]$ is finite.
The details of this argument are easy and left to the reader.
\end{proof}

Consequently, the rooted species we have encountered (i.e., with a single marked atom)
in Section~\ref{sec:appl} can be endowed with efficient random samplers.

\begin{proposition}
In the real-arithmetic complexity model (oracle assumption), the following species
admit an exact-size sampler of expected complexity $O(n^2)$ 
($n$ being the target-size)
and an approximate-size sampler 
of expected complexity $O(n/\epsilon)$ ($\epsilon$ being the tolerance-ratio)
for the uniform distribution on \emph{unlabeled structures}: rooted nonplane (plane, resp.) trees, rooted 
nonplane (plane, resp.) trees whose node degrees are constrained to lie in a
finite integer set $\Omega$, rooted cacti graphs, rooted connected outerplanar graphs. 
\end{proposition}
\begin{proof}
All the rooted tree species stated here are decomposable over the discussed basic species, as we have seen in Section~\ref{sec:appl}.
(For instance, the species $\cF'$ of rooted nonplane trees satisfies $\cF'=\Set\circ(X\bcdot\cF')$.)
Hence, 
by Theorem~\ref{thm:dec-sampling}, the species of rooted trees can be endowed with a linear P\'olya-Boltzmann sampler.
A species $\cG'$ of rooted connected graphs -- provided that it is closed under taking 2-connected components -- is decomposable in terms of the subspecies $\cB'$ of rooted 2-connected graphs:
$$
\cG'=\Set\circ(\cB'\circ(X\bcdot\cG')).
$$ 
As we have explained in Sections~\ref{sec:cacti} and~\ref{sec:outerplanar}, 
there is a decomposition strategy
for rooted 2-connected cacti graphs (rooted polygons) and rooted 2-connected outerplanar
graphs (rooted dissections of a polygon). The linear P\'olya-Boltzmann sampler $\Gamma Z_{\cB'}$ yields
in turn a linear P\'olya-Boltzmann sampler for $\cG'$ (using the specification of $\cG'$ in
terms of $\cB'$ stated above).
Hence, each of the rooted species stated above has a linear P\'olya-Boltzmann sampler, which
becomes a linear ordinary Boltzmann sampler when specialized to $s_i=x^i$.

Moreover, all these rooted species obey the universal asymptotic form $c\ \!\rho^{-n}n^{-3/2}$,
as shown in~\cite{Otter} for trees, in~\cite{Sh07} 
for cacti graphs, and in~\cite{BFKV} for outerplanar graphs.
 Hence, by Proposition~\ref{lem:boltzmann}, 
a Boltzmann sampler run at the dominant singularity
yields an exact-size (approximate-size, respectively) sampler with expected complexity $O(n^2)$
($O(n/\epsilon)$, respectively).
\end{proof}

\subsection{P\'olya-Boltzmann samplers for cycle-pointed species}
\subsubsection{Definition}
Given a cycle-pointed species $\cP$, a vector $\vbar$ of nonnegative
real values is said to be \emph{admissible} if the sum of weight-monomials
defining $Z_{\cP}$ converges when evaluated at this vector. Given a
fixed admissible vector $\vbar$, a \emph{P\'olya-Boltzmann sampler} is
a procedure $\Gamma \oZ_{\cP}\vbar$ that generates a rooted
$c$-symmetry on $\cP$ at random
such that each rooted $c$-symmetry $(P,\sigma,v)$ of $\cR(\cP)$ is drawn with probability 
\begin{equation}\mathbb{P}(P,\sigma,v)=\frac{w_{(P,\sigma,v)}}{\oZ_{\cP}((s_i,t_i)_{i\geq 1})},\end{equation} 
with $w_{(P,\sigma,v)}$ as defined in~\eqref{eq:def_weight_cycle_pointed}.
This probability distribution is called the P\'olya-Boltzmann distribution for $\cP$ at $(s_i,t_i)_{i\geq 1}$.
Similarly as for classical species, the procedure of calling $\Gamma Z_{\cP}(x^i,x^i)_{i\geq 1}$
(where $x$ is admissible for the ordinary generating function $\tilde{\cP}(x)$) and then
returning the underlying unlabeled structure
yields an ordinary Boltzmann sampler $\Gamma \tilde{\cP}(x)$. The following sampling rules 
make it possible to systematically assemble P\'olya-Boltzmann samplers for cycle-pointed species.

\subsubsection{P\'olya-Boltzmann samplers for basic cycle-pointed species}

\begin{figure}\footnotesize
\begin{flushleft}
\begin{tabular}{|l|}
\hline
  \\[-.2cm]
\hspace{-.1cm}{\bf (1) Cycle-pointed sequence.}\\[.1cm]
\hspace{.5cm}{\bf Algorithm } $~~\Gamma \oZ_{\Seq^{\circ}}(s_1,s_2,\ldots)$~:\\
\hspace{1.5cm} $k_1\leftarrow \Geom(s_1)$; $k_2\leftarrow\Geom(s_1)$; \{indep. calls\}\\
\hspace{.5cm}{\bf return} a sequence of $k_1+k_2+1$ atoms (endowed with the identity-automorphism)\\
\hspace{1.5cm} where the atom at position $k_1+1$ is marked.\\[.2cm]
\hline
\\[-.2cm]
\hspace{-.1cm}{\bf (2) Cycle-pointed (symmetric cycle-pointed, resp.) set.}\\[.1cm]
Given $(s_i,t_i)_{i\geq 1}$ such that $\ds\sum_{i\geq 1}t_i$ converges, define the distribution:\\[.1cm]
\hspace{1cm}$\ds\Pr(K)=\frac{t_K}{\sum_{i\geq 1}t_i}$ for $K\geq 1$ \big($\ds\Pr(K)=\frac{t_K}{\sum_{i\geq 2}t_i}$ for $K\geq 2$, resp.\big).\\[.1cm]
Let $\textsc{Root\_Cycle\_Size}(t_1,t_2,\ldots)$ ($\textsc{Root\_Cycle\_Size}_{\geq 2}(t_1,t_2,\ldots)$, resp.) be a  generator for this
distribution.\\[.2cm] 
\hspace{.5cm}{\bf Algorithm } $~~\Gamma \oZ_{\Set^{\circ}}\vecc$ ($~~\Gamma \oZ_{\Set^{\cst}}\vecc$, resp.)~:\\[.1cm]
\hspace{1.5cm}$K\leftarrow \textsc{Root\_Cycle\_Size}(t_1,t_2,\ldots)$ \quad ($K\leftarrow \textsc{Root\_Cycle\_Size}_{\geq 2}(t_1,t_2,\ldots)$, resp.); \\
\hspace{1.5cm}$\gamma\leftarrow\Gamma Z_{\Set}(s_1,s_2,\ldots)$;\\
\hspace{1.5cm}Add to $\gamma$ a marked cycle of length $K$;\\
\hspace{.5cm}{\bf return} $\gamma$.\\[.2cm]
\hline
\\[-.2cm]
\hspace{-.1cm}{\bf (3) Cycle-pointed (symmetric cycle-pointed, resp.) cycle.}\\[.1cm]
Given $(s_i,t_i)_{\geq 1}$ such that $Z:=\oZ_{\Cyc^{\circ}}\vecc$ 
($Z:=\oZ_{\Cyc^{\cst}}\vecc$, resp.) converges,\\ 
consider the probability distribution\\
\hspace{1cm}$\ds\Pr(R=r)=\frac1{Z}\varphi(r)\frac{t_r}{1-s_r}$ for $r\geq 1$ ($r\geq 2$, resp.).\\
Let $\textsc{ReplicOrder}\vecc$ be a generator of this distribution.\\[.2cm]
\hspace{.5cm}{\bf Algorithm } $~~\Gamma \oZ_{\Cyc^{\circ}}\vecc$ ($~~\Gamma \oZ_{\Cyc^{\cst}}\vecc$, resp.)~: \\[.1cm]
\hspace{1.5cm} $r\longleftarrow$ $\textsc{ReplicOrder}\vecc$; \\
\hspace{1.5cm} $j\longleftarrow$ $1+\Geom \left( s_r \right)$; \\
\hspace{1.5cm} Draw an integer $b\in[1..r-1]$ that is relatively prime to $r$  uniformly at random ; \\
\hspace{.5cm}{\bf return} the cycle of length $j\times r$ 
with a marked atom, and endowed with the automorphism:\\
\hspace{1.5cm}``each atom is mapped to the atom that is $j\times b$ units further on the cycle''\\ 
\hspace{1.5cm}(the marked cycle is the automorphism-cycle containing the marked atom).\\[.2cm]
\hline
\hline
 \\[-.2cm]
\hspace{-.1cm}{\bf (1') Cycle-pointed sequence of size $k$; denote $\cE:=\Seq^{[k]}$.}\\[.1cm]
\hspace{.5cm}{\bf Algorithm } $~~\Gamma \oZ_{\cE^{\circ}}\vecc$~:\\
\hspace{.5cm}{\bf return} a sequence of $k$ atoms (with the identity-automorphism)
 where one atom taken u.a.r. is marked\\[.2cm]
\hline
\\[-.2cm]
\hspace{-.1cm}{\bf (2') Cycle-pointed (symmetric cycle-pointed, resp.) set of size $k$; denote
$\cS:=\Set^{[k]}$.}\\[.1cm]
A marked (marked symmetric, resp.) 
partition sequence of order $k$ is a sequence $\overline{\pi}=(\ell,n_1,n_2,\ldots,n_k)$\\ 
such that $\ell\geq 1$ ($\ell\geq 2$, resp.) and $\ell+\sum_iin_i=k$ (one block is marked).\\
The corresponding coefficient $[t^{\ell}s_1^{n_1}\ldots s_k^{n_k}]$ is denoted 
$\mathrm{coef}_{\overline{\pi}}(\oZ_{\cS^{\circ}})$ ($\mathrm{coef}_{\overline{\pi}}(\oZ_{\cS^{\cst}})$, resp.)\\[.1cm]
\hspace{.5cm}{\bf Algorithm } $~~\Gamma \oZ_{\cS^{\circ}}\vecc$ ($~~\Gamma \oZ_{\cS^{\cst}}\vecc$, resp.)~:\\
\hspace{1.5cm}Draw a partition-sequence $\overline{\pi}$ of order $k$ with probability:\\ 
\hspace{1.5cm}$\ds \Pr(\overline{\pi})=\frac{t_{\ell}s_1^{n_1}s_2^{n_2}\ldots s_k^{n_k}\cdot \mathrm{coef}_{\overline{\pi}}(Z_{\cS^{\circ}})}{Z_{\cS^{\circ}}\vecc}$ \Big($\ds \Pr(\overline{\pi})=\frac{t_{\ell}s_1^{n_1}s_2^{n_2}\ldots s_k^{n_k}\cdot \mathrm{coef}_{\overline{\pi}}(Z_{\cS^{\cst}})}{Z_{\cS^{\cst}}\vecc}$, resp.\Big)\\[.3cm]
\hspace{.5cm}{\bf return} a collection of $n_1$ cycles of length $1$, $n_2$ cycles of length $2$, $\ldots$, $n_k$ cycles of length $k$,\\
\hspace{1.7cm}together with a marked cycle of length $\ell$ (with a marked atom on it taken u.a.r.).\\[.2cm]
\hline
\\[-.2cm]
\hspace{-.1cm}{\bf (3') Cycle-pointed (symmetric cycle-pointed, resp.) cycle of size $k$; denote $\cC:=\Cyc^{[k]}$.}\\[.1cm]
\hspace{.5cm}{\bf Algorithm } $~~\Gamma \oZ_{\cC^{\circ}}\vecc$ ($~~\Gamma \oZ_{\cC^{\cst}}\vecc$, resp.)~:\\
\hspace{1.5cm}Draw a divisor $r$ of $k$ (divisor $r\geq 2$ of $k$) with distribution\\
\hspace{1.5cm}$\ds\Pr(r)=\frac{\phi(r)t_rs_r^{-1+k/r}}{\oZ_{\cC^\circ}\vecc}$ \Big($\ds\Pr(r)=\frac{\phi(r)t_rs_r^{-1+k/r}}{\oZ_{\cC^{\cst}}\vecc}$, resp.\Big)\\
\hspace{1.5cm}Draw an integer $b\in[1..r-1]$ that is relatively prime to $r$  uniformly at random; \\
\hspace{.5cm}{\bf return} the cycle of length $k$ with a marked atom and endowed with the automorphism:\\
\hspace{1.5cm}``each atom is mapped to the atom that is $kb/r$ units further on the cycle''.\\[.2cm]
\hline
 \end{tabular}
 \end{flushleft}
\caption{\label{fig:basic-cyc}
P\'olya-Boltzmann sampler for basic cycle-pointed species. 
In all these random samplers, the finally
returned structure is made well-labeled using the procedure \textsc{DistributeLabels}.}
\end{figure}

\begin{figure}
\begin{tabular}{|l|}
\hline
\\[-.2cm]
{\bf Cycle-pointed sum:} $\cR=\cP+\cQ$.\\
Given $(s_i,t_i)_{i\geq 1}$ such that $\oZ_p:=\oZ_{\cP}\vecc$ and 
$\oZ_q:=\oZ_{\cQ}\vecc$ converge:\\[.2cm]
\hspace{.5cm}{\bf Algorithm} $\Gamma \oZ_{\cR}\vecc$~:\\[.1cm]
\hspace{1.5cm}{\bf if} $\ds\Bern\big(\oZ_p/(\oZ_p+\oZ_q)\big)$ {\bf then return} $\Gamma \oZ_{\cP}\vecc$\\[.1cm]
\hspace{1.5cm}{\bf else return} $\Gamma \oZ_{\cQ}\vecc$ {\bf end if}\\[.1cm]
\hline
\\[-.2cm]
 {\bf Cycle-pointed product:} $\cR=\cP\star\cB$ (analog for $\cR=\cB\star\cP$)\\
Given $(s_i,t_i)_{i\geq 1}$ such that $Z_{\cB}(s_1,s_2,\ldots)$ and $\oZ_{\cP}\vecc$  converge:\\[.2cm]
\hspace{.5cm}{\bf Algorithm} $\Gamma \oZ_{\cR}\vecc$~:\\[.1cm]
\hspace{.5cm}{\bf return} $(\Gamma \oZ_{\cP}\vecc, \Gamma Z_{\cB}(s_1,s_2,\ldots))$ \{independent\ calls\}\\[.2cm]
\hline
\\[-.2cm]
{\bf Cycle-pointed substitution:} $\cR=\cP\circledcirc\cB$.\\
Given $(s_i,t_i)_{i\geq 1}$ such that $b_i:=Z_{\cB}(s_i,s_{2i},s_{3i},\ldots)$
and $q_i:=\oZ_{\cB^{\circ}}(s_i,t_i;s_{2i},t_{2i};\ldots)$ converge\\
for each $i\geq 1$, 
and such that $\oZ_{\cP}(b_1,q_1;b_2,q_2;\ldots)$ 
 converges:\\[.2cm]
\hspace{.5cm}{\bf Algorithm} $\Gamma \oZ_{\cR}\vecc $~:\\
\hspace{1.5cm}Compute $(P,\sigma_P,v)\leftarrow\Gamma \oZ_{\cP}(b_1,q_1;b_2,q_2;\ldots)$;\\
\hspace{1.5cm}{\bf for each} {unmarked cycle $C=(u_1,\ldots,u_k)$ of $\sigma_P$} ($u_1$ has smallest label in $C$)\\
\hspace{2cm}Compute 
$(B,\sigma_B) \leftarrow \Gamma Z_{\cB}(s_k,s_{2k},\ldots)$;\\
\hspace{2cm}Replace each atom of $C$ by a copy of $B$;\\
\hspace{2cm}{\bf for each} {cycle $D$ of $\sigma_B$} {\bf do}\\
\hspace{2.5cm}Let $E$ be the cycle composed from the copies of $D$ at $u_1,\ldots,u_k$;\\
\hspace{2cm}{\bf end for}\\
\hspace{1.5cm}{\bf end for}\\
\hspace{1.5cm}Let $F\!=\!(v_1,\ldots,v_{\ell})$ be the marked cycle of $P$ ($v_1$ has smallest label in $F$);\\
\hspace{1.5cm}Compute $(Q,\sigma_Q,q) \leftarrow \Gamma \oZ_{\cBp}(s_{\ell},t_{\ell};s_{2\ell},t_{2\ell};\ldots)$;\\
\hspace{1.5cm}Replace each atom of $F$ by a copy of $Q$;\\
\hspace{1.5cm}{\bf for each} {cycle $G$ of $\sigma_Q$} {\bf do}\\
\hspace{2cm}Let $H$ be the cycle composed from the copies of $G$ at $v_1,\ldots,v_{\ell}$;\\
\hspace{1.5cm}{\bf end for}\\
\hspace{1.5cm}In the resulting structure $R$, mark the cycle composed from the copies of\\ 
\hspace{1.5cm}the marked cycle of $Q$;\\
\hspace{.5cm}{\bf return} $(R, \sigma_R, r)$, where $\sigma_R$ is the automorphism consisting of the cycles $E$\\ 
\hspace{1.8cm}and the cycles $H$, and where $r$ is
the atom $q$ in the copy of the marked cycle\\ 
\hspace{1.7cm}
 of $Q$  substituted at $v$.\\[.2cm]
\hline
\end{tabular}
\caption{The rules to specify a P\'olya-Boltzmann sampler for a cycle-pointed species assembled from other species using the cycle-pointed constructions $\{+,\star,\circledcirc\}$.
 In all these random samplers, the 
output structure is made well-labeled using the procedure \textsc{DistributeLabels}.}
\label{fig:sampling_rules_cyc}
\end{figure}


\begin{proposition}\label{prop:samp_rules_cyc}
The random generators shown in Figure~\ref{fig:basic-cyc} are P\'olya-Boltzmann samplers for the corresponding basic cycle-pointed species.
\end{proposition} 
\begin{proof}
The arguments are very similar to the ones in the proof of Proposition~\ref{prop:basic_samp}.
Observe that $\Seq^{\circ}= (X^{\circ}\star\Seq) \star \Seq$.
The marked atom (the atom bearing the marked cycle, which has length 1 here)
must be preceded by a sequence of $k_1$ atoms and followed by a sequence of $k_2$ atoms such
that $k_1$ and $k_2$ follow independently a geometric law of parameter $s_1$.

Next, we have $\Set^{\circ}_{(\ell)}=\cC_{(\ell)}\star\Set$ where $\cC_{(\ell)}$ is the cycle-pointed species of cycles
of $\ell$ atoms (the cycle being marked), which explains
 the samplers $\Gamma\oZ_{\Set^{\circ}}$ and $\Gamma\oZ_{\Set^{\cst}}$ in Figure~\ref{fig:basic-cyc}.

A cycle-pointed structure in $\Cyc^{\circ}_{\ell}$ consists of $\ell$ isomorphic copies (attached cyclically in a chain) of an object in $X^{\circ}\star\Seq$.
Additionally, one needs to specify the shift of the automorphism; if the cycle has length $n\ell$, the possible shifts are $n\cdot i$ where $i\in[1..\ell]$ is relatively prime to $\ell$, hence there are $\phi(\ell)$ possibilities for the shift.


The proofs for the samplers with $k$ components follow similar arguments.
\end{proof}

\subsubsection{P\'olya-Boltzmann samplers for cycle-pointed constructions}

\begin{proposition}
Let $\cR$ be a species with a recursive specification 
over other species having a P\'olya-Boltzmann sampler. 
Then the random sampler $\Gamma \oZ_{\cR}\vecc$, as 
 given in Figure~\ref{fig:sampling_rules_cyc}, is a P\'olya-Boltzmann sampler for $\cR$.
\end{proposition}
\begin{proof}
The arguments are very similar to the ones in the proof
of Proposition~\ref{prop:cons_samp}.
For the cycle-pointed sum, $\cR=\cP+\cQ$, we have $\RSym(\cR)=\RSym(\cP)+\RSym(\cQ)$.
Therefore the P\'olya-Boltzmann sampler has to be a Bernoulli switch with probability
$\oZ_{\cP}/\oZ_{\cR}$ followed by a call to the P\'olya-Boltzmann sampler of either $\cP$
or $\cQ$ (depending on the Bernoulli ouput to be ``true'' or ``false'').
(We work here with the refined weight  $t^{\ell}s_1^{n_1}s_2^{n_2}\ldots s_k^{n_k}/n!$
instead of the standard weight $x^n/n!$, but the arguments given in~\cite{Boltzmann}
work the same way with these refined weights.)

For the cycle-pointed product, $\cR=\cP\star\cB$, $\RSym(\cR)$ is like a partitional product of $\RSym(\cP)$ and $\Sym(\cB)$.
Therefore, a Boltzmann sampler for $\cR$ consists of two independent calls to P\'olya-Boltzmann samplers for
$\cP$ and for $\cB$. (Again the only difference here with~\cite{Boltzmann} 
is that we consider refined weights:
$t^{\ell}s_1^{n_1}s_2^{n_2}\ldots s_k^{n_k}/n!$ for $\cP$, and $s_1^{n_1}s_2^{n_2}\ldots s_k^{n_k}/n!$ for $\cB$.)
The arguments are the same for $\cB\star\cP$.

For cycle-pointed substitution, $\cR=\cP\circledcirc\cB$, 
recall (from Equation~\eqref{eq:comp_cyc_subs} in the proof of Proposition~\ref{prop:comp_cyc}) 
that for each marked integer partition $\opi$, 
the sum of the weight-monomials over all the symmetries
on $\cR$ whose core has type $\opi=(\ell;n_1,n_2,\ldots,n_k)$ 
satisfies the expression
$$
\oZ_{\cR}^{(\opi)}=\frac{a_{\opi}}{n!}q_{\ell}b_1^{n_1}b_2^{n_2}\ldots b_k^{n_k},\ \ \mathrm{where}\ 
b_i=Z_{\cB}(s_i,s_{2i},\ldots) , \  
q_{\ell}=\oZ_{\cBp}(s_{\ell},t_{\ell};s_{2\ell},t_{2\ell};\ldots).
$$
Hence, a P\'olya-Boltzmann sampler for $\cR$ must draw the core structure following 
the P\'olya-Boltzmann distribution for $\cP$ with parameters $(b_1,q_1;b_2,q_2;\ldots)$.
In addition, once the type $\opi$ of the core symmetry is fixed, 
the structures substituted at the cycles of the core automorphism form a partitional product of the form
{\small
$$
\{\RSym(\cBp)\ \mathrm{replicated}\ \ell\ \mathrm{times}\}\star\Sym(\cB)^{n_1}\star\{\Sym(\cB)\ \mathrm{duplicated}\}^{n_2}\star\cdots\star\{\Sym(\cB)\ \mathrm{replicated}\ k\ \mathrm{times}\}^{n_k}.
$$
}
Recall that a partitional product yields independent Boltzmann samplers.
Hence, once the core symmetry $(P,\sigma)$ is drawn, the symmetries in $\cB$ that are substituted
at each cycle of $\sigma_A$ must be independent calls of a P\'olya-Boltzmann sampler for $\cB$, except
for the marked cycle where we have to call a P\'olya-Boltzmann sampler for $\cB^{\circ}$.
In addition, for an unmarked (marked, resp.) cycle,
the parameters of $\Gamma Z_{\cB}$ (of $\Gamma\oZ_{\cBp}$, resp.) 
must be $(s_i,s_{2i},\ldots)$ ($(s_i,t_i;s_{2i},t_{2i};\ldots)$, resp.) if the cycle has length $i$, as indicated
by the expression of $\oZ_{\cR}^{(\pi)}$ given above.
This is precisely what the generator $\Gamma \oZ_{\cR}$ does.
\end{proof}

\subsubsection{P\'olya-Boltzmann samplers for decomposable cycle-pointed species}
Similarly as for decomposable species,
the random generation rules shown in Figure~\ref{fig:basic-cyc} 
(basic cycle-pointed species) and Figure~\ref{fig:sampling_rules_cyc}
(cycle-pointed constructions) can be combined to design a P\'olya-Boltzmann sampler
for any species with a \emph{cycle-pointed} recursive decomposition
over basic species.
We assume here again that an oracle
provides the required evaluations of cycle-index sums
and pointed cycle-index sums for the species appearing in the decomposition;
and that the cost of drawing $k$ under a specific integer distribution (such as
\textsc{ReplicOrder} in $\Gamma Z_{\Cyc}$) has linear cost in $k$.


\begin{theorem}
Any species $\cP$ with a cycle-pointed recursive specification (Definition~\ref{def:dec_cycle}) over species $\cA_1,\dots,\cA_l$ having a linear P\'olya-Boltzmann sampler can be endowed with a
linear P\'olya-Boltzmann sampler $\Gamma Z_{\cP}\vecc$.
\end{theorem}
\begin{proof}
Analogous to the proof of Theorem~\ref{thm:dec-sampling}.
\end{proof}

Consequently, the \emph{unrooted species} we have encountered 
in Section~\ref{sec:appl} can be endowed with efficient random samplers.

\begin{proposition}
In the real-arithmetic complexity model (oracle assumption), the following \emph{unlabeled} species 
admit an exact-size sampler and an approximate-size sampler 
of expected complexities $O(n^2)$ and $O(n/\epsilon)$ ($n$ being the target-size,
$\epsilon$ the tolerance-ratio): unrooted nonplane (plane, resp.) trees, unrooted 
nonplane (plane, resp.) trees 
whose node degrees are constrained to lie in a
finite integer set $\Omega$, unrooted cacti graphs, unrooted connected outerplanar graphs. 
\end{proposition}

\begin{proof}
The crucial point is that cycle-pointing is unbiased, hence finding 
an exact-size (approx\-i\-mate-size, resp.) sampler for a species $\cA$
is equivalent to finding one for the cycle-pointed species $\cAp$.

For each of the unrooted tree species listed above, we have shown in Section~\ref{sec:appl}
that the corresponding cycle-pointed species is decomposable. If $\cG$ is a species of connected
graphs (closed under taking 2-connected components), the grammar~\eqref{eq:dec_graphs} 
given in Proposition~\ref{prop:gramm_graph}
ensures that the cycle-pointed species $\cG^{\circ}$ is decomposed in terms of the 
2-connected graph species $\cB^{\circ}$, $\cB'$, and $(\cB')^{\circ}$. For cacti graphs and outerplanar graphs,
there is a decomposition strategy for the 2-connected subspecies (polygons for cacti graphs, dissections of a polygon
for outerplanar graphs), which easily yields linear P\'olya-Boltzmann samplers for 
the species $\cB^{\circ}$, $\cB'$, and $(\cB')^{\circ}$. Since $\cG^{\circ}$ is specified over these three species,
there is also a linear P\'olya-Boltzmann sampler for $\cG^{\circ}$.

Hence, for each unrooted species $\cA$ stated above,
there is a linear P\'olya-Boltzmann sampler for $\cA^{\circ}$, which
becomes a linear ordinary Boltzmann sampler when specializing to $(s_i=x^i,t_i=x^i)$.
Moreover, the counting coefficients $|\wt{\cA_n}|$ obey the asymptotic form $c\ \!\rho^{-n}n^{-5/2}$,
as shown in~\cite{Otter} for trees,~\cite{Sh07} for cacti graphs, and~\cite{BFKV} for outerplanar graphs.
Therefore the coefficients $|\wt{\cA^{\circ}_n}|$ obey the asymptotic form $c\ \!\rho^{-n}n^{-3/2}$.

Hence, by Proposition~\ref{lem:boltzmann}, 
running the Boltzmann sampler $\Gamma\oZ_{\cA^{\circ}}$ at $(s_i=\rho^i,t_i=\rho^i)$ -- with 
$\rho$ the dominant singularity of $\wt{\cA}(x)$ -- yields 
an exact-size (approximate-size, resp.) sampler for unlabeled structures from $\cA$ with expected complexity $O(n^2)$
($O(n/\epsilon)$, respectively).
\end{proof}

\subsection{Simplifications}\label{sec:simp}
\subsubsection{Regarding the labels}\label{sec:simp_labels}
Notice that the last step of  all the random generation rules -- as 
written in Figure~\ref{fig:basic},~\ref{fig:sampling_rules},~\ref{fig:basic-cyc}, and~\ref{fig:sampling_rules_cyc} -- is a call to a procedure
that distributes distinct labels uniformly 
at random on the atoms of the output structure.
A P\'olya-Boltzmann sampler for a decomposable species calls each of these random 
generation rules a certain number of times until the structure is completely build.
It is however not necessary to call the label-distribution procedure at each step;
one can just wait until the complete structure is build to distribute the labels.
(If one is only interested in the underlying unlabeled structure of the final output, one can simply forget about any label-distribution procedure.)

\subsubsection{Regarding the cycles in automorphisms}\label{sec:simp_auto}
We want to remark that one can simplify
the random generation rules by forgetting about the cyclic order
of the atoms in each part of the symmetries (automorphisms),
i.e., storing only the partition of the atoms induced by the automorphism.
The composition of cycles then corresponds to merging
 the corresponding parts of the partition;
and the line ``draw an integer that is relatively prime to $r$'' is deleted in the 
P\'olya-Boltzmann samplers for cycle-pointed species.
These simplified random samplers 
output the \emph{profile} of a symmetry drawn under the P\'olya-Boltzmann 
distribution (by profile we mean that the cyclic order of the atoms in each part is forgotten).
This is actually 
enough if one is just interested in the underlying unlabeled structure, which is mostly the case in practice,
since P\'olya-Boltzmann samplers are used as a tool to design ordinary Boltzmann samplers. 

\subsubsection{Specialization to ordinary Boltzmann samplers}\label{sec:simp_spec}
Let $\cF$ be a species 
with a recursive decomposition (cycle-pointed or not) over
finitely many other species.
Remark~\ref{remark:OGS} and Remark~\ref{remark:OGS_point} 
ensure that, to compute $\wt{\cF}(x)$, 
the only species in the recursive specification
for which it is necessary to
know the 
cycle-index sum are the core species of the substitution operations in the recursive specification.
For all the other species that appear in the recursive specification, it is enough to compute just
the ordinary generating function.

A similar remark holds for random generation.
Namely, if one wants an ordinary Boltzmann sampler $\Gamma\wt{\cF}(x)$,
it suffices to have P\'olya-Boltzmann samplers 
for the species that appear as core species
of substitution operations in the recursive specification.
For all the other species, an ordinary
Boltzmann sampler is enough.

\begin{figure}
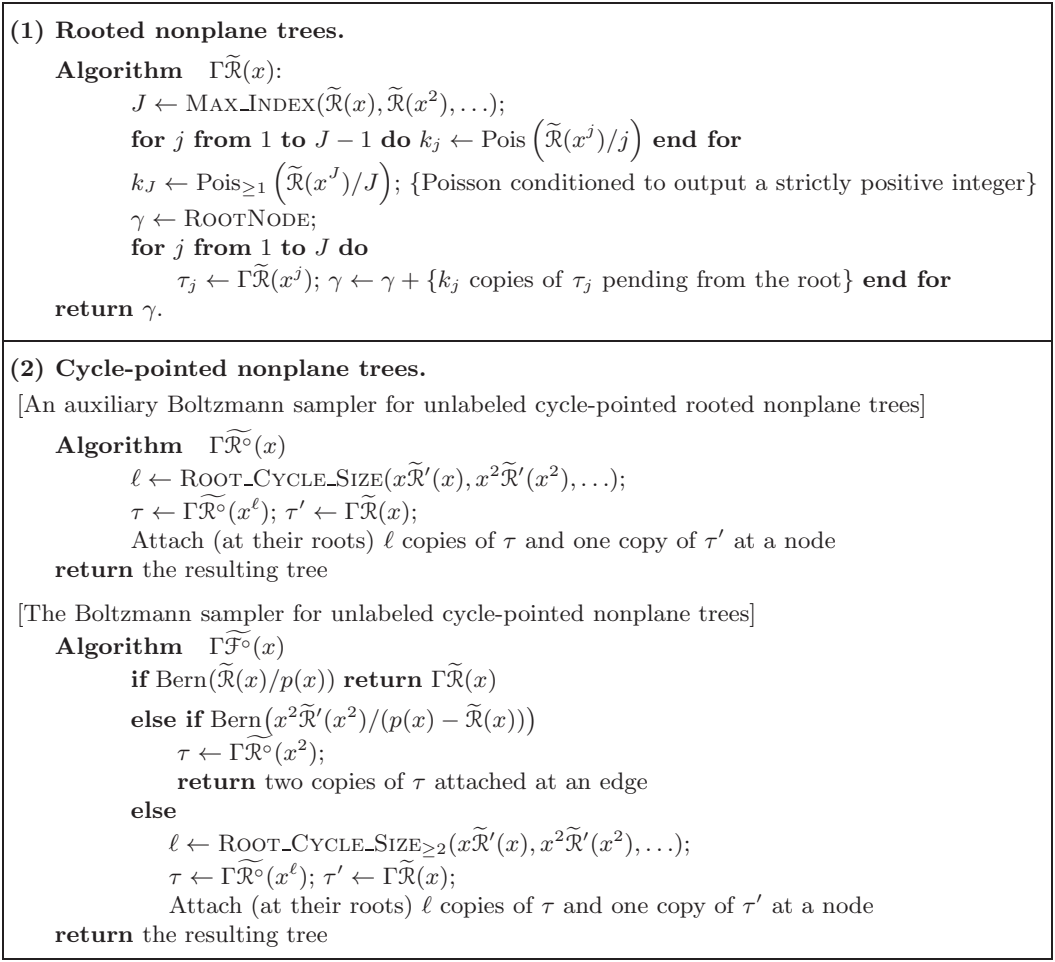
\small
\begin{flushleft}
\phantom{1}\hspace{.5cm}\begin{tabular}{|l|}
\hline
  \\[-.2cm]
\hspace{-.1cm}{\bf (1) Rooted nonplane trees.}\\[.1cm]
\hspace{.5cm}{\bf Algorithm } $~~\Gamma \wt{\cR}(x)$:\\
\hspace{1.5cm}$J\leftarrow \textsc{Max\_Index}(\wt{\cR}(x),\wt\cR(x^2),\ldots)$;\\
\hspace{1.5cm}{\bf for} $j$ {\bf from} $1$ {\bf to} $J-1$ {\bf do} $\ds k_j \leftarrow \Pois\left(\wt{\cR}(x^j)/j\right)$ {\bf end for}\\
\hspace{1.5cm}$\ds k_J \leftarrow \Pois_{\geq 1}\left(\wt{\cR}(x^J)/J\right)$; \{Poisson conditioned to output a 
strictly positive integer\}\\
\hspace{1.5cm}$\gamma\leftarrow\textsc{RootNode}$;\\
\hspace{1.5cm}{\bf for} $j$ {\bf from} $1$ {\bf to} $J$ {\bf do}\\
\hspace{2cm} $\tau_j\leftarrow\Gamma\wt{\cR}(x^j)$;\ 
 $\gamma\leftarrow\gamma+\{k_j\ \mathrm{copies\ of}\ \tau_j\ \mathrm{pending\ from\ the\ root}\}$ {\bf end for}\\
\hspace{.5cm}{\bf return} $\gamma$.\\[.2cm]
\hline
\\[-.2cm]
\hspace{-.1cm}{\bf (2) Cycle-pointed nonplane trees.}\\[.1cm]
[An auxiliary Boltzmann sampler for unlabeled cycle-pointed rooted nonplane trees]\\[.1cm]
\hspace{.5cm}{\bf Algorithm } $~~\Gamma \wt{\cRp}(x)$ \\
\hspace{1.5cm}$\ell\leftarrow\textsc{Root\_Cycle\_Size}(x\wt{\cR}'(x),x^2\wt{\cR}'(x^2),\ldots)$;\\
\hspace{1.5cm}$\tau\leftarrow\Gamma\wt{\cRp}(x^{\ell})$; $\tau'\leftarrow\Gamma\wt{\cR}(x)$;\\
\hspace{1.5cm}Attach (at their roots) $\ell$ copies of $\tau$ and one copy of $\tau'$ at a node\\
\hspace{.5cm}{\bf return} the resulting tree\\[.2cm]

[The Boltzmann sampler for unlabeled cycle-pointed nonplane trees] \\
\hspace{.5cm}{\bf Algorithm } $~~\Gamma \wt{\cFp}(x)$ \\
\hspace{1.5cm}{\bf if} $\Bern(\wt{\cR}(x)/p(x))$ {\bf return} $\Gamma\wt{\cR}(x)$\\[.1cm]
\hspace{1.5cm}{\bf else if} $\Bern\big(x^2\wt{\cR}'(x^2)/(p(x)-\wt{\cR}(x))\big)$\\
\hspace{2cm} $\tau\leftarrow\Gamma\wt{\cRp}(x^2)$; \\
\hspace{2cm} {\bf return} two copies of $\tau$ attached at an edge\\
\hspace{1.5cm}{\bf else}\\
\hspace{2cm}$\ell\leftarrow\textsc{Root\_Cycle\_Size}_{\geq 2}(x\wt{\cR}'(x),x^2\wt{\cR}'(x^2),\ldots)$;\\
\hspace{2cm}$\tau\leftarrow\Gamma\wt{\cRp}(x^{\ell})$; $\tau'\leftarrow\Gamma\wt{\cR}(x)$;\\
\hspace{2cm}Attach (at their roots) $\ell$ copies of $\tau$ and one copy of $\tau'$ at a node\\
\hspace{.5cm}{\bf return} the resulting tree\\[.1cm]
\hline
\end{tabular}
\end{flushleft}
\caption{Ordinary Boltzmann samplers for rooted nonplane trees and
cycle-pointed nonplane trees. Taking $x=\rho$ the singularity of $\wt{\cR}(x)$,
we repeatedly call $\Gamma\wt{\cR}(x)$ ($\Gamma\wt{\cFp}(x)$, respectively) 
until we reach the target size; this yields an exact-size sampler of expected complexity
$O(n^2)$ and an approximate-size sampler of complexity $O(n/\epsilon)$ for 
rooted nonplane trees (for free trees, respectively).}
\label{fig:some_samplers}
\end{figure}

\subsection{Examples}
To illustrate how our samplers (and the simplifications discussed above) operate in practice, 
we give here two examples: the species $\cR$ of
rooted nonplane trees, already treated in~\cite{BoltzmannUnlabeled},
and the species $\cF$ of 
unrooted nonplane trees, also called free trees; here, one has to consider
 the associated cycle-pointed species, which is new.

In order to obtain uniform random samplers for unlabeled structures
from $\cR$, we design an ordinary
Boltzmann sampler for $\cR$ via P\'olya-Boltzmann samplers. Recall
that $\cR$ is specified by $\cR=X\bcdot\Set(\cR)$. Hence, according 
to Section~\ref{sec:simp_spec}, we need a P\'olya-Boltzmann sampler for $\Set$
in order to recursively specify an ordinary Boltzmann sampler $\Gamma\wt{\cR}(x)$. Actually, according 
to the discussion in Section~\ref{sec:simp_auto}, 
we just need the profile of the automorphism computed by $\Gamma Z_{\Set}$.
Given these simplifications and the definition of $\Gamma Z_{\Set}$ (Figure~\ref{fig:basic}),
we obtain the ordinary Boltzmann sampler $\Gamma\wt{\cR}(x)$ shown in Figure~\ref{fig:some_samplers}.

Let us now discuss the species of free trees $\cF$. 
We want to sample unlabeled structures from $\cF$ uniformly at random.
Since cycle-pointing is unbiased, 
it suffices to sample unlabeled structures from the cycle-pointed species $\cFp$. 
From the decomposition grammar for $\cFp$ given in Proposition~\ref{prop:gram_trees},
we can design an ordinary Boltzmann sampler via
P\'olya-Boltzmann samplers. According to Section~\ref{sec:simp_spec},
the only species for which we need P\'olya-Boltzmann samplers
(as refinements of ordinary Boltzmann sampers ) are the species $\Set$,
$\Set^{\circ}$, and $\Set^{\cst}$. Again, only the profile of the 
automorphisms returned by these samplers is necessary. 
Given the definition of $\Gamma Z_{\Set}$, $\Gamma\oZ_{\Set^{\circ}}$, 
and $\Gamma\oZ_{\Set^{\cst}}$  (Figure~\ref{fig:basic}), we obtain the ordinary Boltzmann 
sampler $\Gamma\wt{\cFp}$ for unlabeled structures shown in Figure~\ref{fig:some_samplers}. 

Note that in these Boltzmann samplers, the only evaluations required are
those of the series $p(x)$, and of $r(x^i)$ and $x^ir'(x^i)$ for any $i\geq 1$.
As discussed in Section~\ref{sec:comp_model}, in practice we evaluate 
these series up to the power $i=N/\log_2(1/\rho)$, where $N$ is the precision
(number of bits). 

\bigskip
\noindent\emph{Acknowledgements.} Omid Amini, Olivier Bodini, Philippe Flajolet, and Pierre Leroux are greatly thanked
for fruitful discussions and suggestions.
Further, we thank two anonymous referees for their helpful comments.

\end{document}